\documentclass[format=acmsmall, nonacm]{acmart}
\pdfoutput=1
\usepackage{blindtext, graphicx}
\usepackage{csquotes}
\usepackage{soul}
\usepackage{enumerate}
\usepackage{diagbox}
\usepackage{multirow}
\usepackage{listings}

\newif\iffull
\fullfalse

\usepackage{mathtools}
\usepackage{bbold}
\newcommand{\RNN}{\operatorname{RNN}}

\usepackage{makecell}

\begin{document}

\title[Deep Learning for Source Code Modeling and Generation]{Deep Learning for Source Code Modeling and Generation: Models, Applications and Challenges}

\author{Triet H. M. Le}
\affiliation{\institution{The University of Adelaide}}
\email{triet.h.le@adelaide.edu.au}

\author{Hao Chen}
\affiliation{\institution{The University of Adelaide}}
\email{hao.chen01@adelaide.edu.au}

\author{Muhammad Ali Babar}
\affiliation{\institution{The University of Adelaide}}
\email{ali.babar@adelaide.edu.au}

\begin{abstract}
Deep Learning (DL) techniques for Natural Language Processing have been evolving remarkably fast. Recently, the DL advances in language modeling, machine translation and paragraph understanding are so prominent that the potential of DL in Software Engineering cannot be overlooked, especially in the field of program learning. To facilitate further research and applications of DL in this field, we provide a comprehensive review to categorize and investigate existing DL methods for source code modeling and generation. To address the limitations of the traditional source code models, we formulate common program learning tasks under an encoder-decoder framework. After that, we introduce recent DL mechanisms suitable to solve such problems. Then, we present the state-of-the-art practices and discuss their challenges with some recommendations for practitioners and researchers as well.
\end{abstract}



\maketitle

\lstset{language=Java,keywordstyle={\bfseries}}

\section{Introduction}
\label{sec:introduction}

Deep Learning (DL) has recently emerged as an important branch of Machine Learning (ML) because of its incredible performance in Computer Vision and Natural Language Processing (NLP)~\cite{goodfellow2016deep}. In the field of NLP, it has been shown that DL models can greatly improve the performance of many classic NLP tasks such as semantic role labeling~\cite{he2017deep}, named entity recognition~\cite{peters2017semi}, machine translation~\cite{vaswani2017attention}, and question answering~\cite{mikolov2015roadmap}. Similarly, source code is a special type of structured natural language written by programmers~\cite{hindle2012naturalness}, which can be analyzed by DL models. Machine intelligence that understands and creates complex structure of software has a lot of applications in Software Engineering (SE). Specifically, Big Code is the research area that uses ML and DL for source code modeling\footnote{http://science.dodlive.mil/2014/03/21/darpas-muse-mining-big-code/}. To facilitate the building of ML models, software community offers valuable datasets such as online code corpora like GitHub, question answering forums like Stack Overflow as well as documentation of various software and programming tools that are highly structured and rich in content. There have been many applications of ML in SE thesaurus construction~\cite{chen2017unsupervised}, language models for code~\cite{dam2016deep} and information retrieval of answers, projects and documentation~\cite{xia2015should,ye2016learning,chen2016techland}. Compared to other domains, DL for Big Code is still growing and requiring more research, tasks and well-established datasets.

Among the Big Code tasks, source code generation is an important field to predict explicit code or program structure from multimodal data sources such as incomplete code, programs in another programming language, natural language descriptions or execution examples~\cite{allamanis2017survey}. Code generation tools can assist the development of automatic programming tools to improve programming productivity. Such models can also represent the temporal context and discrete logic of programs. However, traditional source code models were either inflexible~\cite{bielik2016phog, raychev2016probabilistic}, time-consuming to design for specific languages/tasks~\cite{gulwani2010dimensions, gvero2013complete}, or unable to capture long-term dependencies of code statements~\cite{code_completion_ngrams, allamanis2015bimodal}. Listings~\mbox{\ref{listing:code_ex1}} and~\mbox{\ref{listing:code_ex2}} show that the difference of buggy and clean versions of the same code snippet depends on the relative location of the \mbox{\lstinline{close(file)}} function with respect to the \mbox{\lstinline{open(file)}} function, which is an example of long-term dependency in source code. DL models can help address the aforementioned issues because of their superior ability in extracting features from various data formats (e.g., natural language and symbol sequences) and capturing both syntactic and semantic information at various scales~\cite{he2017deep}.

\begin{lstlisting}[language=Java, caption=A buggy code snippet which involves long-term dependencies., label=listing:code_ex1]
try {
    open(file);
    // Do something that can raise exception
    close(file); // Error occurs when the exception is raised.
}
catch (Exception e) {
    // Handle exception
}
\end{lstlisting}

\begin{lstlisting}[language=Java, caption=A clean code snippet which involves long-term dependencies., label=listing:code_ex2]
try {
    open(file);
    // Do something that can raise exception
}
catch (Exception e) {
    // Handle exception
}
finally {
    close(file); // Always close the file, which fixes the bug.
}
\end{lstlisting}

There is an extensive review on ML for Big Code focusing on probabilistic models~\cite{allamanis2017survey}. Compared to this previous work, here are our major differences:
\begin{itemize}
  \item Extensive coverage of state-of-the-art DL models and their extension to source code modeling and generation
  \item A systematic mapping of Big Code tasks based on their inputs (encoder) and outputs (decoder) for applying DL models
  \item List of datasets for source code modeling and generation tasks
  \item Challenges and future directions for deep source code models
\end{itemize}

With such significant differences, our paper gives a more holistic view of applying DL for source code modeling and generation as compared to~\mbox{\cite{allamanis2017survey}}. Recently, there has also been a review ~\cite{pouyanfar2018survey} on DL architectures and its applications in various domains such as computer vision, NLP, and social network analysis. Unlike the previous work, our paper focuses more on many practical applications in source code modeling and generation, and then illustrates how DL models and encoder-decoder framework can be used to address such problems. This work will be a useful guide for practitioners and researchers from both DL and SE fields when working with source code.

The remaining of this literature review is organized as follows. Section~\ref{sec:existingworks} presents existing language models for source code as well as their limitations, which motivates the use of DL models. Section~\ref{sec:seq2seq} formulates source code modeling under encoder-decoder framework and describes the important components of such framework. Section~\ref{sec:deep_language_models} highlights the recent practices for building deep source code models. Section~\ref{sec:applications} reviews DL-based applications for various Big Code tasks. Section~\ref{sec:datasets} presents the available datasets for such tasks. Section~\ref{sec:challenges_directions} discusses the current challenges and proposes some future directions in using DL for source code modeling and generation. Finally, section~\ref{sec:conclusions} summarizes the main contributions of this literature review.

\section{Traditional source code modeling approaches and their limitations}\label{sec:existingworks}

We describe four traditional approaches to handle the syntactic and semantic facets of source code including (\textit{i}) domain-specific language guided models, (\textit{ii}) probabilistic grammars, (\textit{iii}) simple probabilistic language models (i.e., $n$-grams), and (\textit{iv}) simple neural language models. Several serious issues still associate with these traditional approaches, which can be handled effectively using DL models. More details of each approach and its limitations are covered in this section.

\subsection{Domain-specific language guided models}
Domain-specific languages (DSLs) are often used to define the parametrization rules and states for generating the structure of a program. DSL-based models create different grammar rules for common code statements (e.g., control flow, comments, and brackets). Compared to a general-purpose programming language, the grammar size of a DSL is usually smaller, which makes it more efficient for specific code generation tasks. DSL-based model has been studied by Gulwani et al.~\cite{gulwani2010dimensions} and Jha et al.~\cite{jha2010oracle}. Gvero et al.~\cite{gvero2013complete} reduced the search space for Scala expression suggestion using a calculus of ~\textit{succinct types} and designed a higher-order function for code completion. Since the generation process is human interpretable, this type of model could be a promising approach in SE.

In program induction, a DSL specifies the space of candidate programs (\textit{program template}), while the example input-output pairs are known as \textit{specification}. Under this scenario, the problem is known as Inductive Logic Programming (ILP)~\cite{muggleton1994inductive}. Two classic families of solving ILP are the bottom-up approaches, constructing programs from example features, and the top-down approaches, testing examples from generations and adjusting the rules accordingly~\cite{evans2017learning}. Given the precise and combinatorial nature of induction, the induction is commonly cast as a Constraint Satisfaction Problem (CSP)~\cite{solar2013program}. This belongs to the top-down family of ILP~\cite{evans2017learning}. The formal definition of the problem can be found in the work of Pu et al.~\cite{pu2017learning}. Such a CSP problem can be solved by a constraint solver such as Z3~\cite{de2008z3}. The problem with this approach is that the solver is always heuristic and its scalability is poor. Often these systems cannot handle noisy, erroneous, or ambiguous data.

DSL guided models can capture the structure of a specific programming language; however, they require deep domain knowledge to generate the detailed syntactic and semantic rules. One possible way to increase flexibility is to represent a DSL with a probabilistic model.

\subsection{Probabilistic grammars}
In formal languages, production rules can be used to define all possible generations of strings (code statements). Context-Free Grammars (CFGs) is a set of production rules that can be applied regardless of context, i.e., the left-hand side of a production rule only contains a single non-terminal symbol. CFG is a common way to define the structure of a programming language, which can then be used to convert its source code into Abstract Syntax Trees (ASTs)~\cite{hopcroft2008introduction}. Probabilistic Context-Free Grammar (PCFG) is an extension to CFG, in which the production rules of a context-free language are associated with a probabilistic model. Bielik et al.~\cite{bielik2016phog} generalized the idea of PCFG to the task of code completion in other non-context-free languages like JavaScript. Later, Raychev et al.~\mbox{\cite{raychev2016probabilistic}} extended the previous work on PCFGs by learning a decision tree to build a probabilistic model using ASTs of a proposed DSL namely TGen. These works achieved strong results for some code completion tasks.

Another extension of CFG, namely Tree Substitution Grammar (TSG), uses tree fragments instead of a sequence of symbols for the production rules. Such rules of TSG are defined by a Tree Adjoining Grammar~\cite{joshi2003formalism}, which can create more flexibility and better represent complex linguistic structures~\cite{cohn2010inducing}. To limit the model complexity and sparse grammar, researchers often use non-parametric Bayes to infer the distributions~\cite{cohn2010inducing,allamanis2014mining}. These models are suitable for pattern mining since their automatic model selection ability allows the discovery of more complex structures. However, non-parametric Bayesian methods are often extremely slow to compute and hard to scale. Although probabilistic grammars achieve high performance for domain-specific languages, they still require manually-designed rules to model code locality and reuse.

\subsection{$n$-gram language models}
\label{subsection:n-gram}
Besides the two syntactic models above, one simple yet effective statistical language model is $n$-gram. More specifically, this model assumes each token/word is conditionally dependent on the previous $n-1$ tokens/words as described in the following equation:
$
P(\mathbf W) = \prod_tP(w_t|w_{t-(n-1),\dots,t-1})
$, where
$P(w_t|w_{t-(n-1),\dots ,t-1})$ can be simply computed by counting the occurrences of all $n$-grams in the training set. A direct advantage of $n$-grams over syntactic models (e.g., probabilistic grammars and DSL guided models) is that it is easier to generalize since the dependencies and rules of the programming languages are learned automatically from source code. Hindle et al.~\cite{hindle2012naturalness} took the initiative to utilize $n$-gram to build a language model for source code. Since then, besides code completion~\cite{code_completion_ngrams, nguyen2013ngrams}, $n$-gram models have also been applied to other tasks such as idiom mining~\cite{allamanis2013mining}, syntax error detection~\cite{campbell2014syntax}, source code analysis~\cite{hsiao2014using} and code obfuscation~\cite{liu2016towards}. However, simple language models like $n$-gram cannot capture high-level programming paradigms.

To address this issue, a line of work has enhanced language models to be more adaptable to local information. Bielik et al.~\cite{bielik2016program} augmented a DSL-based model with $n$-gram and showed strong empirical results for programming language modeling. The resulting model was also more efficient in training and inference compared to neural models. Hellendoorn et al.~\cite{hellendoorn2017deep} added a local $n$-gram cache and merges the predictions of local and global models. Both models were claimed to surpass DL counterparts (e.g., RNN and LSTM) at the time of their publication, thus these two models would be good baselines for source code modeling.

It is noted that $n$-gram is not good at modeling long-term dependencies (cf. Listings~\ref{listing:code_ex1} and \ref{listing:code_ex2}) between tokens. The $n$-gram truncation discards long-term positional information. Other statistical models, such as Hidden Markov models, also fail to encode long term history~\cite{sutskever2009recurrent}, since its state space becomes exponentially large when encoding several previous history tokens into one state. Besides, another limitation with an $n$-gram model is the sparsity of the vector representation of the word or code token, which is caused by the large vocabulary of source code. This sparsity issue can be solved using the distributed representation of neural language models.

\subsection{Simple neural program models}
\label{subsection:embedding_models}
Instead of incorporating the frequency of each previous token explicitly, neural network embedding models with one hidden layer first convert one-hot encoding of a word into an intermediate word-embedding vector with a much shorter length compared to the vocabulary size (e.g., 100 -- 1000). This idea is also known as \textit{distributed word representation}.
In its original work~\mbox{\cite{schwenk2002connectionist}}, the word embeddings of up to $n$ - 1 previous tokens with respect to the current word were fed to a fully-connected neural network with one hidden layer. At the output layer, a softmax function was applied to calculate the probability of the next word. However, a serious drawback of this original model is the high computational cost of the hidden layer. Therefore, log-bilinear was presented~\cite{mnih2007three} to address this challenge by replacing the non-linear activations with a context matrix to determine the context vector with respect to the current word. Then, the similarity between the context vector of the previous tokens and the current word was computed. Later, several methods were proposed to speed up the training and prediction time using hierarchical architecture~\cite{mnih2009scalable}. In this previous work, the log-bilinear model was demonstrated to outperform the traditional fully-connected neural network and $n$-gram language models for the task of modeling APNews.

Later, the idea of neural network embedding model was adopted by researchers for source code modeling, which can be referred to as \textit{simple neural program models}. More specifically, Maddison et al.~\cite{maddison2014structured} combined log-bilinear with a tree depth-first search traversal technique (i.e., Log-bilinear Tree-Traversal models) to generate human-understandable source code. Allamanis et al.~\cite{allamanis2015bimodal} extended Maddison's approach to retrieve source code snippets from natural language queries and vice versa. Allamanis et al.~\cite{allamanis2015suggesting} also used a log-bilinear model to recommend method and class names for object-oriented programming in Java and this model outperformed an $n$-gram model in both of the tasks. Similar to~\mbox{\textit{n}}-gram models, the knowledge of log-bilinear models is limited to the previous $n$ - 1 tokens. Therefore, the above works needed to define the global and local context explicitly for log-bilinear models to capture the short-term, long-term dependencies and sequential property of source code. The list of contexts is still human-crafted and incomplete, thus limiting its applications in new domains. However, with their simplicity, simple neural program models are being used as (pre-trained) input features (cf. section~\mbox{\ref{subsection:input_embeddings}}) for various Big Code tasks.

\subsection{Advantages of deep learning models over traditional approaches}
Encoder-decoder framework~\mbox{\cite{sutskever2014sequence}} with DL models (cf. section~\mbox{\ref{sec:seq2seq}}) can be used to effectively capture the dependencies and sequential property of an input sequence. To be more specific, DL models are suitable for code modeling and generation since they are good at the following four important aspects: (\mbox{\textit{i}}) automatic feature generation, (\mbox{\textit{ii}}) capturing long-term dependencies as well as sequential property, (\mbox{\textit{iii}}) end-to-end learning, and (\mbox{\textit{iv}}) generalizability. Existing models have to trade-off among these four properties. For example, $n$-gram models can automatically extract features from source code, but cannot capture long-term dependencies well due to the combinatorial explosion of terms. Similarly, simple neural program models (e.g., fully-connected or log-bilinear models) still require human-designed rules to capture the dependencies and structure of source code, which limits its end-to-end training and generalizability. In contrast, with deep domain knowledge incorporated, DSL guided models and probabilistic grammars can effectively capture the dependencies and sequential properties, but the strictly defined rules make the models harder to generalize and automate the learning process in new domains. Next, encoder-decoder framework using DL models for sequence/code modeling is presented.

\section{Deep sequence modeling with encoder-decoder framework}
\label{sec:seq2seq}
\begin{figure*}[h]

  \centering
  \includegraphics[width=13cm,keepaspectratio]{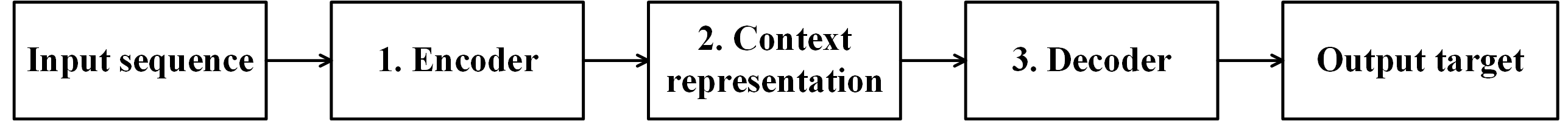}

  \caption{Main steps of an encoder-decoder framework.}
  \label{fig:seq2seq}
\end{figure*}

In NLP, the main objective is to process a large amount of natural language mostly in the form of text or human voice. Code - a means of communication for developers - is similar to natural language~\mbox{\cite{hindle2012naturalness, allamanis2017survey}}, which has syntactic structures and semantic meaning. Specifically, there are various Big Code tasks (cf. section ~\ref{sec:applications}) whose input is a sequence (e.g., source code and/or natural language) and prediction target can be either a sequence or just a simple numeric/categorical value. Inspired by the field of NLP, such Big Code tasks can be formulated under an \mbox{\textit{encoder-decoder framework}}. If both the input and output are sequences, then encoder-decoder framework can also be called seq2seq~\cite{sutskever2014sequence}. The main steps of an encoder-decoder framework are illustrated in Figure~\ref{fig:seq2seq}.


The components for steps 1 and 3 are referred to as encoder and decoder, respectively, which are typically two Deep Learning (DL) models (cf. section ~\ref{subsection:sequence_models}). In the original paper~\cite{sutskever2014sequence}, the final internal state of the encoder is used as the context vector. In more recent works, step 2 is often handled by attention mechanism~\cite{bahdanau2014neural} (cf. section ~\ref{subsubsection:attention_mechanism}) or external memories~\cite{weston2014memory} (cf. section~\ref{subsubsection:memory_nn}) with a richer and position sensitive context sequence. In this section, three main elements of an encoder-decoder learning framework for sequence modeling including (\textit{i}) DL models, (\textit{ii}) input embeddings and (\textit{iii}) stable training of such models are covered. It should be noted that both sequential models (i.e., modeling word-by-word) and structural models (i.e., exploiting syntactic structure of a sentence/code snippet) can be used for sequence modeling. Although section~\mbox{\ref{sec:seq2seq}} mainly reviews different components of an encoder-decoder framework originally proposed for sequence modeling in NLP, this section is still important because of two reasons: (\textit{i}) the presented deep models/techniques can be extended to source code, and (\textit{ii}) only a small portion of such models has been utilized for source code modeling. Based on this section, section~\mbox{\ref{sec:deep_language_models}} subsequently presents the current practices of source code modeling and generation using encoder-decoder framework.

\subsection{Deep learning models for sequence modeling}
\label{subsection:sequence_models}

Two main classes of DL models for sequence modeling, namely (\textit{i}) recurrent and (\textit{ii}) non-recurrent neural networks are first presented. Then, three techniques to build more robust models including (\textit{i}) attention mechanism, (\textit{ii}) external memory and (\textit{iii}) beam search are covered. It is noted that Multi-Layer Perceptron (MLP)~\mbox{\cite{rosenblatt1958perceptron}} (a.k.a. fully-connected/feed-forward neural network) is not reviewed in this section since it is an extension of the neural network described in section~\mbox{\ref{subsection:embedding_models}} (i.e., with more hidden layers). Thus, MLP is still limited in capturing the dependencies and sequential property of a sequence, and not widely used for sequence modeling unless combined with more advanced techniques such as attention mechanism.

\subsubsection{Recurrent Neural Networks}

Recurrent Neural Network (RNN) and its variants have been widely used for building language models~\cite{krause2017dynamic,melis2017state,merity2017regularizing}. RNN is a special type of deep neural network, in which a block of parameters are shared and repeated many times across different parts of a sequence, resulting in a deep computational graph~\cite{goodfellow2016deep}. This architecture helps a network to learn with input/output of various lengths that MLPs cannot.

However, vanilla (plain) RNNs are hard to train~\cite{pascanu2013difficulty} and are not good at keeping past information from different time scales. Gated RNNs, such as Long Short-Term Memory (LSTM)~\cite{hochreiter1997long} and Gated Recurrent Units (GRUs)~\cite{chung2014empirical}, model the keeping and forgetting mechanisms explicitly with sigmoid activations, namely \textit{gates}. An LSTM has three gates to control input, output and forgetting, respectively. In addition, there is a memory cell state to generate the hidden states.

RNN units can be made deep to encode more complex transitions~\cite{pascanu2013construct}. Highway layers~\cite{srivastava2015highway} were introduced to stabilize the training gradients in Recurrent Highway Networks~\cite{zilly2016recurrent}. To capture the long-term dependencies in time series and represent hierarchical information, RNNs layers can be stacked with different update frequencies~\cite{koutnik2014clockwork}. The gated feedback RNNs~\cite{chung2015gated} allow the network to learn its own clock rates by using additional gates. The state-of-the-art RNN for language model is the Fast-Slow RNN~\cite{mujika2017fast} that incorporates the strengths of both deep and multiscale RNNs.

\subsubsection{Non-recurrent neural networks}
Temporal convolution or one-dimensional convolution across time is another type of neural network that can capture long-term relations with hierarchical architecture~\cite{waibel1988phoneme}. It has been applied to sentiment analysis, sentence classification, machine translation and meta-learning~\cite{mishra2017meta}.

Convolutional Neural Networks (CNNs) have been used in several sentence modeling tasks. In 2013, Kalchbrenner and Blunsom ~\cite{kalchbrenner2013recurrent} used a CNN as the encoder and an RNN as the decoder for dialogue generation. One year later, Blunsom et al. proposed Dynamic CNN~\cite{blunsom2014convolutional} for sentence semantic modeling, where variable length and relation discovery were enabled by max pooling. However, these earlier works failed to achieve similar performance of LSTMs.

Recently, non-recurrent structures have emerged again with similar performance as RNN, but they are faster to compute. Masked convolution layers are used as the decoder in a neural machine translation system~\cite{kalchbrenner2016neural}. Gehring et al. ~\cite{gehring2017convolutional} proposed ConvS2S that brought skip connection~\cite{he2016deep} and attention~\cite{bahdanau2014neural} (cf. section~\ref{subsubsection:attention_mechanism}) to sentence modeling and achieved the state-of-the-art translation performance. Combining recurrent and convolutional units is also useful. He et al.~\cite{he2017wider} strengthened the input-to-output correlation by adding \textit{cross-layer convolutions} to stacked RNNs.

Vaswani et al.~\cite{vaswani2017attention} proposed a multi-head attention model called Transformer relying on self-attention and positional encoding to compute the sequence representations. Transformer allows the decoder to attend to information arbitrarily far and reduced training time significantly without quality loss. Like CNN, the causal structure is held by masking later output for the autoregressive factorization. Recently, a deep language representation model, BERT~\mbox{\cite{devlin2018bert}}, utilizing a novel bidirectional training of a Transformer has achieved the state-of-the-art results in eleven NLP tasks such as sentence classification/tagging, question answering, and named entity recognition. Later, BERT was extended to language modeling~\mbox{\cite{dai2019transformer}} and language generation~\mbox{\cite{dong2019unified}} by addressing its limitations of fixed-length contexts and bidirectional nature, respectively. It is noted that Transformers can be leveraged for contextual (same token with varying usages) embeddings of source code.

Speeding up the sequential generation is another interesting direction. It should be noted that all autoregressive models only generate samples sequentially since they use ancestral sampling. Thus, alternative architectures for rapid, parallel sample generation are required. Gu et al.~\cite{gu2017non} enabled non-autoregressive learning by sampling a latent variable representing the \textit{fertilities}, i.e. the usage of each source word in decoding, which required to be supervised by an external alignment system. Inverse-Autoregressive Flows (IAFs)~\cite{kingma2016improved} could generate high-dimensional samples in parallel from latent variables. Oord et al.~\cite{oord2017parallel} incorporated WaveNet with IAF to create parallel WaveNet, which sampled with higher fidelity, but ran 3000 times faster in audio generation.

\subsubsection{Attention mechanism}
\label{subsubsection:attention_mechanism}
One problem of the original encoder-decoder framework is that decoder can only access a single context vector. Human understands text lines by repeatedly attending to different parts of a sequence. To simulate this behavior, Bahdanau et al.~\cite{bahdanau2014neural} used a sequence as the context and propose \textit{attention mechanism} to adapt the weights of context associated with a certain output stage and impose an explicit alignment between input and output tokens. To be more specific, with an encoded sequence $F$, and at each step $t$, the hidden state $\mathbf h_t$ is computed using an RNN model with input source vector $\mathbf c_t$ generated by attention mechanism as additional input:
$
\mathbf h_t = \RNN(\mathbf h_{t-1}, [\mathbf e_{t-1};\mathbf c_t]).
$
This way of incorporating attention is known as \textit{early binding}. Alternatively, attention can be considered just before generating the output token. A typical soft attention similar to Bahdanau et al.'s~\cite{bahdanau2014neural} can be computed as follows:

\begin{enumerate}
    \item With the previous hidden state $\mathbf h_{t-1}$, the \textit{attention energy} is computed with a content-based scoring function $\mathbf u_t = \mbox{score}(F, \mathbf h_{t-1})$.
    \item Exponentiate and normalize $u_t$ to 1: $\mathbf a_t=\mbox{softmax}(\mathbf u_t)$.
    \item Compute the \textit{input source vector} $\mathbf c_t=F\mathbf a_t$.
\end{enumerate}
The simplest way to define the $\mbox{score}$ is dot-product,
$\mbox{score}(F, \mathbf h_{t-1})=F^T\mathbf h_{t-1}$. Or an \textit{expected input embedding} $V$ can be defined so that $\mbox{score}(F, \mathbf h_{t-1})=F^TV\mathbf h_{t-1}$. In the original paper~\cite{bahdanau2014neural}, the attention energy is computed with a Multi-Layer Perceptron (MLP) as follows:
$$
\mbox{score}(F, \mathbf h_{t-1}) = \mathbf v^T\tanh(WF+V\mathbf h_{t-1}).
$$

Content-based attention energy is computed by scoring each element separately, which makes it hard to discriminate the elements with similar content from different locations. Location sensitive attention~\cite{chorowski2015attention} broke through this limitation by generating attentions in an autoregressive fashion. However, unrolling through another RNN during backpropagation can greatly increase the computational time. Vaswani et al.~\mbox{\cite{vaswani2017attention}} presented multi-head self-attention to represent the relevant context of each word in an input sequence at different locations. By combining sequence modeling and attention mechanism, the state-of-the-art results have been achieved for neural machine translation~\cite{bahdanau2014neural, wu2016google, johnson2016google, lample2017unsupervised}.

\subsubsection{Memory-augmented neural networks}
\label{subsubsection:memory_nn}
Attention is closely related to external memory. Together, they have become important building blocks for a neural network. External memories are used as internal states, which can be updated by attention mechanism for selective reading and updating. The classic Memory Network (MemNN)~\cite{weston2014memory} tried to mimic a Random Access Memory (RAM) and use soft attention as a differentiable version of addressing.

A memory network usually takes the following inputs:
\begin{itemize}
  \item A \textit{query} $q$ is the last utterance the speaker said in a general dialogue setting, or the question in a QA setting.
  \item A \textit{memory} vector $\mathbf m$ is the dialogue history of the model. The knowledge can be as large as the whole codebase or documentation given the model is sufficiently powerful.
\end{itemize}

And this type of network has the following modules to handle the inputs:
\begin{itemize}
  \item An \textit{encoder} converts $q$ into a vector using an RNN \cite{cho2014learning} or a simpler word embedding~\cite{mikolov2013efficient}.
  \item A \textit{memory module} $M$ finds the best part of $\mathbf m$ related to $q$. This is the \textit{addressing} stage.
  \item A \textit{controller module} $C$ sends $q$ to $M$ and reads back the relevant memory, adds that to the current state. In practice, we always cycle this process to enable complicated reasoning.
  \item A \textit{decoder} generates output from the final states.
\end{itemize}

The training of MemNN is fully supervised, in which the label of the best part of memory is given at every stage of the memory addressing. Its follow-up End-to-End Memory Network~\cite{sukhbaatar2015end} uses soft attention for memory addressing to train the attention in backpropagation and relax the supervision to only the output. It is noted that using one memory unit to match both the query and the state limits the expressiveness. By splitting the memory into a key-value pair, Millor et al.~\cite{miller2016key} encoded prior knowledge and obtained better results. Recurrent Entity Network~\cite{henaff2016tracking} demonstrates how to enhance the memory unit by letting the agent learn to read and write the memory to track the facts. Weston~\cite{weston2016dialog} generalized this model to unsupervised learning, by adding a new stage to generate answers and predict replies. This kind of model is evaluated on various tasks such as basic reasoning on twenty bAbI tasks~\cite{weston2015towards}, reading children's books~\cite{hill2015goldilocks}, understanding real dialogues from movies~\cite{dodge2015evaluating}. All of the tasks can be found on the website of the bAbI project\footnote{https://research.fb.com/projects/babi/}. Recently, many works have tried to make traditional memory paradigms differentiable so the models can be optimized with SGD. Such end-to-end trainable models have been used to handle algorithmic learning and reasoning tasks such as language understanding and program induction, which are covered in more detail in section~\ref{subsection:program_induction}.

\subsubsection{Beam search}
Searching for the best-decoded result with the highest probability is computationally intractable. In other words, there can be an exponentially large number of generated sentences in NLP or source code in Big Code. One solution would be to choose the word/token with the highest output probability after each time step during the decoding process. However, this greedy process will likely result in a sub-optimum result. Therefore, in machine translation, beam search is widely adopted as the heuristic search technique~\cite{sutskever2014sequence}. Instead of taking the next word with the highest possibility directly, a list of previous most likely partial translations is kept and those selected words/tokens will be extended each translation at the current step and re-rank them. The length of the list at each time step is known as \textit{beam size}. This method often improves the translation, but the performance is closely dependent on the beam size~\cite{koehn2017six}.

\subsection{Input embeddings of deep learning models}
\label{subsection:input_embeddings}
In this section, the input representation for sequence modeling is presented. This is also important for code modeling since keyword representations can vary hugely within the scopes of functions, classes and projects. Such large vocabulary makes traditional representations such as one-hot encoding or $n$-grams form a very sparse embedding vector. As a result, in recent deep language models, input words are often converted into real-valued vectors with distributed representations~\cite{mikolov2013distributed}. Words in the embedded space demonstrate nice emergent properties such as semantic relationship can be represented as vector arithmetic. Sharing the embedding weights with the softmax layer in the decoding process results in notable improvements for language models~\cite{inan2016tying}.

For NLP tasks, it is a common practice to use general-purpose pre-trained word vectors on large corpora such as word2vec~\cite{mikolov2013distributed}, GloVe~\cite{pennington2014glove} and fastText~\cite{bojanowski2016enriching}. The parameters can also be optimized for specific datasets and tasks together with the model. McCann et al.~\cite{mccann2017learned} performed the word vector training based on a machine translation task, which helped their word vectors (i.e., CoVe) to capture more complex contextual information. They showed that replacing traditional word vectors with CoVe could improve the performances of many tasks. The pre-trained word embeddings can also be fine-tuned to adapt to the Big Code task of interest. However, in the domain of source code modeling, the vocabulary size is much larger than that of NLP. As a result, re-training the word embeddings using initialization of the pre-trained parameters is also worth considering. Some treatments of source code representation are presented in more detail in Section~\mbox{\ref{subsection:code_completion}}.

In Big Code applications, sub-word level inference is also substantially important. For example, code completion algorithms should be able to suggest incomplete function or variable names by seeing only a part of the word. To incorporate character-level information, the characters can be combined into word representation. Ling et al.~\cite{ling2015finding} proposed C2W model, which uses Bidirectional LSTM to construct word embedding from character sequences. CNN can also be used for character-level embeddings~\cite{zhang2015character}.

\subsection{Stable training of deep learning models}

Recurrent models (e.g., RNN) for sequence modeling are hard to train~\cite{pascanu2013difficulty}, and similar to other types of neural network, easily prone to overfitting. Like other DL architectures, RNN and its variants are usually trained using a special form of backpropagation, namely \textit{backpropagation through time}. There are several techniques for optimizing the loss function of a model. Among them, Stochastic Gradient Descent (SGD) or mini-batch gradient descent is mostly adopted due to its efficient computation and parallelization on Graphics Processing Units~\cite{goodfellow2016deep}. Merity et al.~\cite{merity2017regularizing} applied a non-monotonically triggered Averaged SGD~\cite{mandt2017stochastic} for language modeling and achieved superior results. It is noted that most of the recent papers on DL report their models trained by modern versions of SGD namely RMSProp~\cite{tieleman2012lecture} or Adam~\cite{kingma2014adam}.

Model regularization also has a significant impact on the generalization performance. We review four common types of regularization for training DL models more effectively including: (\textit{i}) dropout, (\textit{ii}) normalization, (\textit{iii}) activation regularization and (\textit{iv}) structural regularization.

\textit{Dropout}~\cite{srivastava2014dropout} randomly turns off several positions of the activation following a Bernoulli distribution. However, applying it to the RNN hidden states disrupts the model's ability to preserve long-term dependencies~\cite{zaremba2014recurrent}. Two ways of retaining the information have been adopted. The first way is to limit the dropout rate of the hidden states by keeping previous information. Zoneout~\cite{krueger2016zoneout} randomly copies previous values of the activations rather than zeroing them out. Semeniuta et al.~\cite{semeniuta2016recurrent} applied dropout on the input gate to prevent memory loss. The second one is \textit{locked dropout}, i.e., using the same dropout mask for a full forward pass. This method preserves the activation norms instead of gradually dropping information. Gal et al.~\cite{gal2016theoretically} linked locked dropout with variational Bayes inference and used it for embedding dropout. In addition, locked DropConnect~\cite{wan2013regularization} on the hidden weights resulted in substantial improvements~\cite{merity2017regularizing}.

\textit{Normalization} restricts the activations of different time steps to follow a stable distribution. Inspired by batch normalization~\cite{ioffe2015batch}, multiple normalization techniques customized for recurrent structures have been studied, such as recurrent batch normalization~\cite{cooijmans2016recurrent}, weight normalization~\cite{salimans2016weight} and layer normalization~\cite{ba2016layer}. There are also normalization techniques targeting the gradient stability. For example, spectral normalization~\cite{miyato2017spectral} was designed to keep the gradient bounded by constraining the activations to be Lipschitz continuous.

Regularization can be applied to weights and activations of DL models. $L_2$ regularization on the weights is referred to as \textit{weight decay}. \textit{Activation regularization} penalizes activations with $\alpha L_2(m\cdot h_t)$, where $\alpha$ is a regularization term, $m$ is a scaling factor and $h_t$ is the hidden state, respectively. Temporal activation regularization~\cite{merity2017revisiting} penalizes the large changes in the hidden state of a neural model with $\beta L_2(h_t-h_{t-1})$, where $\beta$ is a scaling factor, $h_t$ and $h_{t-1}$ are the hidden states at time $t$ and $t-1$, respectively.

\textit{Structural regularization} prevents exploding or vanishing gradients by restricting the model structure. Model structure restriction can be done by forcing the recurrent matrix to be unitary~\cite{arjovsky2016unitary} or using element-wise interactions. Strongly-typed RNNs~\cite{balduzzi2016strongly} use type-consistent operations for the recurrent units. Other simplifications are Quasi-RNN~\cite{bradbury2016quasi} and Simple Recurrent Unit~\cite{lei2017training}.

These regularization techniques are important to reduce the overfitting and improve the generalization performance of deep source code models, since the learned models can become overly complicated due to the need to represent various types of rules.

\section{Recent practices of building deep learning models for source code modeling and generation}
\label{sec:deep_language_models}

This section focuses on the practices of developing Deep Learning (DL) models for source code under the encoder-decoder framework presented in section~\ref{sec:seq2seq}. Specifically, we present the techniques for the following: (\textit{i}) deep encoder models, (\textit{ii}) deep decoder models, and (\textit{iii}) deep controller models to better generalize the capabilities of DL models to source code modeling and generation.

\subsection{Deep encoder models}
\label{subsection:code_completion}
For many deep source code tasks, the input takes the form of a sequence, such as code snippets, comments or descriptions, where we rely on a deep module to capture the semantic and context of input for further processing. We call this kind of modules deep encoder models. The most widely used encoder models are sequential models such as RNN and its variants. For general-purpose software repository mining, the effectiveness of RNNs has been tested~\cite{white2015toward}. However, sequential models may not be as effective for code modeling and generation due to the following limitations:
\begin{itemize}
    \item Syntactic context is not well represented in a sequential model, which may lead to a violation of the grammar rules of a programming language.
    \item Large vocabulary of code leading to the out-of-vocabulary issue affects the generalizability of a deep code model.
    \item Recurrent models (e.g., RNNs) suffer from the hidden-state bottleneck, in which the size of hidden-state vector limits the information a model can carry through time.
\end{itemize}

Many methods have been proposed to overcome these shortcomings. These are (\textit{i}) structural (tree-/graph-based) representation, (\textit{ii}) open vocabulary model and (\textit{iii}) attention mechanism.

\textbf{Structural representation.} Abstract Syntax Tree (AST) is a natural way to capture the syntactic structure of a program. In an AST, a program is parsed into a hierarchy of non-terminal and terminal (leaf) nodes based on the syntax of a programming language. To utilize AST for code representation, the simplest way would be to use depth-first search to convert an AST into a sequence ~\mbox{\cite{dam2016deep,amodio2017neural,li2017code}}. Other studies proposed DL models (Recursive Neural Networks~\mbox{\cite{white2016deep}}, Tree-LSTM~\mbox{\cite{wei2017supervised}} or CNN ~\mbox{\cite{mou2016convolutional}}) to work directly on the hierarchical structure of a parse tree. Recently, Zhang et al.~\mbox{\cite{zhang2019novel}} have shown that splitting an AST into code-statement subtrees can improve the performance of tree-based representations. Recent works (i.e., code2vec ~\mbox{\cite{alon2019code2vec}} and code2seq ~\mbox{\cite{alon2018code2seq}}) have also proposed to use AST paths as a representation for code, in which the extracted paths would be aggregated using an attention-based deep neural network. Recently, Allamanis et al.~\mbox{\cite{allamanis2018learning}} have presented novel Gated Graph Neural Networks~\mbox{\cite{li2015gated}} to represent source code as a directed graph. Specifically, they incorporated data/control-flow information of variables into AST to capture the syntactic and semantic structures of source code more effectively. It is observed that structural representation of source code has witnessed a growing interest from the Big Code community. There is also a comprehensive review~\mbox{\cite{chen2019literature}} on source code embeddings.

\textbf{Open vocabulary model.}
The vocabulary of source code is open, rather than fixed. Thus, it is impractical to train a classifier using the whole vocabulary, so it is more common to truncate it by keeping only the most frequent 1,000 or 10,000 terms and replacing the others with an Out-of-Vocabulary (OoV) token (i.e., \texttt{<unk>}). The drawback of this truncation is that the OoV tokens (\textit{neologisms}~\cite{allamanis2015suggesting}) from a testing set cannot be predicted. To solve the OoV problem, Karampatsis et al.~\mbox{\cite{karampatsis2019maybe}} have proposed a novel open-vocabulary neural language model for source code modeling. This work used GRU to build a neural language model on top of sub-word (character subsequences of code tokens) units generated by the Byte pair encoding algorithm~\mbox{\cite{gage1994new}}. The large-scale experiments showed that the proposed sub-word neural program model was better than the state-of-the-art \textit{n}-gram model and also more robust against the OoV problem across different programming languages and projects. Character-based DL models~\cite{karpathy2016unreasonable,kim2016character} are also an alternative to subword for addressing the OoV problem. Recently, Cvitkovic et al.~\cite{cvitkovic2018open} extended the graph-based code representation~\cite{allamanis2018learning} to incorporate open vocabulary using a Graph-Structured Cache, in which novel words/tokens would be added as \textit{cached}~\cite{grave2017unbounded} nodes into an existing AST.

\textbf{Attention mechanism.} Attention mechanism can be used for addressing both the OoV issue and the hidden-state bottleneck of RNNs. Bhoopchand et al.~\cite{bhoopchand2016learning} employed a pointer network to copy OoV tokens from the recent past during code completion. The pointer network is a soft attention over previous input embeddings. A controller produces a scalar to decide whether to select from the copying position or language model distribution. Li et al.~\cite{li2017code} recently proposed a similar model but with attention over the previous hidden states. Similar to attention layers of decoder networks, the attention output is concatenated into the input. The attention used in these models is computed with a Multi-Layer Perceptron~\cite{bahdanau2014neural}. The drawback of this approach is that the modification of hidden-state cache and computation of attention are very computationally intensive. We have also observed that adding token copying can lead to a precision drop of token predictions comparing to no pointer network. Splitting the hidden states into separate parts for pointer network and context encoding solves this problem. Attention mechanism is also used by some non-recurrent models, e.g., Das and Shah~\cite{das2015contextual} employed a gated unit over the word embeddings for a feed-forward neural network. This model was used to represent some commonly occurring value types such as iterator variable names for contextual code completion.

\subsection{Deep decoder models}
\label{subsection:program_synthesis}

Given the embedded features produced by an encoder model, a decoder model generates output for a target domain (e.g., code and natural language). Unlike natural languages, source code must adhere to the target syntax of a programming language. Xu et al.~\cite{xu2017sqlnet} used a \textit{sketch} for SQL generation template and trained neural networks to copy certain parts to these slots with a column attention mechanism. Furthermore, the decoding template can be enriched by introducing a DSL. By training a model with direct syntactic information such as ASTs, much better results can be obtained for code generation. Dong and Lapata~\cite{dong2016language} proposed the Seq2Tree model for transferring language into logical forms. A tree structure is decoded by predicting a "nonterminal" token as a root of a subtree. Parent information is fed by incorporating attention mechanism with all previous states. This model makes it easier to keep structural information, but this simple modification only supports limited grammar rules and has no guarantee of syntactic correctness.

Yin et al.~\cite{yin2017syntactic} designed and implemented a syntax-driven neural model to transform natural language statements into corresponding Python ASTs. This model got 10+ BLEU and decent accuracy on several code generation tasks. Instead of directly generating source code $c$, a probabilistic grammar model $g$ representing the distribution of an AST $y$ given input natural language description $x$ is defined. So that the syntax structure is automatically captured. The task is to select the best possible AST
$
\hat y = \underset{y}{\arg\max} g(y|x).
$
Then given the AST, code can be inferred in a definite process.

A finite set of production rules $r\in R$ is the main component of a formal grammar specification. Using a depth-first traversal from left to right, the AST for these rules can be generated with a sequence of two types of actions $a$ on the current AST $y$:
\begin{itemize}
    \item Apply a specific production rule $r$
    \item Add a terminal token $v$ (e.g., \textbackslash n)
\end{itemize}
Now we can map the AST generation to a traditional seq2seq learning task, given an input sequence $\mathbf x$, a sequence of action $\mathbf a$ can be generated as follows,
$
p(\mathbf a|\mathbf x) = \prod_{t=1}^Tp(a_t|\mathbf x, a_{<t})
$
, where $a_t$ is the action taken at time $t$.

In fact, generating highly accurate general-purpose programs is a very challenging task. The performance of program generation relies heavily on the development of the code generative model, which needs to satisfy the following three major requirements:
\begin{itemize}
    \item \textbf{Input representation} of a program should be able to handle OoV tokens, which usually requires learning embedding vector from both word-level and character-level tokens. For example, it is helpful to use compositional word representation such as C2W~\cite{ling2015finding} and copying mechanism~\cite{vinyals2015pointer} to represent newly appeared OoV names in code. Syntactic information is also important as mentioned in section~\ref{subsection:code_completion}. To reconstruct grammatically correct programs from generation, encoders and decoders often accept and yield AST tokens.
    \item \textbf{Addressing mechanisms} (e.g., attention~\cite{bahdanau2014neural} and memory~\cite{graves2014neural}) are important in guiding the decoding process and also used to copy from cache and recent history.
    However, content-based attention~\cite{bahdanau2014neural} and its extension, writable memory~\cite{graves2014neural}, are both not very suitable for implementing this copying mechanism.
    Another structure to keep recent history is associative memory~\cite{ba2016using}. The memory is kept in an associative matrix and updated at every time step. Some variants of associative memory are also incorporated with language models~\cite{dangovski2017rotational}.
    \item \textbf{Discrete structure learning} is the key to representing function logic and control flows of a program. It is important for neural program models to learn a discrete structure and make deterministic decisions based on such structure.
\end{itemize}
We further discuss the solutions and challenges in designing these components in Section~\ref{sec:challenges_directions}.

\subsection{Deep controller models}
\label{subsection:program_induction}

Theoretically, a neural network itself is capable of learning a program~\cite{csaji2001approximation, siegelmann1992computational}. Therefore, to solve more complex problems, deep neural network can be used as a controller to learn the next instruction/operation to execute directly from the input-output examples without a DSL. This class of models is also referred to as \textit{neural abstract machines}. Grefenstette et al. (2015)~\cite{grefenstette2015learning} connected an LSTM with soft-differentiable stack-, queue- and deque-based memories to generalize the ability of RNN in machine translation.
Neural Turing Machine (NTM)~\cite{graves2014neural} and Differentiable Neural Computer (DNC)~\cite{graves2016hybrid} use a recurrent model (RNN or LSTM) to read and write to an external memory matrix in a dynamic manner to simulate the execution of Turing computers. NTM uses a content-based soft attention to control reading and writing and enables access to every memory cell and gradient flow from these units. DNC defines a differentiable free list to track the usage of each memory location to address temporal orders of memory. Yang~\cite{yang2016lie} generalized Turing machines to the continuous setting by storing memory on manifolds and controls memory addressing using Lie group actions that are differentiable. Besides NTM and DNC, there are other neural abstract machines:

\begin{itemize}
  \item Neural Programmer~\cite{neelakantan2015neural}
  \item Neural Programmer-Interpreter~\cite{reed2015neural}
  \item Neural RAM~\cite{kurach2015neural}
  \item Neural Stack~\cite{joulin2015inferring}
  \item Neural Program Lattices~\cite{li2016neural}
  \item Neural GPU~\cite{kaiser2016neural}
\end{itemize}

Neural abstract machine is a flexible and powerful class of DL models that is capable of inferring implicit graph structure, which can be used to simulate modern program executions. Instead of constructing an explicit program representation, they learn the model weights to describe the latent procedure and then simulate the dataflow of program execution. Although neural abstract machines behave similarly to programs and they are also flexible enough to emulate program execution, there is not yet a clear way of extracting program representations directly from these models. Besides program learning, this type of model has also been used in other domains such as question answering~\cite{graves2016hybrid,rae2016scaling}, finding the shortest path~\cite{graves2016hybrid}, navigation in 3D simulation~\cite{parisotto2017neural}, querying a database for entity relations~\cite{graves2016hybrid}, and one-shot recognition~\cite{rae2016scaling}.

\section{Deep Learning for Big Code Applications}
\label{sec:applications}
\begin{figure*}[!t]
  \centering
    \includegraphics[width=14.3cm,keepaspectratio]{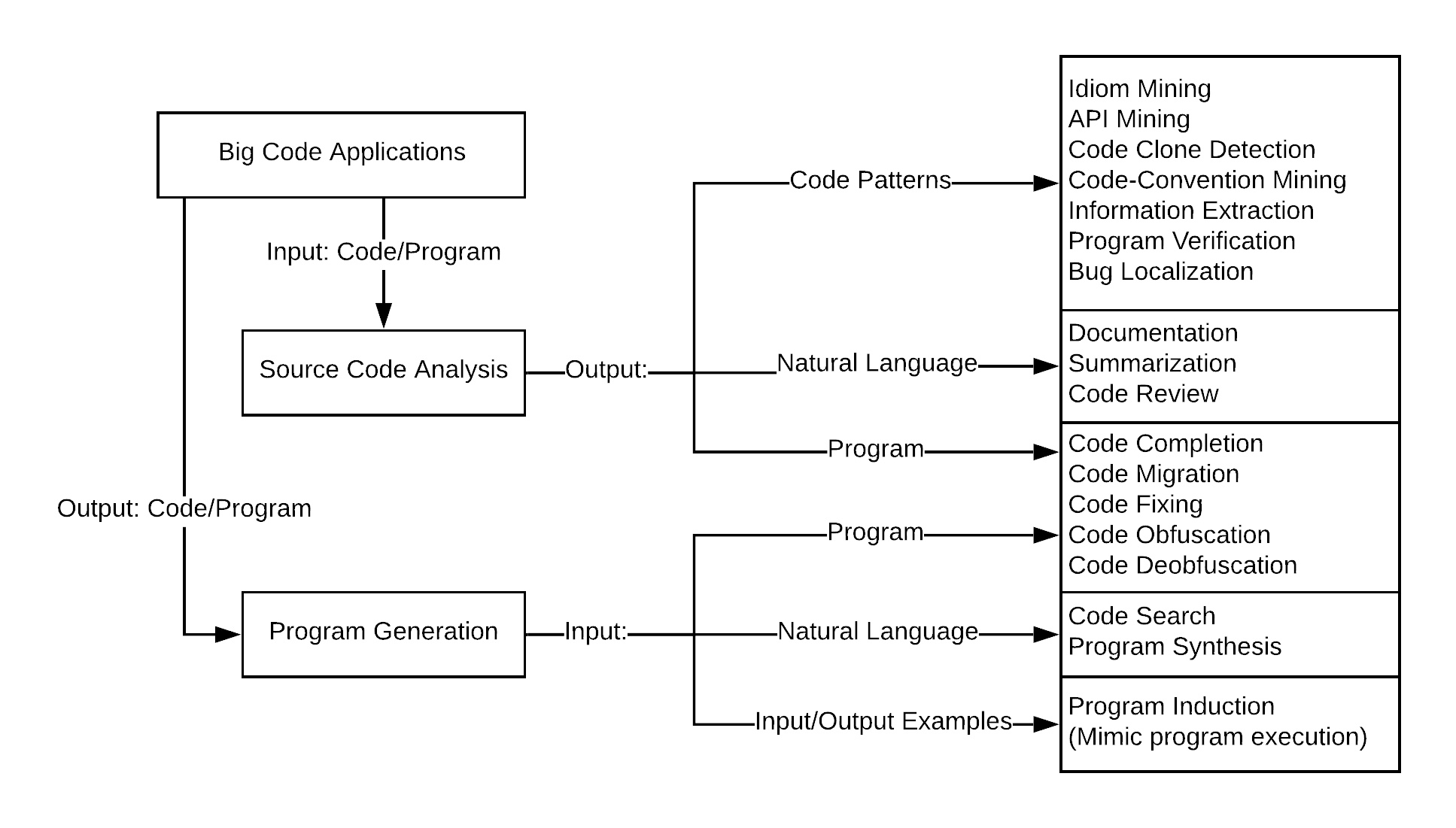}
    \vspace{-15pt}
  \caption{A taxonomy of Big Code applications based on their inputs and outputs.}
  \label{fig:big_code_applications}
  \vspace{-15pt}
\end{figure*}

To review Deep Learning (DL) for Big Code applications, we categorize the input and output formats of different tasks into \textbf{source code analysis} and \textbf{program generation} as shown in Figure~\ref{fig:big_code_applications}. For \textbf{source code analysis}, the input is source code and the output can take on many forms such as natural language, code fragments/pattern or a whole program. For \textbf{program generation}, all tasks have the same output in the format of source code, but different inputs (e.g., code, natural language or input/output examples). The inputs of both source code analysis and program generation in our taxonomy are sequences, which can be modeled by DL models under encoder-decoder framework as introduced in sections~\ref{sec:seq2seq} and
~\ref{sec:deep_language_models}. In addition, this taxonomy covers most of input and output types so any new Big Code tasks can be categorized easily. Then, suitable DL models can be selected accordingly to solve such tasks. DL for Big Code is growing very fast; thus, it is nearly impossible to include every single work for each presented application. Also, direct one-to-one performance comparison between the reviewed models of each application may not be possible due to different datasets used and/or lack of reported results in the original works. However, we still believe that this review is a good starting point for practitioners and researchers working in this emerging area.

\subsection{Source code analysis}
Source code analysis tasks take source code as context and generate outputs in another format. Source code analysis utilizes the distribution learned from large code corpus and performs various kinds of predictions. Firstly, the case when the outputs are code patterns/elements is presented.

\textbf{Idiom mining} extracts the code segments that reappear across projects. Most common code idioms often describe important programming concepts and can be reused across projects~\cite{allamanis2013mining,allamanis2014mining}. A related task to idiom mining is predicting class/method names based on their context/bodies~\cite{allamanis2015suggesting}, which can be generalized to source code classification. One of the first DL work on programming language processing was proposed by Mou et al.~\mbox{\cite{mou2016convolutional}}. This study proposed a tree-based CNN (TBCNN) with dynamic pooling learned directly from an AST of a program. The authors demonstrated that their learned feature vectors of program tokens with TBCNN could be grouped together in terms of functionality. Such representations were also demonstrated to be more effective than \mbox{\textit{n}}-gram (Bag-of-words) methods for identifying programming tasks and detecting bubble sort patterns. Later, several studies utilizing different structural information of code (AST paths~\mbox{\cite{alon2019code2vec}} or data-/control-flow information~\mbox{\cite{allamanis2018learning}}) achieved strong performance for these tasks as well.

\textbf{Application Programming Interface (API) mining} and \textbf{Code clone detection} witness many uses of DL models. More particularly, DeepAPI~\cite{gu2016deep} was devised to learn a distributed representation using a deep RNN seq2seq model for both user queries and associated APIs. This work was found to perform better than the bag-of-word approach for API generation task. After that, many DL works modeling ASTs of source code (e.g., RtNN~\mbox{\cite{white2016deep}}, Tree-LSTM~\mbox{\cite{wei2017supervised}}, ASTNN~\mbox{\cite{zhang2019novel}}) also obtained high detection rates for code clones.

\textbf{Code-convention mining} finds the code practices recommended by a specific programming language but not enforced by compilers. These coding conventions (e.g., indentation, naming conventions, white space) help improve the readability and maintainability of source code. A pioneering work using machine learning in this direction was proposed by Allamanis et al.~\cite{allamanis2014learning} in 2014 using $n$-gram features combined with Support Vector Machine model. Nevertheless, after that, no other work has been reported to directly use DL for this code-convention mining. However, it is observed that recent DL models that effectively capture the usage contexts of source code (e.g.,~\mbox{\cite{allamanis2018learning,cvitkovic2018open}}) can be investigated for this task.

\textbf{Information extraction} aims to identify the existence of some code snippets, program or software-related artifacts from natural language~\cite{cerulo2013hidden, sharma2015nirmal}, images or videos~\cite{yadid2016extracting, ponzanelli2016codetube}. With recent advances of DL models (e.g., CNN) in computer vision, more works~\cite{ott2018learning, ott2018deep} are emerging in this direction to detect code/software elements from image and videos. Parallel to CNN, RNN has also been utilized to separate code fragments from natural language on Stack Overflow forum~\cite{yin2018learning}.

\textbf{Program verification} predicts whether there exist bugs or security issues in a program. \textbf{Bug localization} is closely related to program verification, except that besides code, the locations of specific types of bugs are also identified. Besides formal methods~\cite{madsen2017model} and traditional ML approaches~\cite{chakkrit2016towards,ray2016naturalness,le2015information,wang2014compositional}, DL has been shown to perform quite well for this task. DeepBugs~\cite{pradel2018deepbugs} represented more than 150,000 JavaScript source code files using word2vec and then trained a feed-forward neural network to distinguish between buggy and non-buggy code. This approach was found to achieve an accuracy of more than 90\% and have the potential to perform in real-time. Another study~\mbox{\cite{xiao2019improving}} using a word-embedding CNN architecture for both bug reports and source code files outperformed the existing DL works~\mbox{\cite{xiao2017improving, lam2017bug}} for bug localization. Additionally, VulDeePecker~\cite{li2018vuldeepecker} utilized a bidirectional LSTM (bi-LSTM) model with \textit{code gadgets} as input to identify multiple types of vulnerabilities in source code. There is also a comprehensive review~\cite{ghaffarian2017software} on vulnerability analysis and discovery from source code using ML and DL techniques.

The reviewed DL models for this case have either used sequential and/or structural (i.e., ASTs) code representation. Most of the DL models were better the non-DL ones, while the structural models seemed to have stronger performance than the sequential counterparts.

In other scenarios of source code analysis, the outputs can be natural language, which leads to the following tasks:

\textbf{Documentation} is an important artifact embedded in a software program to help annotate the requirements, operation, and uses of source code as well as make software maintenance easier. Inspired by Neural Machine Translation~\cite{bahdanau2014neural}, Barone et al.~\cite{barone2017parallel} utilized the same model to create the baseline for the task of documentation generation. Later, Hu et al.~\cite{hu2018deep} proposed an LSTM model with attention, DeepCom, to automatically generate the documentation for Java code. Recently, a novel model, code2seq~\mbox{\cite{alon2018code2seq}}, with an attentional decoder to select optimal compositional paths of an AST has been proposed to represent code snippets. Code2seq has achieved better performance than DeepCom for code documentation.

\textbf{Summarization} is a sub-task of documentation, in which the main functionality of source code or a function is briefly described. In 2015, Rush et al.~\cite{rush2015neural} proposed SUM-NN (i.e., a neural attention model) for code summarization, and outperformed existing information retrieval method (i.e., minimizing cosine distance between code and the corresponding summary) and phrase-based systems~\cite{koehn2007moses}. However, this model tended to generate short descriptions. To overcome this issue, Iyer et al.~\cite{iyer2016summarizing} introduced an improved neural attention model namely CODE-NN by replacing feed-forward neural network with LSTM for the decoder in a seq2seq model. Chen et al.~\cite{chen2018neural} presented a bimodal using two Variational AutoEncoders (i.e., one for natural language and one for source code) to support code retrieval and summarization for C\# and SQL languages. To leverage the knowledge of API for summarizing source code, Hu et al.~\mbox{\cite{hu2018summarizing}} combined the representation of API sequences learned from related API summarization tasks with code sequences in a seq2seq model with attention. Like code documentation, the code2seq structural DL model~\mbox{\cite{alon2018code2seq}} has recently outperformed other existing works for code summarization task.

\textbf{Code review} is an important step in software development since it helps identify bad coding practices that may lead to software defects. However, this step is mostly carried out manually, which is time-consuming and prone to error. To automate this process, DeepCodeReviewer~\cite{gupta2018intelligent} aimed to find relevant reviews for the current code snippets/program. Specifically, four separate LSTM models with word2vec embeddings were trained for different parts of source code and reviews. Then, the results were combined using a feed-forward neural network to determine the relevancy of the current review. We found that there is still not much DL work in this area, which can stimulate future efforts to improve the performance of deep code-review models.

We see many modern DL algorithms from neural language processing applied to these tasks, which demonstrates the similarity between these two fields. Some recent techniques such as structural embeddings and attention mechanisms (cf. section~\mbox{\ref{subsection:code_completion}}) have also been integrated into DL models, which enables the models to learn from massive previous knowledge and to be more expressive to capture the underlying structure of the given data. The case when both input and output are code is discussed in the next section along with other \textit{program generation} tasks.

\subsection{Program generation}
Program generation tasks require the inference about the program code or code structure. We classify program generation applications into three categories based on their inputs: (\textit{i}) unconditional program generation, (\textit{ii)} program transduction and (\textit{iii)} multimodal program generation.

In \textbf{unconditional program generation}, the input consists of only code corpus. The goal is to generate the next most likely tokens or draw samples similar to the current input. Code completion/suggestion is a typical task in this category.

\textbf{Code completion} is a useful and common feature that many code editors offer. Although most code completion tools incorporated into the Integrated Development Environments (IDEs) are mostly rule-based\footnote{https://www.jetbrains.com/help/idea/auto-completing-code.html}, learning-based code completion has been an active field of study. Completion algorithms have been developed for different languages with high accuracy and flexibility~\cite{gvero2013complete,raychev2014code,bielik2016phog}. These algorithms help enhance IDEs with better code suggestion, API calls or components based on the current context. \textit{Program generative models} is an extreme case for completion where no input is given and tokens are sampled from the first one to the end as a Markov chain~\cite{maddison2014structured,nguyen2013statistical,nguyen2015graph}. Table~\mbox{\ref{tab:neural_completion_models}} presents several representative works for neural code completion. The deep code completion models have utilized the recent practices presented in section~\mbox{\ref{subsection:code_completion}} including (\textit{i}) structural information (e.g., AST~\mbox{\cite{liu2016neural,li2017code}}), (\textit{ii}) open vocabulary (e.g., character-level features~\mbox{\cite{karpathy2016unreasonable}}), and (\textit{iii}) attention mechanism (e.g., pointer network~\mbox{\cite{bhoopchand2016learning}}). It is noted that these models can also be used as deep encoder models (cf. section~\mbox{\ref{subsection:code_completion}}) for other program generation tasks.

\begin{table}[!t]
\renewcommand{\arraystretch}{1.3}
\caption{Deep Code Completion Models. \textbf{Notes:} MLP: Multi-Layer Perceptron, RNN: Recurrent Neural Network, LSTM: Long Short-Term Memory.}
\label{tab:neural_completion_models}
\centering
\begin{tabular}{|l|l|l|}
\hline
\multicolumn{1}{|c|}{\textbf{Model}} & \makecell{\textbf{Code}\\ \textbf{representation}} & \multicolumn{1}{c|}{\textbf{Study}} \\
\hline
\multirow{3}{*}{LSTM} & word & Dam et al. 2016~\cite{khanh2016deep}\\
\cline{2-3} & character & Karpathy 2015~\cite{karpathy2016unreasonable}\\
\cline{2-3} & AST node & Liu et al. 2016~\cite{liu2016neural}\\
\hline
\multirow{2}{*}{LSTM \& Pointer Net} & word & Bhoopchand et al. 2016~\cite{bhoopchand2016learning}\\
\cline{2-3} & AST node & Li et al. 2017~\cite{li2017code}\\
\hline
\multicolumn{1}{|l|}{RNN} & word & White et al. 2015~\cite{white2015toward}\\
\hline
\multicolumn{1}{|l|}{MLP with attention} & word & Das et al. 2015~\cite{das2015contextual}\\
\hline
\end{tabular}
\end{table}

In \textbf{program transduction}, the goal is to convert source code into another form of code. The following applications fall into this category.

\textbf{Code migration} helps developers to port projects in one language to another~\cite{aggarwal2015using}. It is a common case to upgrade source code for a higher version of language or framework. Automatic transducers to update APIs and code structure are very useful for development and deployment. One such API migration from Java to C\# was carried out using a seq2seq model namely DeepAM~\cite{gu2017deepam}. A similar motivation was proposed by Nguyen et al.~\cite{nguyen2017exploring} to migrate APIs from Java to C\#, yet still preserving the semantic information by learning the word2vec embeddings of pair-wise APIs.

\textbf{Code fixing} or \textbf{Code repair} is the next step after program verification and bug localization, in which bugs need to be fixed. In 2016, sk\_p model using LSTM-based seq2seq framework was proposed to correct seven Python assignments in Massive open online courses, namely MITx, and this model achieved an average 29\% of accuracy for error correcting. SynFix~\cite{bhatia2016automated} and DeepFix~\cite{gupta2017deepfix} with LSTM and GRUs, respectively, were trained on student assignments to fix syntax errors, which obtained roughly a complete fix rate of 30\%. Another work~\cite{santos2018syntax} trained a combined model of LSTM and $n$-gram on a large Java corpus from GitHub, which interestingly reported that 10-gram model slightly outperformed the LSTM counterpart for fixing syntax errors. This finding suggests that a more sophisticated (e.g., incorporating syntactic code structure) and better fine-tuned DL model can be utilized to further improve the result. There is also an extensive review on this topic~\cite{monperrus2018automatic}.

\textbf{Code obfuscation} prevents unauthorized people from analyzing and stealing source code, and thus protects the intellectual properties. There are some off-the-shelf obfuscated code generators such as Allatori\footnote{http://www.allatori.com/} and ProGuard\footnote{https://www.guardsquare.com/en/products/proguard}. Statistical ML language model was also used for this task~\cite{liu2016towards}, but DL models have not been much explored, except for some simple cases~\cite{Ma2014ControlFO}. However, this task is a good fit for encoder-decoder framework since the input is code sequence and the output is obfuscated text sequence of such input code. A different but related task, identifying obfuscated code, has witnessed more applications of DL models~\mbox{\cite{skolka2019anything, wang2016deep}}. The automatically generated features of such works can give some insights into designing effective (DL) methods for code obfuscation.

\textbf{Code deobfuscation} is opposite to obfuscation, in which deobfuscation recovers the original version of source code from the obfuscated one. There are deobfuscation tools for JavaScript~\cite{raychev2015predicting, vasilescu2017recovering,aebersold2016detecting} and Android apps\footnote{http://apk-deguard.com/}. A DL-based approach~\mbox{\cite{bavishi2018context2name}} namely Context2Name was proposed to recover natural variable names for minified code. It first used a deep sequential auto-encoder to extract embeddings for identifiers from their usage contexts. Such embeddings were then fed into an RNN to infer natural variable names. This approach was demonstrated to outperform the state-of-the-art Javascript deobfuscation tools like JSNice\footnote{http://www.jsnice.org/} and JSNaughty\footnote{http://jsnaughty.org/}. Further research on DL models for code deobfuscation can be done since DL for image deblurring, demosaicing and inpainting is being actively investigated~\cite{mcpherson2016defeating}.

As mentioned in section~\mbox{\ref{subsection:code_completion}}, to get more accurate generation of programs, code completion usually takes AST node sequences as input. Attention and copying mechanism also improve the performance of code generative models. However, high-quality code transduction is hard to achieve by sentence-by-sentence transferring, since the differences in programming language properties and designs may require code structure to be changed. Back-translation and generative models can be helpful in these cases, which is similar to image transferring~\mbox{\cite{zhu2017unpaired}}.

A more challenging category is \textbf{multimodal program generation}, where the input type is not restricted. Natural language (e.g., documentation and comments), GUI screenshots~\cite{beltramelli2017pix2code} and speech can all be used. Related applications are listed as follows.

\textbf{Code search} returns the best matching snippets of existing source code based on natural language queries. A log-bilinear neural language model was used to enable retrieving source code with natural language and vice versa~\cite{allamanis2015bimodal}. As explained in Section~\ref{subsection:embedding_models}, log-bilinear model alone, however, is unable to capture long code dependencies, which is essential for code search, especially for long code segments. Later, a deep recurrent model, CODEnn (i.e., bi-LSTM units combined with max-pooling layer), was proposed by Gu et al.~\cite{gu2018deep}. This model was shown to outperform the previous state-of-the-art results of CodeHow~\cite{lv2015codehow} (an extended boolean model) for code search.

\textbf{Program synthesis} extends code completion by generating code based on many forms of information such as natural language, images, and speech. Previously, applications were limited to DSL search~\cite{allamanis2015bimodal}. With DL, general-purpose program synthesis for various tasks can be now tackled as shown in Table~\mbox{\ref{tab:multimodal_models}}. As mentioned in section~\mbox{\ref{subsection:program_synthesis}}, to generate syntactically correct programs, many studies (e.g.,~\mbox{\cite{parisotto2016neuro,dong2016language,rabinovich2017abstract,yin2017syntactic}}) have proposed to utilize AST-based instead of sequential decoder. Such decoders predicted AST nodes sequentially using RNNs (e.g., LSTM), which could be computationally expensive. Later, Sun et al.~\mbox{\cite{sun2019grammar}} proposed a grammar-based structural CNN with attention mechanisms to replace RNN in the decoder for generating code from natural language description. The structural CNN-based decoder generated a grammar rule (sequence of tokens) per step instead of token-by-token, which makes the decoding process more compact and efficient. This approach was also shown to achieve the state-of-the-art results for Python code generation using the HeartStone benchmark dataset (cf. section~\mbox{\ref{subsection:data_sca}}). Recently, Brockschmidt et al.~\mbox{\cite{brockschmidt2018generative}} extended the graph-based code representation~\mbox{\cite{allamanis2018learning}} to code generation by augmenting the AST with \textit{inherited} and \textit{synthesized} nodes to capture the attribute grammars~\mbox{\cite{knuth1968semantics}} during the decoding process. The Graph2Graph model of this work generated more accurate C\# code samples compared to existing methods. With CNNs, graphics programs can be inferred from their output drawings. Ellis et al.~\mbox{\cite{ellis2017learning}} trained a hierarchical neural net to convert simple hand drawings into a DSL, which is then converted into LaTeX code with a bias-optimal search algorithm~\mbox{\cite{schmidhuber2004optimal}}. Beltramelli et al.~\mbox{\cite{beltramelli2017pix2code}} used an encoder-decoder architecture to generate front-end code from GUI screenshots.

\begin{table}[!t]
\renewcommand{\arraystretch}{1.3}
\caption{Deep Program Synthesis Models. \textbf{Notes:} MLP: Multi-Layer Perceptron, LSTM: Long Short-Term Memory, STN: Spatial Transformer Network~\cite{jaderberg2015spatial}.}
\label{tab:multimodal_models}
\centering
\begin{tabular}{|l|l|l|}
\hline
\multicolumn{1}{|c|}{\textbf{Model}} & \multicolumn{1}{c|}{\textbf{Generating Target}} & \multicolumn{1}{c|}{\textbf{Study}} \\
\hline
\multirow{2}{*}{Seq2seq} & Bash shell & Lin et al. 2017~\cite{lin2017program}\\
\cline{2-3} & CSS and HTML & Beltramelli 2017~\cite{beltramelli2017pix2code}\\
\hline
\multicolumn{1}{|l|}{Seq2seq with grammar constraint} & Simple C code & Amodio et al. 2017~\cite{amodio2017neural}\\
\hline
\multirow{3}{*}{Seq2AST} & Domain-specific language & Dong and Lapata 2016~\cite{dong2016language}\\
\cline{2-3} & Card-game code & Rabinovich et al. 2017~\cite{rabinovich2017abstract}\\
\cline{2-3} & General Python code & Yin and Neubig 2017~\cite{yin2017syntactic}\\
\hline
\multicolumn{1}{|l|}{Seq2Set} & SQL queries & Xu et al. 2017~\cite{xu2017sqlnet}\\
\hline
\multicolumn{1}{|l|}{Graph2Graph} & General C\# code & Brockschmidt et al. 2018~\mbox{\cite{brockschmidt2018generative}}\\
\hline
\multicolumn{1}{|l|}{Pointer Network} & Card-game code & Ling et al. 2016~\cite{ling2016latent}\\
\hline
\multicolumn{1}{|l|}{STNs and MLPs} & LaTeX Graphs & Ellis et al. 2017~\cite{ellis2017learning}\\
\hline
\multicolumn{1}{|l|}{Reinforcement learning} & SQL queries & Zhong et al. 2017~\cite{zhong2017seq2sql}\\
\hline
\end{tabular}
\end{table}

\textbf{Program induction} seeks to fit a given pair of input/output examples to mimic program execution, in which the execution correctness is more important than the readability of source
code. Because of the complexity of the problem space, targeted programs are often limited to certain forms of DSL or only some simple problems. Balog et al.~\mbox{\cite{balog2016deepcoder}} used a seq2seq model, namely DeepCoder, to predict the probable DSL functions required to map the given inputs to outputs, which reduced the search space and made the program induction process 10x faster than the corresponding search-based counterparts. Another class of models is \textit{differentiable interpreters}, in which pre-defined grammars of a DSL are continuously parameterized with neural networks. Such parametrization allows a model to search the program for a given input-output pair more efficiently. Evans and Grefenstette~\mbox{\cite{evans2017learning}} made inductive logic programming differentiable by using a neural network to perform inference given generated clauses, their weights and valuation of the axions. The model is end-to-end trainable and generalizable well on small datasets. However, this method is very memory-intensive. Riedel et al.~\mbox{\cite{riedel2016programming}} proposed a differentiable interpreter for Forth programming language to use input-output examples to create a complete program. Differentiable approaches often perform worse than the search-based methods for low-level programming languages (e.g., Assembly)~\mbox{\cite{gaunt2016terpret}}. Also, differentiable interpreters are still limited to solving only simple problems (e.g., accessing, sorting or copying array elements)~\mbox{\cite{feser2016differentiable}}. Neural abstract machines (cf. section~\mbox{\ref{subsection:program_induction}}) are also suitable for program induction. For example, by learning directly from recursions, Cai et al.~\mbox{\cite{cai2017making}} extended the ability of Neural Programmer-Interpreter~\mbox{\cite{reed2015neural}} and provided a generalization proof about the overall system behavior. There is also a recent review~\mbox{\cite{kant2018recent}} on program induction.

We observe that most recent works claiming the state-of-the-art results on various Big Code generation tasks have used some variants of RNN (e.g., LSTM/bi-LSTM/GRU with attention mechanism). Also, embeddings extracted from code tokens and ASTs are common choices for the input of these DL models. However, the proposed DL models have only been investigated for a few applications. There is still lack of extensive ablation studies to test the generalizability of these models for many different Big Code tasks. Next, in section~\mbox{\ref{sec:datasets}}, various datasets are presented to facilitate the building of deep source code models.

\section{Datasets for Big Code Applications}
\label{sec:datasets}
There is a great need of large datasets for Deep Learning (DL) in general and deep code models in particular to exhibit their power~\cite{goodfellow2016deep}. In this section, two types of corpora are presented for source code analysis and program generation (cf. Figure~\ref{fig:big_code_applications}). This section is mainly devoted to highlight the potential, but not yet established development of datasets for Big Code tasks.
\subsection{Datasets for source code analysis}
\label{subsection:data_sca}
There is a growing curated list for various code analysis tasks and datasets\footnote{http://learnbigcode.github.io/}. Currently, there are many unlabeled and large-scale open-source code corpora such as SourceForge\footnote{https://sourceforge.net/} and GitHub. Such datasets are widely used for building a probabilistic model on source code. Some large code corpora and their usefulness for various tasks including code completion and code pattern mining are presented hereafter. These datasets can be utilized for other source code analysis tasks as well.
\begin{itemize}
    \item Karpathy et al.\footnote{http://karpathy.github.io/2015/05/21/rnn-effectiveness/} used the Linux Kernel repository on GitHub\footnote{https://github.com/torvalds/linux} to demonstrate the expressive ability of RNNs. Later, this dataset has also been used to evaluate code completion tasks.
    \item Java projects on SourceForge and GitHub are utilized for various code mining tasks~\cite{nguyen2015graph}.
    \item Another large Java corpus containing all forked projects crawled from GitHub with more than one billion code tokens was also used to design new code complexity metrics and enhance the performance of code suggestion tasks ~\cite{allamanis2013mining}.
    \item Instead of working directly on source code, code can be first converted to ASTs, upon which a code language model is built. More particularly, Bielik et al.~\cite{bielik2016phog} and Raychev et al.~\cite{raychev2016probabilistic}  have used 100,000+ JavaScript\footnote{http://www.srl.inf.ethz.ch/js150.php} and  Python\footnote{http://www.srl.inf.ethz.ch/py150} programs, respectively, to build their language models for code completion tasks.
\end{itemize}
Although the aforementioned corpora are useful for building a robust probabilistic model on source code, there are two major problems with these unlabeled datasets. Firstly, without further annotations, only unsupervised learning can be applied. Secondly, there is no official benchmark dedicated to these tasks, researchers often train their models on their own datasets, which leads to a challenge when comparing the performance between approaches.

There have been some efforts to create datasets with labels for bug identification and code clone detection tasks. For these tasks, issue tracking systems like JIRA or BugZilla and version control systems like Git provide a huge amount of information about software development activities.

\begin{itemize}
    \item Lamkanfi et al.~\cite{lamkanfi2013eclipse} proposed a bug dataset that combined data spanning over the whole bug-triage cycle from both defect tracking systems JIRA and BugZilla into a structured format. Such dataset motivates not only bug analyses but also reproducibility and comparison of the bug detection models in Eclipse and Mozilla.
    \item Later, Just et al. ~\cite{just2014defects4j} prepared a bug dataset namely Defects4J\footnote{https://github.com/rjust/defects4j}, which contains both buggy and fixed source code along with their commit messages for 395 bugs in six Java projects.
    \item A large-scale and continuously growing dataset~\mbox{\cite{tomassi2019bugswarm}} of failed builds and successful fixes was automatically curated from GitHub and Travis-CI\footnote{https://travis-ci.org/}.
    \item BigCloneBench~\cite{svajlenko2014towards}
        is a similar effort for code clone detection tasks. This dataset\footnote{https://github.com/clonebench/BigCloneBench} consists of the clones detected from IJaDataset containing 25,000 open-source Java projects. BigCloneBench has been carefully verified by experts to provide a reliable evaluation benchmark for future research on detecting various types of code clones.
\end{itemize}

\subsection{Datasets for program generation}
For \textit{code completion}, there have been several useful datasets mentioned in the previous section. For other program generation tasks, natural language annotation (or other forms of information) is also required as input. Existing works~\cite{allamanis2015bimodal, iyer2016summarizing, quirk2015language, liu2016latent} have shown that online forums such as Stack Overflow, IFTTT\footnote{https://ifttt.com} can provide an abundant amount of code and related discussions. However, since these sources are user-generated, they are noisy and unaligned with any format. Other existing datasets are carefully annotated manually, but most of them are still limited in quantity. Some examples of such datasets for program synthesis are reviewed here.
\begin{itemize}
    \item Project Euler\footnote{https://projecteuler.net/} is an open platform containing over 600 programming and mathematical problems. New problem is constantly added every two weeks. The description, pseudo-code and some sample inputs and outputs for each problem are included. This dataset can serve as a great source for program synthesis. Although most of the problems have been solved, there is still no public repository that contains completely tested and well-documented code for those problems. Oda et al.~\cite{oda2015learning} utilized this dataset to perform translation between Python and Japanese languages, but unfortunately, the detailed code was not disclosed.
    \item Another large-scale and perpetual dataset~\mbox{\cite{brown2014blackbox}}, namely Blackbox, of code edits and IDE usages have been collected from BlueJ Java IDE. This dataset can be used for various program generation tasks such as code completion, code fixing and program synthesis.
    \item Card2code\footnote{https://github.com/deepmind/card2code} is another high-quality dataset for program generation~\cite{ling2016latent}. The authors collected the card descriptions of two trading card games (i.e., Magic the Gathering and Hearthstone) and their corresponding open-source Python implementations to perform program generation. However, this dataset is too domain-specific and thus limits its applicability.
    \item Later, Oda et al.~\cite{barone2017parallel} presented a parallel Django dataset to address the limitations of the previous datasets. This dataset includes all source code of Django web framework with line-by-line English annotation, including functions for a wide variety of real-world use cases such as I/O operations, string manipulation and exception handling.
    \item Barone et al.~\cite{barone2017parallel} presented a large-scale dataset\footnote{https://github.com/EdinburghNLP/code-docstring-corpus} containing two main corpora. The first corpus contains more than 150,370 triplets of Python function declarations, bodies and their corresponding docstrings for code documentation generation. The second corpus consists of than 161,630 pairs of function declarations and their bodies to facilitate code generation.
    \item WikiSQL~\cite{zhong2017seq2sql} is another crowd-sourced dataset\footnote{https://github.com/salesforce/WikiSQL} for developing natural language interfaces for relational databases from user's queries.
    \item Yin et al.~\mbox{\cite{yin2018learning}} introduced a dataset using Stack Overflow posts, namely CoNaLa, which maps natural language intents of developers and the corresponding code snippets.
\end{itemize}

Besides program synthesis, program induction is another (special) form of program generation, in which the program datasets can always be generative with predefined rules and languages such as arithmetic, lists, group-theory and tree relations. It is noted that the dataset for program induction is still very limited. One of the few related works is the question dataset\footnote{https://github.com/anselmrothe/question\_dataset} collected by Rothe et al.~\cite{rothe2017question}. Such dataset contains 605 natural language questions querying about a board game situation, from which the algorithm/instructions can be learned by a machine to recover the original input-output pairs.

\section{Challenges and Directions of Deep Code Modeling and Generation}
\label{sec:challenges_directions}

Most source code models are limited in their applications since they usually perform significantly worse on corpora different from the training set, especially for complex program generation problems. The models are usually constrained to solve very specific tasks, and thus are not flexible enough to generate code distribution for general purposes. Moreover, a real-world programming problem is often specified in a multimodal way, i.e., described with natural language and demonstrated with example inputs and outputs in text, graphics or action sequences. In this case, a model has to solve program synthesis and induction problems simultaneously. Although it is natural for a human to consider both aspects while programming, there is no existing study using examples and descriptions together to build a neural program model that generates a program non-existent in the database. Tackling a real programming challenge requires a combination of program synthesis and induction models and depends on the development of representation learning and goal-oriented dialogue. However, a proper architecture for this task has yet to come. A model that includes such complex logic and power is hard to create or train because of the following challenges:
\begin{itemize}
    \item Collecting labeled data for this complex task requires massive annotation work. Therefore, the data must be used effectively during the training process.
    \item Deep Learning (DL) models with differentiable variables are not very good at representing discrete features, which is crucial to learn the programming rules and control flows.
    \item A proper evaluation metric that can incorporate both semantic meaning, grammatical and execution correctness is important. However, there is little work in that direction.
    \item Deep source code models tend to be overly complex since a large proportion of parameters are dedicated to solving problems in which real programmers have no difficulty. For example, extensive computation has to be done in memory networks to learn how to copy variable names from previous occurrences. Therefore, we should learn from human experience to devise ways to simplify such procedures.
\end{itemize}

\subsection{Data efficiency}
Real-world programs have a wide range of functionalities, which is difficult to cover with sets of annotated examples. Program context also changes rapidly, e.g., the meaning of local variables changes between function scopes. Thus, we have to find ways to make the best use of data, which in turn asks for a good model and training method that can learn fast and generalize well.

\subsubsection{Unsupervised learning}
To learn translation without access to a parallel corpus,
Lample et al.~\cite{lample2017unsupervised} tried to circumvent the lack of parallel datasets in machine translation with adversarial training. They used Denoising Auto-encoder to learn sentence embedding and adversarial training~\cite{ganin2016domain} on the encoded domain to classify source and target embeddings. We have seen some early signs of using adversarial learning for program repair~\mbox{\cite{harer2018learning}}. However, extraction of generic representations from unsupervised learning is still not the dominant approach~\cite{oord2017neural}.

Reinforcement Learning (RL) is also helpful when the labeled data is scarce. It has been adopted in image captioning to generate more natural photo descriptions. An interesting crossover field for RL and programming is natural language interfaces for system control~\cite{laengle1995kantra,perzanowski2001building}. Automatic analysis of multimodal instructions has the potential to change the way how we develop and use software. Instead of using a rule-based control system, we can use ML and DL to train agents on examples or simulations. Guu et al.~\cite{guu2017language} trained an agent to parse natural instructions to generate code in a stack-based language in SCONE~\cite{dong2016language}. RL also enables intelligent systems to model continuous rewards or policies and interact with its environment to gain new knowledge, which enables grammar induction on even infinite search space.

The optimization target for an RL agent is very flexible. It can be integrated into the state-of-the-art conversation~\cite{zhou2017end} and captioning models~\cite{ren2017deep}. This learning scheme has the potential to be used in source code modeling to improve the generation quality or even guide a model to explore the context of the program. Misra et al.~\cite{misra2017learning} designed a learning model by asking the system for image understanding. By swapping the question selection module with a test selection, an automatic testing framework can be built. A few investigations of RL with DL models have also been conducted for code summarization~\mbox{\cite{wan2018improving}} and code fixing~\mbox{\cite{gupta2018deep}}.

\subsubsection{Weakly and semi-supervised training}
The problem of utilizing partially labeled data is called semi-supervised learning. A very closely related concept is weakly supervised learning, which originally means using fewer training samples by self-training. Recently, this term often refers to the cases of using noisy or partially labeled data, for example, enhancing semantic segmentation with only image-level labeled samples~\cite{shen2017weakly}. A widely adopted semi-supervised learning architecture is student-teacher framework~\cite{rasmus2015semi}. In the state-of-the-art method, Tarvainen et al.~\cite{tarvainen2017mean} constructed a stronger teacher model by taking the exponential moving average of the students. Such models can also be customized for program generation tasks when the labeled dataset is limited.

\subsubsection{Active learning}
Another way of dealing with a large quantity of unlabeled data is active learning, which aims to design an oracle that lets the agent select which samples should be labeled~\cite{settles2010active}. Konyushkova and Raphael~\cite{konyushkova2017learning} demonstrated that their Learning Active Learning algorithm is capable of solving real-world problems in a wide range of domains. Active learning problem can also be considered as a special case of multi-armed bandit~\cite{ganti2013building}. The restriction is that the labeling process totally cannot be abandoned completely. These active learning schemes can be applied in Big Code data collections.

\subsection{Discrete and symbolic representations}
Different from conventional DL models for sequence modeling, a program generation model should pay more attention to representing control flows and discrete rules.
\subsubsection{Representation learning}
Many Big Code tasks require a rapid understanding of source code context and generalization to other source code projects given the limited training data. Attempts to extract structural information from programs date back to grammatical inference in the 90s. Grammatical inference is the process of inducing, learning or inferring grammars~\cite{de2010grammatical}, which generates Finite-State Automata (FSA) of various types. In this field, researchers try to associate complexity theory, formal logic and discrete structures with learning. This field has been connected with many scientific disciplines, including bioinformatics, computational linguistics and lossless compression. Different from the statistical construction of language models~\cite{goodman2001bit,bengio2003neural}, where the probability distribution of transition states of a Markov model~\cite{gagniuc2017markov} is learned, the goal of grammar inference is to determine an explicit representation of a programming language.

However, these methods have limited expressive ability and difficulty handling real-world continuous space. RNNs are powerful enough to learn very complex cases. There are many FSA extraction algorithms from RNN, called \textit{RNN Rule Extraction} (RNN-RE) based on searching and sampling~\cite{jacobsson2005rule}. Sodsong et al.~\cite{sodsong2017spark} trained an RNN with program paths from OpenJDK to detect caller-sensitive method vulnerabilities~\cite{cifuentes2015understanding} and showed that grammar could be learned with CrySSMEx~\cite{jacobsson2006crystallizing}. But the generation process relied on very aggressive quantization and the trained complex RNN structure was hard to understand. RNN-RE has also been criticized to be "Fool's gold"~\cite{kolen1994fool}. Recent deep code models~\mbox{\cite{allamanis2018learning,cvitkovic2018open}} have captured control-/data-flow information of code using graph-based neural networks; however, their usage contexts are still pre-defined and limited to statically typed programming languages.

Deep generative models such as Generative Adversarial Nets ~\cite{goodfellow2014generative}, Variational AutoEncoders (VAEs)~\cite{kingma2013auto} and Autoregressive models~\cite{oord2016wavenet} have been widely used to learn this kind of representations. Recent advances in generative modeling of images~\cite{van2016conditional,goodfellow2014generative,gregor2016towards,kingma2016improved}, audio~\cite{oord2016wavenet,mehri2016samplernn} and videos~\cite{kalchbrenner2016video,finn2016unsupervised} have yielded impressive samples and applications~\cite{ledig2016photo,isola2016image,karras2017progressive}. Such generative models can be utilized to address the \textit{representation learning} issues in program generation. Deep generative models can be used to learn the probabilistic distributions of unlabeled code corpus. Vector representations can then be extracted from the distributions of these models.

\subsubsection{Discrete representation and addressing}
For language and code, more concise and discrete representations are required to conduct complex reasoning or predictions. Discrete learning can also simulate computer operations and improve model interpretability~\mbox{\cite{oord2017neural}}. Recent advances in discrete representation techniques are presented in this section for future adaptation to program modeling and generation.

Sometimes the representation has a desired explicit symbolic form. In this case, we can define the target of representation learning as predicting attributes to fill in the predefined structure. For example, Yang et al.~\cite{yang2017learning} extracted knowledge tuples from question answering datasets with a model similar to \textit{Neural Symbolic Machine}~\cite{liang2016neural}. However, it is more common that the structure is complex and unknown, where the algorithm has to infer the structure by itself.

We have seen some exciting signs of progress in discrete representation in other domains. For example, learning discrete hidden latent vectors in generative models enables interesting applications such as speaker transferring and video prediction. Oord et al.~\cite{oord2017neural} proposed VQ-VAE that used vector quantization to prevent posterior collapse and model discrete representation. A strong autoregressive model is used as the decoder for unsupervised speaker conversion tasks. VQ-VAE gives a great example of how to represent discrete rules with only continuous models and some non-linearity. Also, this model is trained self-supervised and has many potential applications in source code modeling and generation.

\iffull
\subsubsection{Discrete addressing}
\fi
Besides representation, deterministic programming logic can be simulated by discrete addressing in memory-augmented networks. Xu et al. 2015~\cite{xu2015show} proposed stochastic hard attention for image captioning by sampling just a column. During training, one would sample a set of sequences and compute gradient with an estimation similar to REINFORCE~\cite{williams1992simple} used for policy-gradient RL. This technique is also used for other memory-augmented models. Zaremba et al.~\cite{zaremba2015reinforcement} modified Neural Turing Machine to use discrete addressing and trained the model with REINFORCE. Bornschein et al.~\cite{bornschein2017variational} proposed a memory-augmented generative model that used a variational approximation for discrete addressing. They combined discrete memory addressing with continuous latent variables for generative few-shot learning. Discrete addressing has the advantage of definite logic representation; thus, it can model the logic of source code better.

\subsubsection{Training with discrete variables}
Training a discrete latent variable model turns out to be not easy. The gradient estimation based on random sampling such as REINFORCE always introduces high variance. In general, many approaches rely on control variates to reduce the variance.

There are several gradient estimators for discrete latent variable models such as IWAE~\cite{burda2015importance}, NVIL~\cite{mnih2014neural}, RWS~\cite{bornschein2014reweighted}. Tucker et al.~\cite{tucker2017rebar} proposed a low-variance, unbiased gradient estimates for discrete variables using control variates and Gumbel-Softmax~\cite{maddison2016concrete,jang2016categorical}.

\subsection{Real-world application and evaluation}
One major limitation of previous neural code completion models is that they did not provide a complete solution for real-world problems. More details are covered hereafter.
\begin{itemize}
    \item The model is trained on a fixed vocabulary. Thus, to generate new OoV values, the model has to be retrained. In addition, word-level code completion can only predict the next token after the typing of a complete word.
    \item Incomplete code leads to ambiguous parsing results for languages with complex grammars rather than LL(1). Thus, the current state cannot be mapped to a definite training input and this can make the prediction inaccurate.
\end{itemize}

More practical and engineering-friendly frameworks for rapid new concept learning and code generation are required. The following techniques can be adopted to deal with these limitations:
\begin{itemize}
    \item Open-vocabulary learning presented in section~\ref{subsection:code_completion} can address the OoV issue and partial-word code completion. There is a large vocabulary in code and much less regularity compared to natural language. However, code tokens still share similarities on the character and subword levels since variable/function names can contain the same parts of existing words in different order.
    \item PCFG is a natural fit for ambiguous parsing, and it could be integrated into a system. The evaluation process should be changed accordingly since the probabilistic parsing also introduces errors. To better account for such error in training, one can connect PCFG parameters to the learning objective and train the model end-to-end.
\end{itemize}

Furthermore, the evaluation methods being used for the code generation tasks are not perfect. AST node prediction accuracy is not a proper evaluation metric. Although syntactic information is better represented by AST node sequences, this is not the natural order of typing and the precision does not directly reflect the productivity gain of a tool. BLEU~\cite{papineni2002bleu} scores are very sensitive to tokenization. To deal with this problem, there is a standard evaluation code for comparable BLEU scores\footnote{https://github.com/awslabs/sockeye/tree/master/contrib/sacrebleu}. In addition, ROUGE~\cite{lin2004rouge} and BLEU often do not capture linguistic fluency and coherence~\cite{liu2016not}. This problem gets even worse when evaluating code generations. Bielik et al.~\cite{bielik2016phog} used precision and log probability for their probabilistic models. Practically, under a code completion setting, developers would find a tool useful if some of the predicted top-$k$ items hit~\cite{nguyen2015graph}.

A code-in-code-out system for generation and evaluation is more suitable for this situation. Even predicting AST sequences, code completion algorithms should include a converter between AST sequence and incomplete code. The evaluations then can be carried out by measuring the code-level accuracy. Ideally, execution correctness should be used for evaluating code generation, but it can be hard to quantify in most real cases. And the resulting error cannot be directly passed through gradient to the model parameters. Static code analysis and RL techniques can also be adopted.

\subsection{Human-like programming}
Current neural program models are still quite distant from mimicking real programmers. More specifically, developers do not usually implement everything from scratch, but they try to adapt existing code to suit their needs. One way is to divide complicated tasks into simple ones for neural program models to speed up the training and improve performance. For example, Hashimoto et al.~\mbox{\cite{hashimoto2018retrieve}} proposed to solve code generation by retrieving related examples and applying modifications. Other ways to achieve human-like programming would be to use copying mechanism~\cite{vinyals2015pointer} to reuse code as well as utilize both description and examples for program generation.
\subsubsection{Copying mechanism}
Training a model to attend to the right part of information could be tricky. Sometimes, attention tends to repeat itself and thus generates long and meaningless sequences. To deal with this problem, we have to add a structural constraint on the attention mechanism. In most cases, an almost diagonal attention matrix is what we want. On the other hand, complex addressing mechanisms like external memory networks do not perform well on language modeling tasks. Also, these models usually cannot be efficiently trained. The complexity of program generation can be greatly reduced if human-like behavior is considered. See et al.~\mbox{\cite{see2017get}} proposed a technique to copy words from input sequences to generate rare words. Such copying mechanism was adopted by a recent work~\mbox{\cite{chen2018sequencer}} to better handle OoV code tokens to enhance the robustness of automated program repair.

\subsubsection{Programming by description and examples}
In many real-world coding scenarios such as competitive coding\footnote{https://code.google.com/codejam/}, programmers write code by comprehending descriptions in natural language and then test their code with some input/output pairs. There are some search-based methods for program generation with both description and examples as context~\cite{polosukhin2018neural}. However, most models are not end-to-end trainable and they also have very limited capacity, which requires further research.

\section{Conclusions}
\label{sec:conclusions}

Artificial Intelligence (AI) in general and Deep Learning (DL) in particular have been increasingly leveraged for source code modeling. To facilitate more DL uses for practitioners and researchers in this area, our literature review first presented the limitations of existing source code models. We highlighted the potential of DL models to address these challenges and provided a more general solution for a wide range of problems. We then described important elements of encoder-decoder framework using DL models. We gave some recommendations on applying encoder-decoder framework with DL models to source code modeling and generation. Various Big Code applications (i.e., source code analysis and program generation) following encoder-decoder framework were then presented. We also identified the gaps between the state-of-the-art DL models and their applicability in source code modeling and generation. For some of these gaps, we proposed several suggestions on how to address them in future research.

We also want to provide some final thoughts about AI safety for source code modeling and generation. Recently, despite the impressive performance of DL in various domains, another trend of research is emerging to point out the vulnerability in the DL models that may result in bad consequences, namely \textit{adversarial DL}. For instance, an image can be perturbed to change the output of a DL model completely~\cite{nguyen2015deep,moosavi2016deepfool}, but a human cannot distinguish such a subtle change. Later, the idea of \textit{adversarial DL} was also adapted to models used for sequence modeling (e.g., RNN)~\cite{papernot2016crafting}. Such finding indicates that it is totally possible to fool source code models, e.g., turning a predicted vulnerable code snippet into a benign one. Therefore, besides aiming for high performance, practitioners and researchers in this area should also be aware of the robustness of a DL source code model to input data changes.

\bibliographystyle{ACM-Reference-Format}
\bibliography{reference}


\begin{thebibliography}{289}


\ifx \showCODEN    \undefined \def \showCODEN     #1{\unskip}     \fi
\ifx \showDOI      \undefined \def \showDOI       #1{#1}\fi
\ifx \showISBNx    \undefined \def \showISBNx     #1{\unskip}     \fi
\ifx \showISBNxiii \undefined \def \showISBNxiii  #1{\unskip}     \fi
\ifx \showISSN     \undefined \def \showISSN      #1{\unskip}     \fi
\ifx \showLCCN     \undefined \def \showLCCN      #1{\unskip}     \fi
\ifx \shownote     \undefined \def \shownote      #1{#1}          \fi
\ifx \showarticletitle \undefined \def \showarticletitle #1{#1}   \fi
\ifx \showURL      \undefined \def \showURL       {\relax}        \fi
\providecommand\bibfield[2]{#2}
\providecommand\bibinfo[2]{#2}
\providecommand\natexlab[1]{#1}
\providecommand\showeprint[2][]{arXiv:#2}

\bibitem[\protect\citeauthoryear{Aebersold, Kryszczuk, Paganoni, Tellenbach,
  and Trowbridge}{Aebersold et~al\mbox{.}}{2016}]%
        {aebersold2016detecting}
\bibfield{author}{\bibinfo{person}{Simon Aebersold}, \bibinfo{person}{Krzysztof
  Kryszczuk}, \bibinfo{person}{Sergio Paganoni}, \bibinfo{person}{Bernhard
  Tellenbach}, {and} \bibinfo{person}{Timothy Trowbridge}.}
  \bibinfo{year}{2016}\natexlab{}.
\newblock \showarticletitle{Detecting obfuscated javascripts using machine
  learning}. In \bibinfo{booktitle}{\emph{The 11th International Conference on
  Internet Monitoring and Protection (ICIMP). IARIA}}.
\newblock


\bibitem[\protect\citeauthoryear{Aggarwal, Salameh, and Hindle}{Aggarwal
  et~al\mbox{.}}{2015}]%
        {aggarwal2015using}
\bibfield{author}{\bibinfo{person}{Karan Aggarwal}, \bibinfo{person}{Mohammad
  Salameh}, {and} \bibinfo{person}{Abram Hindle}.}
  \bibinfo{year}{2015}\natexlab{}.
\newblock \bibinfo{booktitle}{\emph{Using machine translation for converting
  Python 2 to Python 3 code}}.
\newblock \bibinfo{type}{{T}echnical {R}eport}. \bibinfo{institution}{PeerJ
  PrePrints}.
\newblock


\bibitem[\protect\citeauthoryear{Allamanis, Barr, Bird, and Sutton}{Allamanis
  et~al\mbox{.}}{2014}]%
        {allamanis2014learning}
\bibfield{author}{\bibinfo{person}{Miltiadis Allamanis},
  \bibinfo{person}{Earl~T Barr}, \bibinfo{person}{Christian Bird}, {and}
  \bibinfo{person}{Charles Sutton}.} \bibinfo{year}{2014}\natexlab{}.
\newblock \showarticletitle{Learning natural coding conventions}. In
  \bibinfo{booktitle}{\emph{Proceedings of the 22nd ACM SIGSOFT International
  Symposium on Foundations of Software Engineering}}. ACM,
  \bibinfo{pages}{281--293}.
\newblock


\bibitem[\protect\citeauthoryear{Allamanis, Barr, Bird, and Sutton}{Allamanis
  et~al\mbox{.}}{2015a}]%
        {allamanis2015suggesting}
\bibfield{author}{\bibinfo{person}{Miltiadis Allamanis},
  \bibinfo{person}{Earl~T Barr}, \bibinfo{person}{Christian Bird}, {and}
  \bibinfo{person}{Charles Sutton}.} \bibinfo{year}{2015}\natexlab{a}.
\newblock \showarticletitle{Suggesting accurate method and class names}. In
  \bibinfo{booktitle}{\emph{Proceedings of the 2015 10th Joint Meeting on
  Foundations of Software Engineering}}. ACM, \bibinfo{pages}{38--49}.
\newblock


\bibitem[\protect\citeauthoryear{Allamanis, Barr, Devanbu, and
  Sutton}{Allamanis et~al\mbox{.}}{2018a}]%
        {allamanis2017survey}
\bibfield{author}{\bibinfo{person}{Miltiadis Allamanis},
  \bibinfo{person}{Earl~T. Barr}, \bibinfo{person}{Premkumar Devanbu}, {and}
  \bibinfo{person}{Charles Sutton}.} \bibinfo{year}{2018}\natexlab{a}.
\newblock \showarticletitle{A Survey of Machine Learning for Big Code and
  Naturalness}.
\newblock \bibinfo{journal}{\emph{ACM Comput. Surv.}} \bibinfo{volume}{51},
  \bibinfo{number}{4}, Article \bibinfo{articleno}{81} (\bibinfo{date}{July}
  \bibinfo{year}{2018}), \bibinfo{numpages}{37}~pages.
\newblock
\showISSN{0360-0300}


\bibitem[\protect\citeauthoryear{Allamanis, Brockschmidt, and
  Khademi}{Allamanis et~al\mbox{.}}{2018b}]%
        {allamanis2018learning}
\bibfield{author}{\bibinfo{person}{Miltiadis Allamanis}, \bibinfo{person}{Marc
  Brockschmidt}, {and} \bibinfo{person}{Mahmoud Khademi}.}
  \bibinfo{year}{2018}\natexlab{b}.
\newblock \showarticletitle{Learning to Represent Programs with Graphs}. In
  \bibinfo{booktitle}{\emph{Proceedings of the International Conference on
  Learning Representations (ICLR)}}.
\newblock


\bibitem[\protect\citeauthoryear{Allamanis and Sutton}{Allamanis and
  Sutton}{2013}]%
        {allamanis2013mining}
\bibfield{author}{\bibinfo{person}{Miltiadis Allamanis} {and}
  \bibinfo{person}{Charles Sutton}.} \bibinfo{year}{2013}\natexlab{}.
\newblock \showarticletitle{Mining source code repositories at massive scale
  using language modeling}. In \bibinfo{booktitle}{\emph{Proceedings of the
  10th Working Conference on Mining Software Repositories}}. IEEE Press,
  \bibinfo{pages}{207--216}.
\newblock


\bibitem[\protect\citeauthoryear{Allamanis and Sutton}{Allamanis and
  Sutton}{2014}]%
        {allamanis2014mining}
\bibfield{author}{\bibinfo{person}{Miltiadis Allamanis} {and}
  \bibinfo{person}{Charles Sutton}.} \bibinfo{year}{2014}\natexlab{}.
\newblock \showarticletitle{Mining idioms from source code}. In
  \bibinfo{booktitle}{\emph{Proceedings of the 22nd ACM SIGSOFT International
  Symposium on Foundations of Software Engineering}}. ACM,
  \bibinfo{pages}{472--483}.
\newblock


\bibitem[\protect\citeauthoryear{Allamanis, Tarlow, Gordon, and Wei}{Allamanis
  et~al\mbox{.}}{2015b}]%
        {allamanis2015bimodal}
\bibfield{author}{\bibinfo{person}{Miltos Allamanis}, \bibinfo{person}{Daniel
  Tarlow}, \bibinfo{person}{Andrew Gordon}, {and} \bibinfo{person}{Yi Wei}.}
  \bibinfo{year}{2015}\natexlab{b}.
\newblock \showarticletitle{Bimodal modelling of source code and natural
  language}. In \bibinfo{booktitle}{\emph{Proceedings of the 32nd International
  Conference on Machine Learning (ICML-15)}}. \bibinfo{pages}{2123--2132}.
\newblock


\bibitem[\protect\citeauthoryear{Alon, Brody, Levy, and Yahav}{Alon
  et~al\mbox{.}}{2018}]%
        {alon2018code2seq}
\bibfield{author}{\bibinfo{person}{Uri Alon}, \bibinfo{person}{Shaked Brody},
  \bibinfo{person}{Omer Levy}, {and} \bibinfo{person}{Eran Yahav}.}
  \bibinfo{year}{2018}\natexlab{}.
\newblock \showarticletitle{code2seq: Generating sequences from structured
  representations of code}.
\newblock \bibinfo{journal}{\emph{arXiv preprint arXiv:1808.01400}}
  (\bibinfo{year}{2018}).
\newblock


\bibitem[\protect\citeauthoryear{Alon, Zilberstein, Levy, and Yahav}{Alon
  et~al\mbox{.}}{2019}]%
        {alon2019code2vec}
\bibfield{author}{\bibinfo{person}{Uri Alon}, \bibinfo{person}{Meital
  Zilberstein}, \bibinfo{person}{Omer Levy}, {and} \bibinfo{person}{Eran
  Yahav}.} \bibinfo{year}{2019}\natexlab{}.
\newblock \showarticletitle{code2vec: Learning distributed representations of
  code}.
\newblock \bibinfo{journal}{\emph{Proceedings of the ACM on Programming
  Languages}} \bibinfo{volume}{3}, \bibinfo{number}{POPL}
  (\bibinfo{year}{2019}), \bibinfo{pages}{40}.
\newblock


\bibitem[\protect\citeauthoryear{Amodio, Chaudhuri, and Reps}{Amodio
  et~al\mbox{.}}{2017}]%
        {amodio2017neural}
\bibfield{author}{\bibinfo{person}{Matthew Amodio}, \bibinfo{person}{Swarat
  Chaudhuri}, {and} \bibinfo{person}{Thomas Reps}.}
  \bibinfo{year}{2017}\natexlab{}.
\newblock \showarticletitle{Neural Attribute Machines for Program Generation}.
\newblock \bibinfo{journal}{\emph{arXiv preprint arXiv:1705.09231}}
  (\bibinfo{year}{2017}).
\newblock


\bibitem[\protect\citeauthoryear{Arjovsky, Shah, and Bengio}{Arjovsky
  et~al\mbox{.}}{2016}]%
        {arjovsky2016unitary}
\bibfield{author}{\bibinfo{person}{Martin Arjovsky}, \bibinfo{person}{Amar
  Shah}, {and} \bibinfo{person}{Yoshua Bengio}.}
  \bibinfo{year}{2016}\natexlab{}.
\newblock \showarticletitle{Unitary evolution recurrent neural networks}. In
  \bibinfo{booktitle}{\emph{International Conference on Machine Learning}}.
  \bibinfo{pages}{1120--1128}.
\newblock


\bibitem[\protect\citeauthoryear{Ba, Hinton, Mnih, Leibo, and Ionescu}{Ba
  et~al\mbox{.}}{2016a}]%
        {ba2016using}
\bibfield{author}{\bibinfo{person}{Jimmy Ba}, \bibinfo{person}{Geoffrey~E
  Hinton}, \bibinfo{person}{Volodymyr Mnih}, \bibinfo{person}{Joel~Z Leibo},
  {and} \bibinfo{person}{Catalin Ionescu}.} \bibinfo{year}{2016}\natexlab{a}.
\newblock \showarticletitle{Using fast weights to attend to the recent past}.
  In \bibinfo{booktitle}{\emph{Advances in Neural Information Processing
  Systems}}. \bibinfo{pages}{4331--4339}.
\newblock


\bibitem[\protect\citeauthoryear{Ba, Kiros, and Hinton}{Ba
  et~al\mbox{.}}{2016b}]%
        {ba2016layer}
\bibfield{author}{\bibinfo{person}{Jimmy~Lei Ba}, \bibinfo{person}{Jamie~Ryan
  Kiros}, {and} \bibinfo{person}{Geoffrey~E Hinton}.}
  \bibinfo{year}{2016}\natexlab{b}.
\newblock \showarticletitle{Layer normalization}.
\newblock \bibinfo{journal}{\emph{arXiv preprint arXiv:1607.06450}}
  (\bibinfo{year}{2016}).
\newblock


\bibitem[\protect\citeauthoryear{Bahdanau, Cho, and Bengio}{Bahdanau
  et~al\mbox{.}}{2014}]%
        {bahdanau2014neural}
\bibfield{author}{\bibinfo{person}{Dzmitry Bahdanau},
  \bibinfo{person}{Kyunghyun Cho}, {and} \bibinfo{person}{Yoshua Bengio}.}
  \bibinfo{year}{2014}\natexlab{}.
\newblock \showarticletitle{Neural machine translation by jointly learning to
  align and translate}.
\newblock \bibinfo{journal}{\emph{arXiv preprint arXiv:1409.0473}}
  (\bibinfo{year}{2014}).
\newblock


\bibitem[\protect\citeauthoryear{Balduzzi and Ghifary}{Balduzzi and
  Ghifary}{2016}]%
        {balduzzi2016strongly}
\bibfield{author}{\bibinfo{person}{David Balduzzi} {and}
  \bibinfo{person}{Muhammad Ghifary}.} \bibinfo{year}{2016}\natexlab{}.
\newblock \showarticletitle{Strongly-typed recurrent neural networks}.
\newblock \bibinfo{journal}{\emph{arXiv preprint arXiv:1602.02218}}
  (\bibinfo{year}{2016}).
\newblock


\bibitem[\protect\citeauthoryear{Balog, Gaunt, Brockschmidt, Nowozin, and
  Tarlow}{Balog et~al\mbox{.}}{2016}]%
        {balog2016deepcoder}
\bibfield{author}{\bibinfo{person}{Matej Balog}, \bibinfo{person}{Alexander~L
  Gaunt}, \bibinfo{person}{Marc Brockschmidt}, \bibinfo{person}{Sebastian
  Nowozin}, {and} \bibinfo{person}{Daniel Tarlow}.}
  \bibinfo{year}{2016}\natexlab{}.
\newblock \showarticletitle{Deepcoder: Learning to write programs}.
\newblock \bibinfo{journal}{\emph{arXiv preprint arXiv:1611.01989}}
  (\bibinfo{year}{2016}).
\newblock


\bibitem[\protect\citeauthoryear{Barone and Sennrich}{Barone and
  Sennrich}{2017}]%
        {barone2017parallel}
\bibfield{author}{\bibinfo{person}{Antonio Valerio~Miceli Barone} {and}
  \bibinfo{person}{Rico Sennrich}.} \bibinfo{year}{2017}\natexlab{}.
\newblock \showarticletitle{A parallel corpus of Python functions and
  documentation strings for automated code documentation and code generation}.
\newblock \bibinfo{journal}{\emph{arXiv preprint arXiv:1707.02275}}
  (\bibinfo{year}{2017}).
\newblock


\bibitem[\protect\citeauthoryear{Bavishi, Pradel, and Sen}{Bavishi
  et~al\mbox{.}}{2018}]%
        {bavishi2018context2name}
\bibfield{author}{\bibinfo{person}{Rohan Bavishi}, \bibinfo{person}{Michael
  Pradel}, {and} \bibinfo{person}{Koushik Sen}.}
  \bibinfo{year}{2018}\natexlab{}.
\newblock \showarticletitle{Context2Name: A deep learning-based approach to
  infer natural variable names from usage contexts}.
\newblock \bibinfo{journal}{\emph{arXiv preprint arXiv:1809.05193}}
  (\bibinfo{year}{2018}).
\newblock


\bibitem[\protect\citeauthoryear{Beltramelli}{Beltramelli}{2017}]%
        {beltramelli2017pix2code}
\bibfield{author}{\bibinfo{person}{Tony Beltramelli}.}
  \bibinfo{year}{2017}\natexlab{}.
\newblock \showarticletitle{pix2code: Generating Code from a Graphical User
  Interface Screenshot}.
\newblock \bibinfo{journal}{\emph{arXiv preprint arXiv:1705.07962}}
  (\bibinfo{year}{2017}).
\newblock


\bibitem[\protect\citeauthoryear{Bengio, Ducharme, Vincent, and Jauvin}{Bengio
  et~al\mbox{.}}{2003}]%
        {bengio2003neural}
\bibfield{author}{\bibinfo{person}{Yoshua Bengio}, \bibinfo{person}{R{\'e}jean
  Ducharme}, \bibinfo{person}{Pascal Vincent}, {and} \bibinfo{person}{Christian
  Jauvin}.} \bibinfo{year}{2003}\natexlab{}.
\newblock \showarticletitle{A neural probabilistic language model}.
\newblock \bibinfo{journal}{\emph{Journal of machine learning research}}
  \bibinfo{volume}{3}, \bibinfo{number}{Feb} (\bibinfo{year}{2003}),
  \bibinfo{pages}{1137--1155}.
\newblock


\bibitem[\protect\citeauthoryear{Bhatia and Singh}{Bhatia and Singh}{2016}]%
        {bhatia2016automated}
\bibfield{author}{\bibinfo{person}{Sahil Bhatia} {and} \bibinfo{person}{Rishabh
  Singh}.} \bibinfo{year}{2016}\natexlab{}.
\newblock \showarticletitle{Automated correction for syntax errors in
  programming assignments using recurrent neural networks}.
\newblock \bibinfo{journal}{\emph{arXiv preprint arXiv:1603.06129}}
  (\bibinfo{year}{2016}).
\newblock


\bibitem[\protect\citeauthoryear{Bhoopchand, Rockt{\"a}schel, Barr, and
  Riedel}{Bhoopchand et~al\mbox{.}}{2016}]%
        {bhoopchand2016learning}
\bibfield{author}{\bibinfo{person}{Avishkar Bhoopchand}, \bibinfo{person}{Tim
  Rockt{\"a}schel}, \bibinfo{person}{Earl Barr}, {and}
  \bibinfo{person}{Sebastian Riedel}.} \bibinfo{year}{2016}\natexlab{}.
\newblock \showarticletitle{Learning Python Code Suggestion with a Sparse
  Pointer Network}.
\newblock \bibinfo{journal}{\emph{arXiv preprint arXiv:1611.08307}}
  (\bibinfo{year}{2016}).
\newblock


\bibitem[\protect\citeauthoryear{Bielik, Raychev, and Vechev}{Bielik
  et~al\mbox{.}}{2016a}]%
        {bielik2016phog}
\bibfield{author}{\bibinfo{person}{Pavol Bielik}, \bibinfo{person}{Veselin
  Raychev}, {and} \bibinfo{person}{Martin Vechev}.}
  \bibinfo{year}{2016}\natexlab{a}.
\newblock \showarticletitle{PHOG: probabilistic model for code}. In
  \bibinfo{booktitle}{\emph{International Conference on Machine Learning}}.
  \bibinfo{pages}{2933--2942}.
\newblock


\bibitem[\protect\citeauthoryear{Bielik, Raychev, and Vechev}{Bielik
  et~al\mbox{.}}{2016b}]%
        {bielik2016program}
\bibfield{author}{\bibinfo{person}{Pavol Bielik}, \bibinfo{person}{Veselin
  Raychev}, {and} \bibinfo{person}{Martin Vechev}.}
  \bibinfo{year}{2016}\natexlab{b}.
\newblock \showarticletitle{Program Synthesis for Character Level Language
  Modeling}. In \bibinfo{booktitle}{\emph{Proceedings of the International
  Conference on Learning Representations (ICLR)}}.
\newblock


\bibitem[\protect\citeauthoryear{Blunsom, Grefenstette, and
  Kalchbrenner}{Blunsom et~al\mbox{.}}{2014}]%
        {blunsom2014convolutional}
\bibfield{author}{\bibinfo{person}{Phil Blunsom}, \bibinfo{person}{Edward
  Grefenstette}, {and} \bibinfo{person}{Nal Kalchbrenner}.}
  \bibinfo{year}{2014}\natexlab{}.
\newblock \showarticletitle{A convolutional neural network for modelling
  sentences}. In \bibinfo{booktitle}{\emph{Proceedings of the 52nd Annual
  Meeting of the Association for Computational Linguistics}}. Proceedings of
  the 52nd Annual Meeting of the Association for Computational Linguistics.
\newblock


\bibitem[\protect\citeauthoryear{Bojanowski, Grave, Joulin, and
  Mikolov}{Bojanowski et~al\mbox{.}}{2016}]%
        {bojanowski2016enriching}
\bibfield{author}{\bibinfo{person}{Piotr Bojanowski}, \bibinfo{person}{Edouard
  Grave}, \bibinfo{person}{Armand Joulin}, {and} \bibinfo{person}{Tomas
  Mikolov}.} \bibinfo{year}{2016}\natexlab{}.
\newblock \showarticletitle{Enriching Word Vectors with Subword Information}.
\newblock \bibinfo{journal}{\emph{arXiv preprint arXiv:1607.04606}}
  (\bibinfo{year}{2016}).
\newblock


\bibitem[\protect\citeauthoryear{Bornschein and Bengio}{Bornschein and
  Bengio}{2014}]%
        {bornschein2014reweighted}
\bibfield{author}{\bibinfo{person}{J{\"o}rg Bornschein} {and}
  \bibinfo{person}{Yoshua Bengio}.} \bibinfo{year}{2014}\natexlab{}.
\newblock \showarticletitle{Reweighted wake-sleep}.
\newblock \bibinfo{journal}{\emph{arXiv preprint arXiv:1406.2751}}
  (\bibinfo{year}{2014}).
\newblock


\bibitem[\protect\citeauthoryear{Bornschein, Mnih, Zoran, and
  Rezende}{Bornschein et~al\mbox{.}}{2017}]%
        {bornschein2017variational}
\bibfield{author}{\bibinfo{person}{J{\"o}rg Bornschein},
  \bibinfo{person}{Andriy Mnih}, \bibinfo{person}{Daniel Zoran}, {and}
  \bibinfo{person}{Danilo~Jimenez Rezende}.} \bibinfo{year}{2017}\natexlab{}.
\newblock \showarticletitle{Variational Memory Addressing in Generative
  Models}. In \bibinfo{booktitle}{\emph{Advances in Neural Information
  Processing Systems}}. \bibinfo{pages}{3921--3930}.
\newblock


\bibitem[\protect\citeauthoryear{Bradbury, Merity, Xiong, and Socher}{Bradbury
  et~al\mbox{.}}{2016}]%
        {bradbury2016quasi}
\bibfield{author}{\bibinfo{person}{James Bradbury}, \bibinfo{person}{Stephen
  Merity}, \bibinfo{person}{Caiming Xiong}, {and} \bibinfo{person}{Richard
  Socher}.} \bibinfo{year}{2016}\natexlab{}.
\newblock \showarticletitle{Quasi-recurrent neural networks}.
\newblock \bibinfo{journal}{\emph{arXiv preprint arXiv:1611.01576}}
  (\bibinfo{year}{2016}).
\newblock


\bibitem[\protect\citeauthoryear{Brockschmidt, Allamanis, Gaunt, and
  Polozov}{Brockschmidt et~al\mbox{.}}{2018}]%
        {brockschmidt2018generative}
\bibfield{author}{\bibinfo{person}{Marc Brockschmidt},
  \bibinfo{person}{Miltiadis Allamanis}, \bibinfo{person}{Alexander~L Gaunt},
  {and} \bibinfo{person}{Oleksandr Polozov}.} \bibinfo{year}{2018}\natexlab{}.
\newblock \showarticletitle{Generative code modeling with graphs}.
\newblock \bibinfo{journal}{\emph{arXiv preprint arXiv:1805.08490}}
  (\bibinfo{year}{2018}).
\newblock


\bibitem[\protect\citeauthoryear{Brown, K{\"o}lling, McCall, and Utting}{Brown
  et~al\mbox{.}}{2014}]%
        {brown2014blackbox}
\bibfield{author}{\bibinfo{person}{Neil Christopher~Charles Brown},
  \bibinfo{person}{Michael K{\"o}lling}, \bibinfo{person}{Davin McCall}, {and}
  \bibinfo{person}{Ian Utting}.} \bibinfo{year}{2014}\natexlab{}.
\newblock \showarticletitle{Blackbox: a large scale repository of novice
  programmers' activity}. In \bibinfo{booktitle}{\emph{Proceedings of the 45th
  ACM technical symposium on Computer science education}}. ACM,
  \bibinfo{pages}{223--228}.
\newblock


\bibitem[\protect\citeauthoryear{Burda, Grosse, and Salakhutdinov}{Burda
  et~al\mbox{.}}{2015}]%
        {burda2015importance}
\bibfield{author}{\bibinfo{person}{Yuri Burda}, \bibinfo{person}{Roger Grosse},
  {and} \bibinfo{person}{Ruslan Salakhutdinov}.}
  \bibinfo{year}{2015}\natexlab{}.
\newblock \showarticletitle{Importance weighted autoencoders}.
\newblock \bibinfo{journal}{\emph{arXiv preprint arXiv:1509.00519}}
  (\bibinfo{year}{2015}).
\newblock


\bibitem[\protect\citeauthoryear{Cai, Shin, and Song}{Cai
  et~al\mbox{.}}{2017}]%
        {cai2017making}
\bibfield{author}{\bibinfo{person}{Jonathon Cai}, \bibinfo{person}{Richard
  Shin}, {and} \bibinfo{person}{Dawn Song}.} \bibinfo{year}{2017}\natexlab{}.
\newblock \showarticletitle{Making neural programming architectures generalize
  via recursion}.
\newblock \bibinfo{journal}{\emph{arXiv preprint arXiv:1704.06611}}
  (\bibinfo{year}{2017}).
\newblock


\bibitem[\protect\citeauthoryear{Campbell, Hindle, and Amaral}{Campbell
  et~al\mbox{.}}{2014}]%
        {campbell2014syntax}
\bibfield{author}{\bibinfo{person}{Joshua~Charles Campbell},
  \bibinfo{person}{Abram Hindle}, {and} \bibinfo{person}{Jos{\'e}~Nelson
  Amaral}.} \bibinfo{year}{2014}\natexlab{}.
\newblock \showarticletitle{Syntax errors just aren't natural: improving error
  reporting with language models}. In \bibinfo{booktitle}{\emph{Proceedings of
  the 11th Working Conference on Mining Software Repositories}}. ACM,
  \bibinfo{pages}{252--261}.
\newblock


\bibitem[\protect\citeauthoryear{Cerulo, Ceccarelli, Di~Penta, and
  Canfora}{Cerulo et~al\mbox{.}}{2013}]%
        {cerulo2013hidden}
\bibfield{author}{\bibinfo{person}{Luigi Cerulo}, \bibinfo{person}{Michele
  Ceccarelli}, \bibinfo{person}{Massimiliano Di~Penta}, {and}
  \bibinfo{person}{Gerardo Canfora}.} \bibinfo{year}{2013}\natexlab{}.
\newblock \showarticletitle{A hidden markov model to detect coded information
  islands in free text}. In \bibinfo{booktitle}{\emph{Source Code Analysis and
  Manipulation (SCAM), 2013 IEEE 13th International Working Conference on}}.
  IEEE, \bibinfo{pages}{157--166}.
\newblock


\bibitem[\protect\citeauthoryear{Chakkrit}{Chakkrit}{2016}]%
        {chakkrit2016towards}
\bibfield{author}{\bibinfo{person}{Tantithamthavorn Chakkrit}.}
  \bibinfo{year}{2016}\natexlab{}.
\newblock \showarticletitle{Towards a Better Understanding of the Impact of
  Experimental Components on Defect Prediction Models}.
\newblock  (\bibinfo{year}{2016}).
\newblock


\bibitem[\protect\citeauthoryear{Chen, Xing, and Han}{Chen
  et~al\mbox{.}}{2016}]%
        {chen2016techland}
\bibfield{author}{\bibinfo{person}{Chunyang Chen}, \bibinfo{person}{Zhenchang
  Xing}, {and} \bibinfo{person}{Lei Han}.} \bibinfo{year}{2016}\natexlab{}.
\newblock \showarticletitle{Techland: Assisting technology landscape inquiries
  with insights from stack overflow}. In \bibinfo{booktitle}{\emph{Software
  Maintenance and Evolution (ICSME), 2016 IEEE International Conference on}}.
  IEEE, \bibinfo{pages}{356--366}.
\newblock


\bibitem[\protect\citeauthoryear{Chen, Xing, and Wang}{Chen
  et~al\mbox{.}}{2017}]%
        {chen2017unsupervised}
\bibfield{author}{\bibinfo{person}{Chunyang Chen}, \bibinfo{person}{Zhenchang
  Xing}, {and} \bibinfo{person}{Ximing Wang}.} \bibinfo{year}{2017}\natexlab{}.
\newblock \showarticletitle{Unsupervised software-specific morphological forms
  inference from informal discussions}. In
  \bibinfo{booktitle}{\emph{Proceedings of the 39th International Conference on
  Software Engineering}}. IEEE Press, \bibinfo{pages}{450--461}.
\newblock


\bibitem[\protect\citeauthoryear{Chen and Zhou}{Chen and Zhou}{2018}]%
        {chen2018neural}
\bibfield{author}{\bibinfo{person}{Qingying Chen} {and}
  \bibinfo{person}{Minghui Zhou}.} \bibinfo{year}{2018}\natexlab{}.
\newblock \showarticletitle{A neural framework for retrieval and summarization
  of source code}. In \bibinfo{booktitle}{\emph{Proceedings of the 33rd
  ACM/IEEE International Conference on Automated Software Engineering}}. ACM,
  \bibinfo{pages}{826--831}.
\newblock


\bibitem[\protect\citeauthoryear{Chen, Kommrusch, Tufano, Pouchet, Poshyvanyk,
  and Monperrus}{Chen et~al\mbox{.}}{2018}]%
        {chen2018sequencer}
\bibfield{author}{\bibinfo{person}{Zimin Chen}, \bibinfo{person}{Steve
  Kommrusch}, \bibinfo{person}{Michele Tufano}, \bibinfo{person}{Louis-No{\"e}l
  Pouchet}, \bibinfo{person}{Denys Poshyvanyk}, {and} \bibinfo{person}{Martin
  Monperrus}.} \bibinfo{year}{2018}\natexlab{}.
\newblock \showarticletitle{Sequencer: Sequence-to-sequence learning for
  end-to-end program repair}.
\newblock \bibinfo{journal}{\emph{arXiv preprint arXiv:1901.01808}}
  (\bibinfo{year}{2018}).
\newblock


\bibitem[\protect\citeauthoryear{Chen and Monperrus}{Chen and
  Monperrus}{2019}]%
        {chen2019literature}
\bibfield{author}{\bibinfo{person}{Zimin Chen} {and} \bibinfo{person}{Martin
  Monperrus}.} \bibinfo{year}{2019}\natexlab{}.
\newblock \showarticletitle{A Literature Study of Embeddings on Source Code}.
\newblock \bibinfo{journal}{\emph{arXiv preprint arXiv:1904.03061}}
  (\bibinfo{year}{2019}).
\newblock


\bibitem[\protect\citeauthoryear{Cho, Van~Merri{\"e}nboer, Gulcehre, Bahdanau,
  Bougares, Schwenk, and Bengio}{Cho et~al\mbox{.}}{2014}]%
        {cho2014learning}
\bibfield{author}{\bibinfo{person}{Kyunghyun Cho}, \bibinfo{person}{Bart
  Van~Merri{\"e}nboer}, \bibinfo{person}{Caglar Gulcehre},
  \bibinfo{person}{Dzmitry Bahdanau}, \bibinfo{person}{Fethi Bougares},
  \bibinfo{person}{Holger Schwenk}, {and} \bibinfo{person}{Yoshua Bengio}.}
  \bibinfo{year}{2014}\natexlab{}.
\newblock \showarticletitle{Learning phrase representations using RNN
  encoder-decoder for statistical machine translation}.
\newblock \bibinfo{journal}{\emph{arXiv preprint arXiv:1406.1078}}
  (\bibinfo{year}{2014}).
\newblock


\bibitem[\protect\citeauthoryear{Chorowski, Bahdanau, Serdyuk, Cho, and
  Bengio}{Chorowski et~al\mbox{.}}{2015}]%
        {chorowski2015attention}
\bibfield{author}{\bibinfo{person}{Jan~K Chorowski}, \bibinfo{person}{Dzmitry
  Bahdanau}, \bibinfo{person}{Dmitriy Serdyuk}, \bibinfo{person}{Kyunghyun
  Cho}, {and} \bibinfo{person}{Yoshua Bengio}.}
  \bibinfo{year}{2015}\natexlab{}.
\newblock \showarticletitle{Attention-based models for speech recognition}. In
  \bibinfo{booktitle}{\emph{Advances in Neural Information Processing
  Systems}}. \bibinfo{pages}{577--585}.
\newblock


\bibitem[\protect\citeauthoryear{Chung, Gulcehre, Cho, and Bengio}{Chung
  et~al\mbox{.}}{2014}]%
        {chung2014empirical}
\bibfield{author}{\bibinfo{person}{Junyoung Chung}, \bibinfo{person}{Caglar
  Gulcehre}, \bibinfo{person}{KyungHyun Cho}, {and} \bibinfo{person}{Yoshua
  Bengio}.} \bibinfo{year}{2014}\natexlab{}.
\newblock \showarticletitle{Empirical evaluation of gated recurrent neural
  networks on sequence modeling}.
\newblock \bibinfo{journal}{\emph{arXiv preprint arXiv:1412.3555}}
  (\bibinfo{year}{2014}).
\newblock


\bibitem[\protect\citeauthoryear{Chung, Gulcehre, Cho, and Bengio}{Chung
  et~al\mbox{.}}{2015}]%
        {chung2015gated}
\bibfield{author}{\bibinfo{person}{Junyoung Chung}, \bibinfo{person}{Caglar
  Gulcehre}, \bibinfo{person}{Kyunghyun Cho}, {and} \bibinfo{person}{Yoshua
  Bengio}.} \bibinfo{year}{2015}\natexlab{}.
\newblock \showarticletitle{Gated feedback recurrent neural networks}. In
  \bibinfo{booktitle}{\emph{International Conference on Machine Learning}}.
  \bibinfo{pages}{2067--2075}.
\newblock


\bibitem[\protect\citeauthoryear{Cifuentes, Gross, and Keynes}{Cifuentes
  et~al\mbox{.}}{2015}]%
        {cifuentes2015understanding}
\bibfield{author}{\bibinfo{person}{Cristina Cifuentes}, \bibinfo{person}{Andrew
  Gross}, {and} \bibinfo{person}{Nathan Keynes}.}
  \bibinfo{year}{2015}\natexlab{}.
\newblock \showarticletitle{Understanding caller-sensitive method
  vulnerabilities: A class of access control vulnerabilities in the java
  platform}. In \bibinfo{booktitle}{\emph{Proceedings of the 4th ACM SIGPLAN
  International Workshop on State Of the Art in Program Analysis}}. ACM,
  \bibinfo{pages}{7--12}.
\newblock


\bibitem[\protect\citeauthoryear{Cohn, Blunsom, and Goldwater}{Cohn
  et~al\mbox{.}}{2010}]%
        {cohn2010inducing}
\bibfield{author}{\bibinfo{person}{Trevor Cohn}, \bibinfo{person}{Phil
  Blunsom}, {and} \bibinfo{person}{Sharon Goldwater}.}
  \bibinfo{year}{2010}\natexlab{}.
\newblock \showarticletitle{Inducing tree-substitution grammars}.
\newblock \bibinfo{journal}{\emph{Journal of Machine Learning Research}}
  \bibinfo{volume}{11}, \bibinfo{number}{Nov} (\bibinfo{year}{2010}),
  \bibinfo{pages}{3053--3096}.
\newblock


\bibitem[\protect\citeauthoryear{Cooijmans, Ballas, Laurent,
  G{\"u}l{\c{c}}ehre, and Courville}{Cooijmans et~al\mbox{.}}{2016}]%
        {cooijmans2016recurrent}
\bibfield{author}{\bibinfo{person}{Tim Cooijmans}, \bibinfo{person}{Nicolas
  Ballas}, \bibinfo{person}{C{\'e}sar Laurent},
  \bibinfo{person}{{\c{C}}a{\u{g}}lar G{\"u}l{\c{c}}ehre}, {and}
  \bibinfo{person}{Aaron Courville}.} \bibinfo{year}{2016}\natexlab{}.
\newblock \showarticletitle{Recurrent batch normalization}.
\newblock \bibinfo{journal}{\emph{arXiv preprint arXiv:1603.09025}}
  (\bibinfo{year}{2016}).
\newblock


\bibitem[\protect\citeauthoryear{Cs{\'a}ji}{Cs{\'a}ji}{2001}]%
        {csaji2001approximation}
\bibfield{author}{\bibinfo{person}{Bal{\'a}zs~Csan{\'a}d Cs{\'a}ji}.}
  \bibinfo{year}{2001}\natexlab{}.
\newblock \showarticletitle{Approximation with artificial neural networks}.
\newblock \bibinfo{journal}{\emph{Faculty of Sciences, Etvs Lornd University,
  Hungary}}  \bibinfo{volume}{24} (\bibinfo{year}{2001}), \bibinfo{pages}{48}.
\newblock


\bibitem[\protect\citeauthoryear{Cvitkovic, Singh, and Anandkumar}{Cvitkovic
  et~al\mbox{.}}{2018}]%
        {cvitkovic2018open}
\bibfield{author}{\bibinfo{person}{Milan Cvitkovic}, \bibinfo{person}{Badal
  Singh}, {and} \bibinfo{person}{Anima Anandkumar}.}
  \bibinfo{year}{2018}\natexlab{}.
\newblock \showarticletitle{Open vocabulary learning on source code with a
  graph-structured cache}.
\newblock \bibinfo{journal}{\emph{arXiv preprint arXiv:1810.08305}}
  (\bibinfo{year}{2018}).
\newblock


\bibitem[\protect\citeauthoryear{Dai, Yang, Yang, Carbonell, Le, and
  Salakhutdinov}{Dai et~al\mbox{.}}{2019}]%
        {dai2019transformer}
\bibfield{author}{\bibinfo{person}{Zihang Dai}, \bibinfo{person}{Zhilin Yang},
  \bibinfo{person}{Yiming Yang}, \bibinfo{person}{Jaime Carbonell},
  \bibinfo{person}{Quoc~V Le}, {and} \bibinfo{person}{Ruslan Salakhutdinov}.}
  \bibinfo{year}{2019}\natexlab{}.
\newblock \showarticletitle{Transformer-xl: Attentive language models beyond a
  fixed-length context}.
\newblock \bibinfo{journal}{\emph{arXiv preprint arXiv:1901.02860}}
  (\bibinfo{year}{2019}).
\newblock


\bibitem[\protect\citeauthoryear{Dam, Tran, and Pham}{Dam
  et~al\mbox{.}}{2016}]%
        {dam2016deep}
\bibfield{author}{\bibinfo{person}{Hoa~Khanh Dam}, \bibinfo{person}{Truyen
  Tran}, {and} \bibinfo{person}{Trang Pham}.} \bibinfo{year}{2016}\natexlab{}.
\newblock \showarticletitle{A deep language model for software code}.
\newblock \bibinfo{journal}{\emph{arXiv preprint arXiv:1608.02715}}
  (\bibinfo{year}{2016}).
\newblock


\bibitem[\protect\citeauthoryear{Dangovski, Jing, and Soljacic}{Dangovski
  et~al\mbox{.}}{2017}]%
        {dangovski2017rotational}
\bibfield{author}{\bibinfo{person}{Rumen Dangovski}, \bibinfo{person}{Li Jing},
  {and} \bibinfo{person}{Marin Soljacic}.} \bibinfo{year}{2017}\natexlab{}.
\newblock \showarticletitle{Rotational Unit of Memory}.
\newblock \bibinfo{journal}{\emph{arXiv preprint arXiv:1710.09537}}
  (\bibinfo{year}{2017}).
\newblock


\bibitem[\protect\citeauthoryear{Das and Shah}{Das and Shah}{2015}]%
        {das2015contextual}
\bibfield{author}{\bibinfo{person}{Subhasis Das} {and}
  \bibinfo{person}{Chinmayee Shah}.} \bibinfo{year}{2015}\natexlab{}.
\newblock \showarticletitle{Contextual Code Completion Using Machine Learning}.
\newblock  (\bibinfo{year}{2015}).
\newblock


\bibitem[\protect\citeauthoryear{De~la Higuera}{De~la Higuera}{2010}]%
        {de2010grammatical}
\bibfield{author}{\bibinfo{person}{Colin De~la Higuera}.}
  \bibinfo{year}{2010}\natexlab{}.
\newblock \bibinfo{booktitle}{\emph{Grammatical inference: learning automata
  and grammars}}.
\newblock \bibinfo{publisher}{Cambridge University Press}.
\newblock


\bibitem[\protect\citeauthoryear{De~Moura and Bj{\o}rner}{De~Moura and
  Bj{\o}rner}{2008}]%
        {de2008z3}
\bibfield{author}{\bibinfo{person}{Leonardo De~Moura} {and}
  \bibinfo{person}{Nikolaj Bj{\o}rner}.} \bibinfo{year}{2008}\natexlab{}.
\newblock \showarticletitle{Z3: An efficient SMT solver}.
\newblock \bibinfo{journal}{\emph{Tools and Algorithms for the Construction and
  Analysis of Systems}} (\bibinfo{year}{2008}), \bibinfo{pages}{337--340}.
\newblock


\bibitem[\protect\citeauthoryear{Devlin, Chang, Lee, and Toutanova}{Devlin
  et~al\mbox{.}}{2018}]%
        {devlin2018bert}
\bibfield{author}{\bibinfo{person}{Jacob Devlin}, \bibinfo{person}{Ming-Wei
  Chang}, \bibinfo{person}{Kenton Lee}, {and} \bibinfo{person}{Kristina
  Toutanova}.} \bibinfo{year}{2018}\natexlab{}.
\newblock \showarticletitle{Bert: Pre-training of deep bidirectional
  transformers for language understanding}.
\newblock \bibinfo{journal}{\emph{arXiv preprint arXiv:1810.04805}}
  (\bibinfo{year}{2018}).
\newblock


\bibitem[\protect\citeauthoryear{Dodge, Gane, Zhang, Bordes, Chopra, Miller,
  Szlam, and Weston}{Dodge et~al\mbox{.}}{2015}]%
        {dodge2015evaluating}
\bibfield{author}{\bibinfo{person}{Jesse Dodge}, \bibinfo{person}{Andreea
  Gane}, \bibinfo{person}{Xiang Zhang}, \bibinfo{person}{Antoine Bordes},
  \bibinfo{person}{Sumit Chopra}, \bibinfo{person}{Alexander Miller},
  \bibinfo{person}{Arthur Szlam}, {and} \bibinfo{person}{Jason Weston}.}
  \bibinfo{year}{2015}\natexlab{}.
\newblock \showarticletitle{Evaluating prerequisite qualities for learning
  end-to-end dialog systems}.
\newblock \bibinfo{journal}{\emph{arXiv preprint arXiv:1511.06931}}
  (\bibinfo{year}{2015}).
\newblock


\bibitem[\protect\citeauthoryear{Dong and Lapata}{Dong and Lapata}{2016}]%
        {dong2016language}
\bibfield{author}{\bibinfo{person}{Li Dong} {and} \bibinfo{person}{Mirella
  Lapata}.} \bibinfo{year}{2016}\natexlab{}.
\newblock \showarticletitle{Language to logical form with neural attention}.
\newblock \bibinfo{journal}{\emph{arXiv preprint arXiv:1601.01280}}
  (\bibinfo{year}{2016}).
\newblock


\bibitem[\protect\citeauthoryear{Dong, Yang, Wang, Wei, Liu, Wang, Gao, Zhou,
  and Hon}{Dong et~al\mbox{.}}{2019}]%
        {dong2019unified}
\bibfield{author}{\bibinfo{person}{Li Dong}, \bibinfo{person}{Nan Yang},
  \bibinfo{person}{Wenhui Wang}, \bibinfo{person}{Furu Wei},
  \bibinfo{person}{Xiaodong Liu}, \bibinfo{person}{Yu Wang},
  \bibinfo{person}{Jianfeng Gao}, \bibinfo{person}{Ming Zhou}, {and}
  \bibinfo{person}{Hsiao-Wuen Hon}.} \bibinfo{year}{2019}\natexlab{}.
\newblock \showarticletitle{Unified language model pre-training for natural
  language understanding and generation}. In \bibinfo{booktitle}{\emph{Advances
  in Neural Information Processing Systems}}. \bibinfo{pages}{13042--13054}.
\newblock


\bibitem[\protect\citeauthoryear{Ellis, Ritchie, Solar-Lezama, and
  Tenenbaum}{Ellis et~al\mbox{.}}{2017}]%
        {ellis2017learning}
\bibfield{author}{\bibinfo{person}{Kevin Ellis}, \bibinfo{person}{Daniel
  Ritchie}, \bibinfo{person}{Armando Solar-Lezama}, {and}
  \bibinfo{person}{Joshua~B Tenenbaum}.} \bibinfo{year}{2017}\natexlab{}.
\newblock \showarticletitle{Learning to Infer Graphics Programs from Hand-Drawn
  Images}.
\newblock \bibinfo{journal}{\emph{arXiv preprint arXiv:1707.09627}}
  (\bibinfo{year}{2017}).
\newblock


\bibitem[\protect\citeauthoryear{Evans and Grefenstette}{Evans and
  Grefenstette}{2017}]%
        {evans2017learning}
\bibfield{author}{\bibinfo{person}{Richard Evans} {and} \bibinfo{person}{Edward
  Grefenstette}.} \bibinfo{year}{2017}\natexlab{}.
\newblock \showarticletitle{Learning Explanatory Rules from Noisy Data}.
\newblock \bibinfo{journal}{\emph{arXiv preprint arXiv:1711.04574}}
  (\bibinfo{year}{2017}).
\newblock


\bibitem[\protect\citeauthoryear{Feser, Brockschmidt, Gaunt, and Tarlow}{Feser
  et~al\mbox{.}}{2016}]%
        {feser2016differentiable}
\bibfield{author}{\bibinfo{person}{John~K Feser}, \bibinfo{person}{Marc
  Brockschmidt}, \bibinfo{person}{Alexander~L Gaunt}, {and}
  \bibinfo{person}{Daniel Tarlow}.} \bibinfo{year}{2016}\natexlab{}.
\newblock \showarticletitle{Differentiable Functional Program Interpreters}.
\newblock \bibinfo{journal}{\emph{arXiv preprint arXiv:1611.01988}}
  (\bibinfo{year}{2016}).
\newblock


\bibitem[\protect\citeauthoryear{Finn, Goodfellow, and Levine}{Finn
  et~al\mbox{.}}{2016}]%
        {finn2016unsupervised}
\bibfield{author}{\bibinfo{person}{Chelsea Finn}, \bibinfo{person}{Ian
  Goodfellow}, {and} \bibinfo{person}{Sergey Levine}.}
  \bibinfo{year}{2016}\natexlab{}.
\newblock \showarticletitle{Unsupervised learning for physical interaction
  through video prediction}. In \bibinfo{booktitle}{\emph{Advances in Neural
  Information Processing Systems}}. \bibinfo{pages}{64--72}.
\newblock


\bibitem[\protect\citeauthoryear{Gage}{Gage}{1994}]%
        {gage1994new}
\bibfield{author}{\bibinfo{person}{Philip Gage}.}
  \bibinfo{year}{1994}\natexlab{}.
\newblock \showarticletitle{A new algorithm for data compression}.
\newblock \bibinfo{journal}{\emph{The C Users Journal}} \bibinfo{volume}{12},
  \bibinfo{number}{2} (\bibinfo{year}{1994}), \bibinfo{pages}{23--38}.
\newblock


\bibitem[\protect\citeauthoryear{Gagniuc}{Gagniuc}{2017}]%
        {gagniuc2017markov}
\bibfield{author}{\bibinfo{person}{Paul~A Gagniuc}.}
  \bibinfo{year}{2017}\natexlab{}.
\newblock \bibinfo{booktitle}{\emph{Markov Chains: From Theory to
  Implementation and Experimentation}}.
\newblock \bibinfo{publisher}{John Wiley \& Sons}.
\newblock


\bibitem[\protect\citeauthoryear{Gal and Ghahramani}{Gal and
  Ghahramani}{2016}]%
        {gal2016theoretically}
\bibfield{author}{\bibinfo{person}{Yarin Gal} {and} \bibinfo{person}{Zoubin
  Ghahramani}.} \bibinfo{year}{2016}\natexlab{}.
\newblock \showarticletitle{A theoretically grounded application of dropout in
  recurrent neural networks}. In \bibinfo{booktitle}{\emph{Advances in neural
  information processing systems}}. \bibinfo{pages}{1019--1027}.
\newblock


\bibitem[\protect\citeauthoryear{Ganin, Ustinova, Ajakan, Germain, Larochelle,
  Laviolette, Marchand, and Lempitsky}{Ganin et~al\mbox{.}}{2016}]%
        {ganin2016domain}
\bibfield{author}{\bibinfo{person}{Yaroslav Ganin}, \bibinfo{person}{Evgeniya
  Ustinova}, \bibinfo{person}{Hana Ajakan}, \bibinfo{person}{Pascal Germain},
  \bibinfo{person}{Hugo Larochelle}, \bibinfo{person}{Fran{\c{c}}ois
  Laviolette}, \bibinfo{person}{Mario Marchand}, {and} \bibinfo{person}{Victor
  Lempitsky}.} \bibinfo{year}{2016}\natexlab{}.
\newblock \showarticletitle{Domain-adversarial training of neural networks}.
\newblock \bibinfo{journal}{\emph{Journal of Machine Learning Research}}
  \bibinfo{volume}{17}, \bibinfo{number}{59} (\bibinfo{year}{2016}),
  \bibinfo{pages}{1--35}.
\newblock


\bibitem[\protect\citeauthoryear{Ganti and Gray}{Ganti and Gray}{2013}]%
        {ganti2013building}
\bibfield{author}{\bibinfo{person}{Ravi Ganti} {and}
  \bibinfo{person}{Alexander~G Gray}.} \bibinfo{year}{2013}\natexlab{}.
\newblock \showarticletitle{Building bridges: Viewing active learning from the
  multi-armed bandit lens}.
\newblock \bibinfo{journal}{\emph{arXiv preprint arXiv:1309.6830}}
  (\bibinfo{year}{2013}).
\newblock


\bibitem[\protect\citeauthoryear{Gaunt, Brockschmidt, Singh, Kushman, Kohli,
  Taylor, and Tarlow}{Gaunt et~al\mbox{.}}{2016}]%
        {gaunt2016terpret}
\bibfield{author}{\bibinfo{person}{Alexander~L Gaunt}, \bibinfo{person}{Marc
  Brockschmidt}, \bibinfo{person}{Rishabh Singh}, \bibinfo{person}{Nate
  Kushman}, \bibinfo{person}{Pushmeet Kohli}, \bibinfo{person}{Jonathan
  Taylor}, {and} \bibinfo{person}{Daniel Tarlow}.}
  \bibinfo{year}{2016}\natexlab{}.
\newblock \showarticletitle{Terpret: A probabilistic programming language for
  program induction}.
\newblock \bibinfo{journal}{\emph{arXiv preprint arXiv:1608.04428}}
  (\bibinfo{year}{2016}).
\newblock


\bibitem[\protect\citeauthoryear{Gehring, Auli, Grangier, Yarats, and
  Dauphin}{Gehring et~al\mbox{.}}{2017}]%
        {gehring2017convolutional}
\bibfield{author}{\bibinfo{person}{Jonas Gehring}, \bibinfo{person}{Michael
  Auli}, \bibinfo{person}{David Grangier}, \bibinfo{person}{Denis Yarats},
  {and} \bibinfo{person}{Yann~N Dauphin}.} \bibinfo{year}{2017}\natexlab{}.
\newblock \showarticletitle{Convolutional Sequence to Sequence Learning}.
\newblock \bibinfo{journal}{\emph{arXiv preprint arXiv:1705.03122}}
  (\bibinfo{year}{2017}).
\newblock


\bibitem[\protect\citeauthoryear{Ghaffarian and Shahriari}{Ghaffarian and
  Shahriari}{2017}]%
        {ghaffarian2017software}
\bibfield{author}{\bibinfo{person}{Seyed~Mohammad Ghaffarian} {and}
  \bibinfo{person}{Hamid~Reza Shahriari}.} \bibinfo{year}{2017}\natexlab{}.
\newblock \showarticletitle{Software vulnerability analysis and discovery using
  machine-learning and data-mining techniques: a survey}.
\newblock \bibinfo{journal}{\emph{ACM Computing Surveys (CSUR)}}
  \bibinfo{volume}{50}, \bibinfo{number}{4} (\bibinfo{year}{2017}),
  \bibinfo{pages}{56}.
\newblock


\bibitem[\protect\citeauthoryear{Goodfellow, Bengio, Courville, and
  Bengio}{Goodfellow et~al\mbox{.}}{2016}]%
        {goodfellow2016deep}
\bibfield{author}{\bibinfo{person}{Ian Goodfellow}, \bibinfo{person}{Yoshua
  Bengio}, \bibinfo{person}{Aaron Courville}, {and} \bibinfo{person}{Yoshua
  Bengio}.} \bibinfo{year}{2016}\natexlab{}.
\newblock \bibinfo{booktitle}{\emph{Deep learning}}. Vol.~\bibinfo{volume}{1}.
\newblock \bibinfo{publisher}{MIT press Cambridge}.
\newblock


\bibitem[\protect\citeauthoryear{Goodfellow, Pouget-Abadie, Mirza, Xu,
  Warde-Farley, Ozair, Courville, and Bengio}{Goodfellow et~al\mbox{.}}{2014}]%
        {goodfellow2014generative}
\bibfield{author}{\bibinfo{person}{Ian Goodfellow}, \bibinfo{person}{Jean
  Pouget-Abadie}, \bibinfo{person}{Mehdi Mirza}, \bibinfo{person}{Bing Xu},
  \bibinfo{person}{David Warde-Farley}, \bibinfo{person}{Sherjil Ozair},
  \bibinfo{person}{Aaron Courville}, {and} \bibinfo{person}{Yoshua Bengio}.}
  \bibinfo{year}{2014}\natexlab{}.
\newblock \showarticletitle{Generative adversarial nets}. In
  \bibinfo{booktitle}{\emph{Advances in neural information processing
  systems}}. \bibinfo{pages}{2672--2680}.
\newblock


\bibitem[\protect\citeauthoryear{Goodman}{Goodman}{2001}]%
        {goodman2001bit}
\bibfield{author}{\bibinfo{person}{Joshua~T Goodman}.}
  \bibinfo{year}{2001}\natexlab{}.
\newblock \showarticletitle{A bit of progress in language modeling}.
\newblock \bibinfo{journal}{\emph{Computer Speech \& Language}}
  \bibinfo{volume}{15}, \bibinfo{number}{4} (\bibinfo{year}{2001}),
  \bibinfo{pages}{403--434}.
\newblock


\bibitem[\protect\citeauthoryear{Grave, Cisse, and Joulin}{Grave
  et~al\mbox{.}}{2017}]%
        {grave2017unbounded}
\bibfield{author}{\bibinfo{person}{Edouard Grave}, \bibinfo{person}{Moustapha~M
  Cisse}, {and} \bibinfo{person}{Armand Joulin}.}
  \bibinfo{year}{2017}\natexlab{}.
\newblock \showarticletitle{Unbounded cache model for online language modeling
  with open vocabulary}. In \bibinfo{booktitle}{\emph{Advances in Neural
  Information Processing Systems}}. \bibinfo{pages}{6042--6052}.
\newblock


\bibitem[\protect\citeauthoryear{Graves, Wayne, and Danihelka}{Graves
  et~al\mbox{.}}{2014}]%
        {graves2014neural}
\bibfield{author}{\bibinfo{person}{Alex Graves}, \bibinfo{person}{Greg Wayne},
  {and} \bibinfo{person}{Ivo Danihelka}.} \bibinfo{year}{2014}\natexlab{}.
\newblock \showarticletitle{Neural turing machines}.
\newblock \bibinfo{journal}{\emph{arXiv preprint arXiv:1410.5401}}
  (\bibinfo{year}{2014}).
\newblock


\bibitem[\protect\citeauthoryear{Graves, Wayne, Reynolds, Harley, Danihelka,
  Grabska-Barwi{\'n}ska, Colmenarejo, Grefenstette, Ramalho, Agapiou,
  et~al\mbox{.}}{Graves et~al\mbox{.}}{2016}]%
        {graves2016hybrid}
\bibfield{author}{\bibinfo{person}{Alex Graves}, \bibinfo{person}{Greg Wayne},
  \bibinfo{person}{Malcolm Reynolds}, \bibinfo{person}{Tim Harley},
  \bibinfo{person}{Ivo Danihelka}, \bibinfo{person}{Agnieszka
  Grabska-Barwi{\'n}ska}, \bibinfo{person}{Sergio~G{\'o}mez Colmenarejo},
  \bibinfo{person}{Edward Grefenstette}, \bibinfo{person}{Tiago Ramalho},
  \bibinfo{person}{John Agapiou}, {et~al\mbox{.}}}
  \bibinfo{year}{2016}\natexlab{}.
\newblock \showarticletitle{Hybrid computing using a neural network with
  dynamic external memory}.
\newblock \bibinfo{journal}{\emph{Nature}} \bibinfo{volume}{538},
  \bibinfo{number}{7626} (\bibinfo{year}{2016}), \bibinfo{pages}{471--476}.
\newblock


\bibitem[\protect\citeauthoryear{Grefenstette, Hermann, Suleyman, and
  Blunsom}{Grefenstette et~al\mbox{.}}{2015}]%
        {grefenstette2015learning}
\bibfield{author}{\bibinfo{person}{Edward Grefenstette},
  \bibinfo{person}{Karl~Moritz Hermann}, \bibinfo{person}{Mustafa Suleyman},
  {and} \bibinfo{person}{Phil Blunsom}.} \bibinfo{year}{2015}\natexlab{}.
\newblock \showarticletitle{Learning to transduce with unbounded memory}. In
  \bibinfo{booktitle}{\emph{Advances in Neural Information Processing
  Systems}}. \bibinfo{pages}{1828--1836}.
\newblock


\bibitem[\protect\citeauthoryear{Gregor, Besse, Rezende, Danihelka, and
  Wierstra}{Gregor et~al\mbox{.}}{2016}]%
        {gregor2016towards}
\bibfield{author}{\bibinfo{person}{Karol Gregor}, \bibinfo{person}{Frederic
  Besse}, \bibinfo{person}{Danilo~Jimenez Rezende}, \bibinfo{person}{Ivo
  Danihelka}, {and} \bibinfo{person}{Daan Wierstra}.}
  \bibinfo{year}{2016}\natexlab{}.
\newblock \showarticletitle{Towards conceptual compression}. In
  \bibinfo{booktitle}{\emph{Advances In Neural Information Processing
  Systems}}. \bibinfo{pages}{3549--3557}.
\newblock


\bibitem[\protect\citeauthoryear{Gu, Bradbury, Xiong, Li, and Socher}{Gu
  et~al\mbox{.}}{2017a}]%
        {gu2017non}
\bibfield{author}{\bibinfo{person}{Jiatao Gu}, \bibinfo{person}{James
  Bradbury}, \bibinfo{person}{Caiming Xiong}, \bibinfo{person}{Victor~O.K. Li},
  {and} \bibinfo{person}{Richard Socher}.} \bibinfo{year}{2017}\natexlab{a}.
\newblock \showarticletitle{Non-Autoregressive Neural Machine Translation}.
\newblock \bibinfo{journal}{\emph{arXiv preprint arXiv:1711.02281}}
  (\bibinfo{year}{2017}).
\newblock


\bibitem[\protect\citeauthoryear{Gu, Zhang, and Kim}{Gu et~al\mbox{.}}{2018}]%
        {gu2018deep}
\bibfield{author}{\bibinfo{person}{Xiaodong Gu}, \bibinfo{person}{Hongyu
  Zhang}, {and} \bibinfo{person}{Sunghun Kim}.}
  \bibinfo{year}{2018}\natexlab{}.
\newblock \showarticletitle{Deep code search}. In
  \bibinfo{booktitle}{\emph{Proceedings of the 40th International Conference on
  Software Engineering}}. ACM, \bibinfo{pages}{933--944}.
\newblock


\bibitem[\protect\citeauthoryear{Gu, Zhang, Zhang, and Kim}{Gu
  et~al\mbox{.}}{2016}]%
        {gu2016deep}
\bibfield{author}{\bibinfo{person}{Xiaodong Gu}, \bibinfo{person}{Hongyu
  Zhang}, \bibinfo{person}{Dongmei Zhang}, {and} \bibinfo{person}{Sunghun
  Kim}.} \bibinfo{year}{2016}\natexlab{}.
\newblock \showarticletitle{Deep API learning}. In
  \bibinfo{booktitle}{\emph{Proceedings of the 2016 24th ACM SIGSOFT
  International Symposium on Foundations of Software Engineering}}. ACM,
  \bibinfo{pages}{631--642}.
\newblock


\bibitem[\protect\citeauthoryear{Gu, Zhang, Zhang, and Kim}{Gu
  et~al\mbox{.}}{2017b}]%
        {gu2017deepam}
\bibfield{author}{\bibinfo{person}{Xiaodong Gu}, \bibinfo{person}{Hongyu
  Zhang}, \bibinfo{person}{Dongmei Zhang}, {and} \bibinfo{person}{Sunghun
  Kim}.} \bibinfo{year}{2017}\natexlab{b}.
\newblock \showarticletitle{DeepAM: Migrate APIs with multi-modal sequence to
  sequence learning}.
\newblock \bibinfo{journal}{\emph{arXiv preprint arXiv:1704.07734}}
  (\bibinfo{year}{2017}).
\newblock


\bibitem[\protect\citeauthoryear{Gulwani}{Gulwani}{2010}]%
        {gulwani2010dimensions}
\bibfield{author}{\bibinfo{person}{Sumit Gulwani}.}
  \bibinfo{year}{2010}\natexlab{}.
\newblock \showarticletitle{Dimensions in program synthesis}. In
  \bibinfo{booktitle}{\emph{Proceedings of the 12th international ACM SIGPLAN
  symposium on Principles and practice of declarative programming}}. ACM,
  \bibinfo{pages}{13--24}.
\newblock


\bibitem[\protect\citeauthoryear{Gupta and Sundaresan}{Gupta and
  Sundaresan}{2018}]%
        {gupta2018intelligent}
\bibfield{author}{\bibinfo{person}{Anshul Gupta} {and} \bibinfo{person}{Neel
  Sundaresan}.} \bibinfo{year}{2018}\natexlab{}.
\newblock \showarticletitle{Intelligent code reviews using deep learning}.
\newblock  (\bibinfo{year}{2018}).
\newblock


\bibitem[\protect\citeauthoryear{Gupta, Kanade, and Shevade}{Gupta
  et~al\mbox{.}}{2018}]%
        {gupta2018deep}
\bibfield{author}{\bibinfo{person}{Rahul Gupta}, \bibinfo{person}{Aditya
  Kanade}, {and} \bibinfo{person}{Shirish Shevade}.}
  \bibinfo{year}{2018}\natexlab{}.
\newblock \showarticletitle{Deep reinforcement learning for programming
  language correction}.
\newblock \bibinfo{journal}{\emph{arXiv preprint arXiv:1801.10467}}
  (\bibinfo{year}{2018}).
\newblock


\bibitem[\protect\citeauthoryear{Gupta, Pal, Kanade, and Shevade}{Gupta
  et~al\mbox{.}}{2017}]%
        {gupta2017deepfix}
\bibfield{author}{\bibinfo{person}{Rahul Gupta}, \bibinfo{person}{Soham Pal},
  \bibinfo{person}{Aditya Kanade}, {and} \bibinfo{person}{Shirish Shevade}.}
  \bibinfo{year}{2017}\natexlab{}.
\newblock \showarticletitle{DeepFix: Fixing Common C Language Errors by Deep
  Learning.}. In \bibinfo{booktitle}{\emph{AAAI}}. \bibinfo{pages}{1345--1351}.
\newblock


\bibitem[\protect\citeauthoryear{Guu, Pasupat, Liu, and Liang}{Guu
  et~al\mbox{.}}{2017}]%
        {guu2017language}
\bibfield{author}{\bibinfo{person}{Kelvin Guu}, \bibinfo{person}{Panupong
  Pasupat}, \bibinfo{person}{Evan~Zheran Liu}, {and} \bibinfo{person}{Percy
  Liang}.} \bibinfo{year}{2017}\natexlab{}.
\newblock \showarticletitle{From Language to Programs: Bridging Reinforcement
  Learning and Maximum Marginal Likelihood}.
\newblock \bibinfo{journal}{\emph{arXiv preprint arXiv:1704.07926}}
  (\bibinfo{year}{2017}).
\newblock


\bibitem[\protect\citeauthoryear{Gvero, Kuncak, Kuraj, and Piskac}{Gvero
  et~al\mbox{.}}{2013}]%
        {gvero2013complete}
\bibfield{author}{\bibinfo{person}{Tihomir Gvero}, \bibinfo{person}{Viktor
  Kuncak}, \bibinfo{person}{Ivan Kuraj}, {and} \bibinfo{person}{Ruzica
  Piskac}.} \bibinfo{year}{2013}\natexlab{}.
\newblock \showarticletitle{Complete completion using types and weights}. In
  \bibinfo{booktitle}{\emph{ACM SIGPLAN Notices}}, Vol.~\bibinfo{volume}{48}.
  ACM, \bibinfo{pages}{27--38}.
\newblock


\bibitem[\protect\citeauthoryear{Harer, Ozdemir, Lazovich, Reale, Russell, Kim,
  et~al\mbox{.}}{Harer et~al\mbox{.}}{2018}]%
        {harer2018learning}
\bibfield{author}{\bibinfo{person}{Jacob Harer}, \bibinfo{person}{Onur
  Ozdemir}, \bibinfo{person}{Tomo Lazovich}, \bibinfo{person}{Christopher
  Reale}, \bibinfo{person}{Rebecca Russell}, \bibinfo{person}{Louis Kim},
  {et~al\mbox{.}}} \bibinfo{year}{2018}\natexlab{}.
\newblock \showarticletitle{Learning to repair software vulnerabilities with
  generative adversarial networks}. In \bibinfo{booktitle}{\emph{Advances in
  Neural Information Processing Systems}}. \bibinfo{pages}{7933--7943}.
\newblock


\bibitem[\protect\citeauthoryear{Hashimoto, Guu, Oren, and Liang}{Hashimoto
  et~al\mbox{.}}{2018}]%
        {hashimoto2018retrieve}
\bibfield{author}{\bibinfo{person}{Tatsunori~B Hashimoto},
  \bibinfo{person}{Kelvin Guu}, \bibinfo{person}{Yonatan Oren}, {and}
  \bibinfo{person}{Percy~S Liang}.} \bibinfo{year}{2018}\natexlab{}.
\newblock \showarticletitle{A Retrieve-and-Edit Framework for Predicting
  Structured Outputs}. In \bibinfo{booktitle}{\emph{Advances in Neural
  Information Processing Systems}}. \bibinfo{pages}{10073--10083}.
\newblock


\bibitem[\protect\citeauthoryear{He, Zhang, Ren, and Sun}{He
  et~al\mbox{.}}{2016}]%
        {he2016deep}
\bibfield{author}{\bibinfo{person}{Kaiming He}, \bibinfo{person}{Xiangyu
  Zhang}, \bibinfo{person}{Shaoqing Ren}, {and} \bibinfo{person}{Jian Sun}.}
  \bibinfo{year}{2016}\natexlab{}.
\newblock \showarticletitle{Deep residual learning for image recognition}. In
  \bibinfo{booktitle}{\emph{Proceedings of the IEEE conference on computer
  vision and pattern recognition}}. \bibinfo{pages}{770--778}.
\newblock


\bibitem[\protect\citeauthoryear{He, Lee, Lewis, and Zettlemoyer}{He
  et~al\mbox{.}}{2017b}]%
        {he2017deep}
\bibfield{author}{\bibinfo{person}{Luheng He}, \bibinfo{person}{Kenton Lee},
  \bibinfo{person}{Mike Lewis}, {and} \bibinfo{person}{Luke Zettlemoyer}.}
  \bibinfo{year}{2017}\natexlab{b}.
\newblock \showarticletitle{Deep semantic role labeling: What works and
  what’s next}. In \bibinfo{booktitle}{\emph{Proceedings of the Annual
  Meeting of the Association for Computational Linguistics}}.
\newblock


\bibitem[\protect\citeauthoryear{He, Gao, Xiao, and Barber}{He
  et~al\mbox{.}}{2017a}]%
        {he2017wider}
\bibfield{author}{\bibinfo{person}{Zhen He}, \bibinfo{person}{Shaobing Gao},
  \bibinfo{person}{Liang Xiao}, {and} \bibinfo{person}{David Barber}.}
  \bibinfo{year}{2017}\natexlab{a}.
\newblock \showarticletitle{Wider and Deeper, Cheaper and Faster: Tensorized
  LSTMs for Sequence Learning}. In \bibinfo{booktitle}{\emph{Advances in Neural
  Information Processing Systems}}. \bibinfo{pages}{1--11}.
\newblock


\bibitem[\protect\citeauthoryear{Hellendoorn and Devanbu}{Hellendoorn and
  Devanbu}{2017}]%
        {hellendoorn2017deep}
\bibfield{author}{\bibinfo{person}{Vincent~J Hellendoorn} {and}
  \bibinfo{person}{Premkumar Devanbu}.} \bibinfo{year}{2017}\natexlab{}.
\newblock \showarticletitle{Are deep neural networks the best choice for
  modeling source code?}. In \bibinfo{booktitle}{\emph{Proceedings of the 2017
  11th Joint Meeting on Foundations of Software Engineering}}. ACM,
  \bibinfo{pages}{763--773}.
\newblock


\bibitem[\protect\citeauthoryear{Henaff, Weston, Szlam, Bordes, and
  LeCun}{Henaff et~al\mbox{.}}{2016}]%
        {henaff2016tracking}
\bibfield{author}{\bibinfo{person}{Mikael Henaff}, \bibinfo{person}{Jason
  Weston}, \bibinfo{person}{Arthur Szlam}, \bibinfo{person}{Antoine Bordes},
  {and} \bibinfo{person}{Yann LeCun}.} \bibinfo{year}{2016}\natexlab{}.
\newblock \showarticletitle{Tracking the World State with Recurrent Entity
  Networks}.
\newblock \bibinfo{journal}{\emph{arXiv preprint arXiv:1612.03969}}
  (\bibinfo{year}{2016}).
\newblock


\bibitem[\protect\citeauthoryear{Hill, Bordes, Chopra, and Weston}{Hill
  et~al\mbox{.}}{2015}]%
        {hill2015goldilocks}
\bibfield{author}{\bibinfo{person}{Felix Hill}, \bibinfo{person}{Antoine
  Bordes}, \bibinfo{person}{Sumit Chopra}, {and} \bibinfo{person}{Jason
  Weston}.} \bibinfo{year}{2015}\natexlab{}.
\newblock \showarticletitle{The Goldilocks Principle: Reading Children's Books
  with Explicit Memory Representations}.
\newblock \bibinfo{journal}{\emph{arXiv preprint arXiv:1511.02301}}
  (\bibinfo{year}{2015}).
\newblock


\bibitem[\protect\citeauthoryear{Hindle, Barr, Su, Gabel, and Devanbu}{Hindle
  et~al\mbox{.}}{2012}]%
        {hindle2012naturalness}
\bibfield{author}{\bibinfo{person}{Abram Hindle}, \bibinfo{person}{Earl~T
  Barr}, \bibinfo{person}{Zhendong Su}, \bibinfo{person}{Mark Gabel}, {and}
  \bibinfo{person}{Premkumar Devanbu}.} \bibinfo{year}{2012}\natexlab{}.
\newblock \showarticletitle{On the naturalness of software}. In
  \bibinfo{booktitle}{\emph{Software Engineering (ICSE), 2012 34th
  International Conference on}}. IEEE, \bibinfo{pages}{837--847}.
\newblock


\bibitem[\protect\citeauthoryear{Hochreiter and Schmidhuber}{Hochreiter and
  Schmidhuber}{1997}]%
        {hochreiter1997long}
\bibfield{author}{\bibinfo{person}{Sepp Hochreiter} {and}
  \bibinfo{person}{J{\"u}rgen Schmidhuber}.} \bibinfo{year}{1997}\natexlab{}.
\newblock \showarticletitle{Long short-term memory}.
\newblock \bibinfo{journal}{\emph{Neural computation}} \bibinfo{volume}{9},
  \bibinfo{number}{8} (\bibinfo{year}{1997}), \bibinfo{pages}{1735--1780}.
\newblock


\bibitem[\protect\citeauthoryear{Hopcroft}{Hopcroft}{2008}]%
        {hopcroft2008introduction}
\bibfield{author}{\bibinfo{person}{John~E Hopcroft}.}
  \bibinfo{year}{2008}\natexlab{}.
\newblock \bibinfo{booktitle}{\emph{Introduction to automata theory, languages,
  and computation}}.
\newblock \bibinfo{publisher}{Pearson Education India}. 77--106 pages.
\newblock


\bibitem[\protect\citeauthoryear{Hsiao, Cafarella, and Narayanasamy}{Hsiao
  et~al\mbox{.}}{2014}]%
        {hsiao2014using}
\bibfield{author}{\bibinfo{person}{Chun-Hung Hsiao}, \bibinfo{person}{Michael
  Cafarella}, {and} \bibinfo{person}{Satish Narayanasamy}.}
  \bibinfo{year}{2014}\natexlab{}.
\newblock \showarticletitle{Using web corpus statistics for program analysis}.
  In \bibinfo{booktitle}{\emph{ACM SIGPLAN Notices}},
  Vol.~\bibinfo{volume}{49}. ACM, \bibinfo{pages}{49--65}.
\newblock


\bibitem[\protect\citeauthoryear{Hu, Li, Xia, Lo, and Jin}{Hu
  et~al\mbox{.}}{2018a}]%
        {hu2018deep}
\bibfield{author}{\bibinfo{person}{Xing Hu}, \bibinfo{person}{Ge Li},
  \bibinfo{person}{Xin Xia}, \bibinfo{person}{David Lo}, {and}
  \bibinfo{person}{Zhi Jin}.} \bibinfo{year}{2018}\natexlab{a}.
\newblock \showarticletitle{Deep code comment generation}. In
  \bibinfo{booktitle}{\emph{Proceedings of the 26th Conference on Program
  Comprehension}}. ACM, \bibinfo{pages}{200--210}.
\newblock


\bibitem[\protect\citeauthoryear{Hu, Li, Xia, Lo, Lu, and Jin}{Hu
  et~al\mbox{.}}{2018b}]%
        {hu2018summarizing}
\bibfield{author}{\bibinfo{person}{Xing Hu}, \bibinfo{person}{Ge Li},
  \bibinfo{person}{Xin Xia}, \bibinfo{person}{David Lo}, \bibinfo{person}{Shuai
  Lu}, {and} \bibinfo{person}{Zhi Jin}.} \bibinfo{year}{2018}\natexlab{b}.
\newblock \showarticletitle{Summarizing source code with transferred api
  knowledge}.
\newblock  (\bibinfo{year}{2018}).
\newblock


\bibitem[\protect\citeauthoryear{Inan, Khosravi, and Socher}{Inan
  et~al\mbox{.}}{2016}]%
        {inan2016tying}
\bibfield{author}{\bibinfo{person}{Hakan Inan}, \bibinfo{person}{Khashayar
  Khosravi}, {and} \bibinfo{person}{Richard Socher}.}
  \bibinfo{year}{2016}\natexlab{}.
\newblock \showarticletitle{Tying word vectors and word classifiers: A loss
  framework for language modeling}.
\newblock \bibinfo{journal}{\emph{arXiv preprint arXiv:1611.01462}}
  (\bibinfo{year}{2016}).
\newblock


\bibitem[\protect\citeauthoryear{Ioffe and Szegedy}{Ioffe and Szegedy}{2015}]%
        {ioffe2015batch}
\bibfield{author}{\bibinfo{person}{Sergey Ioffe} {and}
  \bibinfo{person}{Christian Szegedy}.} \bibinfo{year}{2015}\natexlab{}.
\newblock \showarticletitle{Batch normalization: Accelerating deep network
  training by reducing internal covariate shift}. In
  \bibinfo{booktitle}{\emph{International Conference on Machine Learning}}.
  \bibinfo{pages}{448--456}.
\newblock


\bibitem[\protect\citeauthoryear{Isola, Zhu, Zhou, and Efros}{Isola
  et~al\mbox{.}}{2016}]%
        {isola2016image}
\bibfield{author}{\bibinfo{person}{Phillip Isola}, \bibinfo{person}{Jun-Yan
  Zhu}, \bibinfo{person}{Tinghui Zhou}, {and} \bibinfo{person}{Alexei~A
  Efros}.} \bibinfo{year}{2016}\natexlab{}.
\newblock \showarticletitle{Image-to-image translation with conditional
  adversarial networks}.
\newblock \bibinfo{journal}{\emph{arXiv preprint arXiv:1611.07004}}
  (\bibinfo{year}{2016}).
\newblock


\bibitem[\protect\citeauthoryear{Iyer, Konstas, Cheung, and Zettlemoyer}{Iyer
  et~al\mbox{.}}{2016}]%
        {iyer2016summarizing}
\bibfield{author}{\bibinfo{person}{Srinivasan Iyer}, \bibinfo{person}{Ioannis
  Konstas}, \bibinfo{person}{Alvin Cheung}, {and} \bibinfo{person}{Luke
  Zettlemoyer}.} \bibinfo{year}{2016}\natexlab{}.
\newblock \showarticletitle{Summarizing source code using a neural attention
  model}. In \bibinfo{booktitle}{\emph{Proceedings of the 54th Annual Meeting
  of the Association for Computational Linguistics (Volume 1: Long Papers)}},
  Vol.~\bibinfo{volume}{1}. \bibinfo{pages}{2073--2083}.
\newblock


\bibitem[\protect\citeauthoryear{Jacobsson}{Jacobsson}{2005}]%
        {jacobsson2005rule}
\bibfield{author}{\bibinfo{person}{Henrik Jacobsson}.}
  \bibinfo{year}{2005}\natexlab{}.
\newblock \showarticletitle{Rule extraction from recurrent neural networks: A
  taxonomy and review}.
\newblock \bibinfo{journal}{\emph{Neural Computation}} \bibinfo{volume}{17},
  \bibinfo{number}{6} (\bibinfo{year}{2005}), \bibinfo{pages}{1223--1263}.
\newblock


\bibitem[\protect\citeauthoryear{Jacobsson}{Jacobsson}{2006}]%
        {jacobsson2006crystallizing}
\bibfield{author}{\bibinfo{person}{Henrik Jacobsson}.}
  \bibinfo{year}{2006}\natexlab{}.
\newblock \showarticletitle{The crystallizing substochastic sequential machine
  extractor: CrySSMEx}.
\newblock \bibinfo{journal}{\emph{Neural Computation}} \bibinfo{volume}{18},
  \bibinfo{number}{9} (\bibinfo{year}{2006}), \bibinfo{pages}{2211--2255}.
\newblock


\bibitem[\protect\citeauthoryear{Jaderberg, Simonyan, Zisserman,
  et~al\mbox{.}}{Jaderberg et~al\mbox{.}}{2015}]%
        {jaderberg2015spatial}
\bibfield{author}{\bibinfo{person}{Max Jaderberg}, \bibinfo{person}{Karen
  Simonyan}, \bibinfo{person}{Andrew Zisserman}, {et~al\mbox{.}}}
  \bibinfo{year}{2015}\natexlab{}.
\newblock \showarticletitle{Spatial transformer networks}. In
  \bibinfo{booktitle}{\emph{Advances in neural information processing
  systems}}. \bibinfo{pages}{2017--2025}.
\newblock


\bibitem[\protect\citeauthoryear{Jang, Gu, and Poole}{Jang
  et~al\mbox{.}}{2016}]%
        {jang2016categorical}
\bibfield{author}{\bibinfo{person}{Eric Jang}, \bibinfo{person}{Shixiang Gu},
  {and} \bibinfo{person}{Ben Poole}.} \bibinfo{year}{2016}\natexlab{}.
\newblock \showarticletitle{Categorical reparameterization with
  gumbel-softmax}.
\newblock \bibinfo{journal}{\emph{arXiv preprint arXiv:1611.01144}}
  (\bibinfo{year}{2016}).
\newblock


\bibitem[\protect\citeauthoryear{Jha, Gulwani, Seshia, and Tiwari}{Jha
  et~al\mbox{.}}{2010}]%
        {jha2010oracle}
\bibfield{author}{\bibinfo{person}{Susmit Jha}, \bibinfo{person}{Sumit
  Gulwani}, \bibinfo{person}{Sanjit~A Seshia}, {and} \bibinfo{person}{Ashish
  Tiwari}.} \bibinfo{year}{2010}\natexlab{}.
\newblock \showarticletitle{Oracle-guided component-based program synthesis}.
  In \bibinfo{booktitle}{\emph{Proceedings of the 32nd ACM/IEEE International
  Conference on Software Engineering-Volume 1}}. ACM,
  \bibinfo{pages}{215--224}.
\newblock


\bibitem[\protect\citeauthoryear{Johnson, Schuster, Le, Krikun, Wu, Chen,
  Thorat, Vi{\'e}gas, Wattenberg, Corrado, et~al\mbox{.}}{Johnson
  et~al\mbox{.}}{2016}]%
        {johnson2016google}
\bibfield{author}{\bibinfo{person}{Melvin Johnson}, \bibinfo{person}{Mike
  Schuster}, \bibinfo{person}{Quoc~V Le}, \bibinfo{person}{Maxim Krikun},
  \bibinfo{person}{Yonghui Wu}, \bibinfo{person}{Zhifeng Chen},
  \bibinfo{person}{Nikhil Thorat}, \bibinfo{person}{Fernanda Vi{\'e}gas},
  \bibinfo{person}{Martin Wattenberg}, \bibinfo{person}{Greg Corrado},
  {et~al\mbox{.}}} \bibinfo{year}{2016}\natexlab{}.
\newblock \showarticletitle{Google's multilingual neural machine translation
  system: enabling zero-shot translation}.
\newblock \bibinfo{journal}{\emph{arXiv preprint arXiv:1611.04558}}
  (\bibinfo{year}{2016}).
\newblock


\bibitem[\protect\citeauthoryear{Joshi and Rambow}{Joshi and Rambow}{2003}]%
        {joshi2003formalism}
\bibfield{author}{\bibinfo{person}{Aravind Joshi} {and} \bibinfo{person}{Owen
  Rambow}.} \bibinfo{year}{2003}\natexlab{}.
\newblock \showarticletitle{A formalism for dependency grammar based on tree
  adjoining grammar}. In \bibinfo{booktitle}{\emph{Proceedings of the
  Conference on Meaning-text Theory}}. \bibinfo{pages}{207--216}.
\newblock


\bibitem[\protect\citeauthoryear{Joulin and Mikolov}{Joulin and
  Mikolov}{2015}]%
        {joulin2015inferring}
\bibfield{author}{\bibinfo{person}{Armand Joulin} {and} \bibinfo{person}{Tomas
  Mikolov}.} \bibinfo{year}{2015}\natexlab{}.
\newblock \showarticletitle{Inferring algorithmic patterns with stack-augmented
  recurrent nets}. In \bibinfo{booktitle}{\emph{Advances in neural information
  processing systems}}. \bibinfo{pages}{190--198}.
\newblock


\bibitem[\protect\citeauthoryear{Just, Jalali, and Ernst}{Just
  et~al\mbox{.}}{2014}]%
        {just2014defects4j}
\bibfield{author}{\bibinfo{person}{Ren{\'e} Just}, \bibinfo{person}{Darioush
  Jalali}, {and} \bibinfo{person}{Michael~D Ernst}.}
  \bibinfo{year}{2014}\natexlab{}.
\newblock \showarticletitle{Defects4J: A database of existing faults to enable
  controlled testing studies for Java programs}. In
  \bibinfo{booktitle}{\emph{Proceedings of the 2014 International Symposium on
  Software Testing and Analysis}}. ACM, \bibinfo{pages}{437--440}.
\newblock


\bibitem[\protect\citeauthoryear{Kaiser and Sutskever}{Kaiser and
  Sutskever}{2016}]%
        {kaiser2016neural}
\bibfield{author}{\bibinfo{person}{{\L}ukasz Kaiser} {and}
  \bibinfo{person}{Ilya Sutskever}.} \bibinfo{year}{2016}\natexlab{}.
\newblock \showarticletitle{Neural gpus learn algorithms}.
\newblock \bibinfo{journal}{\emph{ICLR}} (\bibinfo{year}{2016}).
\newblock


\bibitem[\protect\citeauthoryear{Kalchbrenner and Blunsom}{Kalchbrenner and
  Blunsom}{2013}]%
        {kalchbrenner2013recurrent}
\bibfield{author}{\bibinfo{person}{Nal Kalchbrenner} {and}
  \bibinfo{person}{Phil Blunsom}.} \bibinfo{year}{2013}\natexlab{}.
\newblock \showarticletitle{Recurrent Continuous Translation Models.}. In
  \bibinfo{booktitle}{\emph{EMNLP}}, Vol.~\bibinfo{volume}{3}.
  \bibinfo{pages}{413}.
\newblock


\bibitem[\protect\citeauthoryear{Kalchbrenner, Espeholt, Simonyan, Oord,
  Graves, and Kavukcuoglu}{Kalchbrenner et~al\mbox{.}}{2016a}]%
        {kalchbrenner2016neural}
\bibfield{author}{\bibinfo{person}{Nal Kalchbrenner}, \bibinfo{person}{Lasse
  Espeholt}, \bibinfo{person}{Karen Simonyan}, \bibinfo{person}{Aaron van~den
  Oord}, \bibinfo{person}{Alex Graves}, {and} \bibinfo{person}{Koray
  Kavukcuoglu}.} \bibinfo{year}{2016}\natexlab{a}.
\newblock \showarticletitle{Neural machine translation in linear time}.
\newblock \bibinfo{journal}{\emph{arXiv preprint arXiv:1610.10099}}
  (\bibinfo{year}{2016}).
\newblock


\bibitem[\protect\citeauthoryear{Kalchbrenner, Oord, Simonyan, Danihelka,
  Vinyals, Graves, and Kavukcuoglu}{Kalchbrenner et~al\mbox{.}}{2016b}]%
        {kalchbrenner2016video}
\bibfield{author}{\bibinfo{person}{Nal Kalchbrenner}, \bibinfo{person}{Aaron
  van~den Oord}, \bibinfo{person}{Karen Simonyan}, \bibinfo{person}{Ivo
  Danihelka}, \bibinfo{person}{Oriol Vinyals}, \bibinfo{person}{Alex Graves},
  {and} \bibinfo{person}{Koray Kavukcuoglu}.} \bibinfo{year}{2016}\natexlab{b}.
\newblock \showarticletitle{Video pixel networks}.
\newblock \bibinfo{journal}{\emph{arXiv preprint arXiv:1610.00527}}
  (\bibinfo{year}{2016}).
\newblock


\bibitem[\protect\citeauthoryear{Karampatsis and Sutton}{Karampatsis and
  Sutton}{2019}]%
        {karampatsis2019maybe}
\bibfield{author}{\bibinfo{person}{Rafael-Michael Karampatsis} {and}
  \bibinfo{person}{Charles Sutton}.} \bibinfo{year}{2019}\natexlab{}.
\newblock \showarticletitle{Maybe Deep Neural Networks are the Best Choice for
  Modeling Source Code}.
\newblock \bibinfo{journal}{\emph{arXiv preprint arXiv:1903.05734}}
  (\bibinfo{year}{2019}).
\newblock


\bibitem[\protect\citeauthoryear{Karpathy}{Karpathy}{2016}]%
        {karpathy2016unreasonable}
\bibfield{author}{\bibinfo{person}{Andrej Karpathy}.}
  \bibinfo{year}{2016}\natexlab{}.
\newblock \showarticletitle{The unreasonable effectiveness of recurrent neural
  networks, 2015}.
\newblock \bibinfo{journal}{\emph{URL
  http://karpathy.github.io/2015/05/21/rnn-effectiveness}}
  (\bibinfo{year}{2016}).
\newblock


\bibitem[\protect\citeauthoryear{Karras, Aila, Laine, and Lehtinen}{Karras
  et~al\mbox{.}}{2017}]%
        {karras2017progressive}
\bibfield{author}{\bibinfo{person}{Tero Karras}, \bibinfo{person}{Timo Aila},
  \bibinfo{person}{Samuli Laine}, {and} \bibinfo{person}{Jaakko Lehtinen}.}
  \bibinfo{year}{2017}\natexlab{}.
\newblock \showarticletitle{Progressive Growing of GANs for Improved Quality,
  Stability, and Variation}.
\newblock \bibinfo{journal}{\emph{arXiv preprint arXiv:1710.10196}}
  (\bibinfo{year}{2017}).
\newblock


\bibitem[\protect\citeauthoryear{Khanh~Dam, Tran, and Pham}{Khanh~Dam
  et~al\mbox{.}}{2016}]%
        {khanh2016deep}
\bibfield{author}{\bibinfo{person}{Hoa Khanh~Dam}, \bibinfo{person}{Truyen
  Tran}, {and} \bibinfo{person}{Trang Pham}.} \bibinfo{year}{2016}\natexlab{}.
\newblock \showarticletitle{A deep language model for software code}.
\newblock \bibinfo{journal}{\emph{arXiv preprint arXiv:1608.02715}}
  (\bibinfo{year}{2016}).
\newblock


\bibitem[\protect\citeauthoryear{Kim, Jernite, Sontag, and Rush}{Kim
  et~al\mbox{.}}{2016}]%
        {kim2016character}
\bibfield{author}{\bibinfo{person}{Yoon Kim}, \bibinfo{person}{Yacine Jernite},
  \bibinfo{person}{David Sontag}, {and} \bibinfo{person}{Alexander~M Rush}.}
  \bibinfo{year}{2016}\natexlab{}.
\newblock \showarticletitle{Character-Aware Neural Language Models.}. In
  \bibinfo{booktitle}{\emph{AAAI}}. \bibinfo{pages}{2741--2749}.
\newblock


\bibitem[\protect\citeauthoryear{Kingma and Ba}{Kingma and Ba}{2014}]%
        {kingma2014adam}
\bibfield{author}{\bibinfo{person}{Diederik~P Kingma} {and}
  \bibinfo{person}{Jimmy Ba}.} \bibinfo{year}{2014}\natexlab{}.
\newblock \showarticletitle{Adam: A method for stochastic optimization}.
\newblock \bibinfo{journal}{\emph{arXiv preprint arXiv:1412.6980}}
  (\bibinfo{year}{2014}).
\newblock


\bibitem[\protect\citeauthoryear{Kingma, Salimans, Jozefowicz, Chen, Sutskever,
  and Welling}{Kingma et~al\mbox{.}}{2016}]%
        {kingma2016improved}
\bibfield{author}{\bibinfo{person}{Diederik~P Kingma}, \bibinfo{person}{Tim
  Salimans}, \bibinfo{person}{Rafal Jozefowicz}, \bibinfo{person}{Xi Chen},
  \bibinfo{person}{Ilya Sutskever}, {and} \bibinfo{person}{Max Welling}.}
  \bibinfo{year}{2016}\natexlab{}.
\newblock \showarticletitle{Improved variational inference with inverse
  autoregressive flow}. In \bibinfo{booktitle}{\emph{Advances in Neural
  Information Processing Systems}}. \bibinfo{pages}{4743--4751}.
\newblock


\bibitem[\protect\citeauthoryear{Kingma and Welling}{Kingma and
  Welling}{2013}]%
        {kingma2013auto}
\bibfield{author}{\bibinfo{person}{Diederik~P Kingma} {and}
  \bibinfo{person}{Max Welling}.} \bibinfo{year}{2013}\natexlab{}.
\newblock \showarticletitle{Auto-encoding variational bayes}.
\newblock \bibinfo{journal}{\emph{arXiv preprint arXiv:1312.6114}}
  (\bibinfo{year}{2013}).
\newblock


\bibitem[\protect\citeauthoryear{Knuth}{Knuth}{1968}]%
        {knuth1968semantics}
\bibfield{author}{\bibinfo{person}{Donald~E Knuth}.}
  \bibinfo{year}{1968}\natexlab{}.
\newblock \showarticletitle{Semantics of context-free languages}.
\newblock \bibinfo{journal}{\emph{Mathematical systems theory}}
  \bibinfo{volume}{2}, \bibinfo{number}{2} (\bibinfo{year}{1968}),
  \bibinfo{pages}{127--145}.
\newblock


\bibitem[\protect\citeauthoryear{Koehn, Hoang, Birch, Callison-Burch, Federico,
  Bertoldi, Cowan, Shen, Moran, Zens, et~al\mbox{.}}{Koehn
  et~al\mbox{.}}{2007}]%
        {koehn2007moses}
\bibfield{author}{\bibinfo{person}{Philipp Koehn}, \bibinfo{person}{Hieu
  Hoang}, \bibinfo{person}{Alexandra Birch}, \bibinfo{person}{Chris
  Callison-Burch}, \bibinfo{person}{Marcello Federico}, \bibinfo{person}{Nicola
  Bertoldi}, \bibinfo{person}{Brooke Cowan}, \bibinfo{person}{Wade Shen},
  \bibinfo{person}{Christine Moran}, \bibinfo{person}{Richard Zens},
  {et~al\mbox{.}}} \bibinfo{year}{2007}\natexlab{}.
\newblock \showarticletitle{Moses: Open source toolkit for statistical machine
  translation}. In \bibinfo{booktitle}{\emph{Proceedings of the 45th annual
  meeting of the ACL on interactive poster and demonstration sessions}}.
  Association for Computational Linguistics, \bibinfo{pages}{177--180}.
\newblock


\bibitem[\protect\citeauthoryear{Koehn and Knowles}{Koehn and Knowles}{2017}]%
        {koehn2017six}
\bibfield{author}{\bibinfo{person}{Philipp Koehn} {and}
  \bibinfo{person}{Rebecca Knowles}.} \bibinfo{year}{2017}\natexlab{}.
\newblock \showarticletitle{Six Challenges for Neural Machine Translation}.
\newblock \bibinfo{journal}{\emph{arXiv preprint arXiv:1706.03872}}
  (\bibinfo{year}{2017}).
\newblock


\bibitem[\protect\citeauthoryear{Kolen}{Kolen}{1994}]%
        {kolen1994fool}
\bibfield{author}{\bibinfo{person}{John~F Kolen}.}
  \bibinfo{year}{1994}\natexlab{}.
\newblock \showarticletitle{Fool's gold: Extracting finite state machines from
  recurrent network dynamics}. In \bibinfo{booktitle}{\emph{Advances in neural
  information processing systems}}. \bibinfo{pages}{501--508}.
\newblock


\bibitem[\protect\citeauthoryear{Konyushkova, Sznitman, and Fua}{Konyushkova
  et~al\mbox{.}}{2017}]%
        {konyushkova2017learning}
\bibfield{author}{\bibinfo{person}{Ksenia Konyushkova},
  \bibinfo{person}{Raphael Sznitman}, {and} \bibinfo{person}{Pascal Fua}.}
  \bibinfo{year}{2017}\natexlab{}.
\newblock \showarticletitle{Learning Active Learning from Data}. In
  \bibinfo{booktitle}{\emph{Advances in Neural Information Processing
  Systems}}. \bibinfo{pages}{4226--4236}.
\newblock


\bibitem[\protect\citeauthoryear{Koutnik, Greff, Gomez, and
  Schmidhuber}{Koutnik et~al\mbox{.}}{2014}]%
        {koutnik2014clockwork}
\bibfield{author}{\bibinfo{person}{Jan Koutnik}, \bibinfo{person}{Klaus Greff},
  \bibinfo{person}{Faustino Gomez}, {and} \bibinfo{person}{Juergen
  Schmidhuber}.} \bibinfo{year}{2014}\natexlab{}.
\newblock \showarticletitle{A clockwork rnn}. In
  \bibinfo{booktitle}{\emph{International Conference on Machine Learning}}.
  \bibinfo{pages}{1863--1871}.
\newblock


\bibitem[\protect\citeauthoryear{Krause, Kahembwe, Murray, and Renals}{Krause
  et~al\mbox{.}}{2017}]%
        {krause2017dynamic}
\bibfield{author}{\bibinfo{person}{Ben Krause}, \bibinfo{person}{Emmanuel
  Kahembwe}, \bibinfo{person}{Iain Murray}, {and} \bibinfo{person}{Steve
  Renals}.} \bibinfo{year}{2017}\natexlab{}.
\newblock \showarticletitle{Dynamic evaluation of neural sequence models}.
\newblock \bibinfo{journal}{\emph{arXiv preprint arXiv:1709.07432}}
  (\bibinfo{year}{2017}).
\newblock


\bibitem[\protect\citeauthoryear{Krueger, Maharaj, Kram{\'a}r, Pezeshki,
  Ballas, Ke, Goyal, Bengio, Courville, and Pal}{Krueger et~al\mbox{.}}{2016}]%
        {krueger2016zoneout}
\bibfield{author}{\bibinfo{person}{David Krueger}, \bibinfo{person}{Tegan
  Maharaj}, \bibinfo{person}{J{\'a}nos Kram{\'a}r}, \bibinfo{person}{Mohammad
  Pezeshki}, \bibinfo{person}{Nicolas Ballas}, \bibinfo{person}{Nan~Rosemary
  Ke}, \bibinfo{person}{Anirudh Goyal}, \bibinfo{person}{Yoshua Bengio},
  \bibinfo{person}{Aaron Courville}, {and} \bibinfo{person}{Chris Pal}.}
  \bibinfo{year}{2016}\natexlab{}.
\newblock \showarticletitle{Zoneout: Regularizing rnns by randomly preserving
  hidden activations}.
\newblock \bibinfo{journal}{\emph{arXiv preprint arXiv:1606.01305}}
  (\bibinfo{year}{2016}).
\newblock


\bibitem[\protect\citeauthoryear{Kurach, Andrychowicz, and Sutskever}{Kurach
  et~al\mbox{.}}{2015}]%
        {kurach2015neural}
\bibfield{author}{\bibinfo{person}{Karol Kurach}, \bibinfo{person}{Marcin
  Andrychowicz}, {and} \bibinfo{person}{Ilya Sutskever}.}
  \bibinfo{year}{2015}\natexlab{}.
\newblock \showarticletitle{Neural random-access machines}.
\newblock \bibinfo{journal}{\emph{arXiv preprint arXiv:1511.06392}}
  (\bibinfo{year}{2015}).
\newblock


\bibitem[\protect\citeauthoryear{Laengle, Lueth, Stopp, Herzog, and
  Kamstrup}{Laengle et~al\mbox{.}}{1995}]%
        {laengle1995kantra}
\bibfield{author}{\bibinfo{person}{Thomas Laengle}, \bibinfo{person}{Tim~C
  Lueth}, \bibinfo{person}{Eva Stopp}, \bibinfo{person}{Gerd Herzog}, {and}
  \bibinfo{person}{Gjertrud Kamstrup}.} \bibinfo{year}{1995}\natexlab{}.
\newblock \showarticletitle{Kantra-a natural language interface for intelligent
  robots}. In \bibinfo{booktitle}{\emph{Intelligent Autonomous Systems (IAS
  4)}}. \bibinfo{pages}{357--364}.
\newblock


\bibitem[\protect\citeauthoryear{Lam, Nguyen, Nguyen, and Nguyen}{Lam
  et~al\mbox{.}}{2017}]%
        {lam2017bug}
\bibfield{author}{\bibinfo{person}{An~Ngoc Lam}, \bibinfo{person}{Anh~Tuan
  Nguyen}, \bibinfo{person}{Hoan~Anh Nguyen}, {and} \bibinfo{person}{Tien~N
  Nguyen}.} \bibinfo{year}{2017}\natexlab{}.
\newblock \showarticletitle{Bug localization with combination of deep learning
  and information retrieval}. In \bibinfo{booktitle}{\emph{2017 IEEE/ACM 25th
  International Conference on Program Comprehension (ICPC)}}. IEEE,
  \bibinfo{pages}{218--229}.
\newblock


\bibitem[\protect\citeauthoryear{Lamkanfi, P{\'e}rez, and Demeyer}{Lamkanfi
  et~al\mbox{.}}{2013}]%
        {lamkanfi2013eclipse}
\bibfield{author}{\bibinfo{person}{Ahmed Lamkanfi}, \bibinfo{person}{Javier
  P{\'e}rez}, {and} \bibinfo{person}{Serge Demeyer}.}
  \bibinfo{year}{2013}\natexlab{}.
\newblock \showarticletitle{The eclipse and mozilla defect tracking dataset: a
  genuine dataset for mining bug information}. In
  \bibinfo{booktitle}{\emph{Mining Software Repositories (MSR), 2013 10th IEEE
  Working Conference on}}. IEEE, \bibinfo{pages}{203--206}.
\newblock


\bibitem[\protect\citeauthoryear{Lample, Denoyer, and Ranzato}{Lample
  et~al\mbox{.}}{2017}]%
        {lample2017unsupervised}
\bibfield{author}{\bibinfo{person}{Guillaume Lample}, \bibinfo{person}{Ludovic
  Denoyer}, {and} \bibinfo{person}{Marc'Aurelio Ranzato}.}
  \bibinfo{year}{2017}\natexlab{}.
\newblock \showarticletitle{Unsupervised Machine Translation Using Monolingual
  Corpora Only}.
\newblock \bibinfo{journal}{\emph{arXiv preprint arXiv:1711.00043}}
  (\bibinfo{year}{2017}).
\newblock


\bibitem[\protect\citeauthoryear{Le, Oentaryo, and Lo}{Le
  et~al\mbox{.}}{2015}]%
        {le2015information}
\bibfield{author}{\bibinfo{person}{Tien-Duy~B Le}, \bibinfo{person}{Richard~J
  Oentaryo}, {and} \bibinfo{person}{David Lo}.}
  \bibinfo{year}{2015}\natexlab{}.
\newblock \showarticletitle{Information retrieval and spectrum based bug
  localization: Better together}. In \bibinfo{booktitle}{\emph{Proceedings of
  the 2015 10th Joint Meeting on Foundations of Software Engineering}}. ACM,
  \bibinfo{pages}{579--590}.
\newblock


\bibitem[\protect\citeauthoryear{Ledig, Theis, Husz{\'a}r, Caballero,
  Cunningham, Acosta, Aitken, Tejani, Totz, Wang, et~al\mbox{.}}{Ledig
  et~al\mbox{.}}{2016}]%
        {ledig2016photo}
\bibfield{author}{\bibinfo{person}{Christian Ledig}, \bibinfo{person}{Lucas
  Theis}, \bibinfo{person}{Ferenc Husz{\'a}r}, \bibinfo{person}{Jose
  Caballero}, \bibinfo{person}{Andrew Cunningham}, \bibinfo{person}{Alejandro
  Acosta}, \bibinfo{person}{Andrew Aitken}, \bibinfo{person}{Alykhan Tejani},
  \bibinfo{person}{Johannes Totz}, \bibinfo{person}{Zehan Wang},
  {et~al\mbox{.}}} \bibinfo{year}{2016}\natexlab{}.
\newblock \showarticletitle{Photo-realistic single image super-resolution using
  a generative adversarial network}.
\newblock \bibinfo{journal}{\emph{arXiv preprint arXiv:1609.04802}}
  (\bibinfo{year}{2016}).
\newblock


\bibitem[\protect\citeauthoryear{Lei and Zhang}{Lei and Zhang}{2017}]%
        {lei2017training}
\bibfield{author}{\bibinfo{person}{Tao Lei} {and} \bibinfo{person}{Yu Zhang}.}
  \bibinfo{year}{2017}\natexlab{}.
\newblock \showarticletitle{Training RNNs as Fast as CNNs}.
\newblock \bibinfo{journal}{\emph{arXiv preprint arXiv:1709.02755}}
  (\bibinfo{year}{2017}).
\newblock


\bibitem[\protect\citeauthoryear{Li, Tarlow, Gaunt, Brockschmidt, and
  Kushman}{Li et~al\mbox{.}}{2017a}]%
        {li2016neural}
\bibfield{author}{\bibinfo{person}{Chengtao Li}, \bibinfo{person}{Daniel
  Tarlow}, \bibinfo{person}{Alexander~L Gaunt}, \bibinfo{person}{Marc
  Brockschmidt}, {and} \bibinfo{person}{Nate Kushman}.}
  \bibinfo{year}{2017}\natexlab{a}.
\newblock \showarticletitle{Neural program lattices}. In
  \bibinfo{booktitle}{\emph{Proceedings of the International Conference on
  Learning Representations (ICLR)}}.
\newblock


\bibitem[\protect\citeauthoryear{Li, Wang, King, and Lyu}{Li
  et~al\mbox{.}}{2017b}]%
        {li2017code}
\bibfield{author}{\bibinfo{person}{Jian Li}, \bibinfo{person}{Yue Wang},
  \bibinfo{person}{Irwin King}, {and} \bibinfo{person}{Michael~R. Lyu}.}
  \bibinfo{year}{2017}\natexlab{b}.
\newblock \showarticletitle{Code Completion with Neural Attention and Pointer
  Networks}.
\newblock \bibinfo{journal}{\emph{arXiv preprint arXiv:1711.09573}}
  (\bibinfo{year}{2017}).
\newblock


\bibitem[\protect\citeauthoryear{Li, Tarlow, Brockschmidt, and Zemel}{Li
  et~al\mbox{.}}{2015}]%
        {li2015gated}
\bibfield{author}{\bibinfo{person}{Yujia Li}, \bibinfo{person}{Daniel Tarlow},
  \bibinfo{person}{Marc Brockschmidt}, {and} \bibinfo{person}{Richard Zemel}.}
  \bibinfo{year}{2015}\natexlab{}.
\newblock \showarticletitle{Gated graph sequence neural networks}.
\newblock \bibinfo{journal}{\emph{arXiv preprint arXiv:1511.05493}}
  (\bibinfo{year}{2015}).
\newblock


\bibitem[\protect\citeauthoryear{Li, Zou, Xu, Ou, Jin, Wang, Deng, and
  Zhong}{Li et~al\mbox{.}}{2018}]%
        {li2018vuldeepecker}
\bibfield{author}{\bibinfo{person}{Zhen Li}, \bibinfo{person}{Deqing Zou},
  \bibinfo{person}{Shouhuai Xu}, \bibinfo{person}{Xinyu Ou},
  \bibinfo{person}{Hai Jin}, \bibinfo{person}{Sujuan Wang},
  \bibinfo{person}{Zhijun Deng}, {and} \bibinfo{person}{Yuyi Zhong}.}
  \bibinfo{year}{2018}\natexlab{}.
\newblock \showarticletitle{VulDeePecker: A Deep Learning-Based System for
  Vulnerability Detection}.
\newblock \bibinfo{journal}{\emph{arXiv preprint arXiv:1801.01681}}
  (\bibinfo{year}{2018}).
\newblock


\bibitem[\protect\citeauthoryear{Liang, Berant, Le, Forbus, and Lao}{Liang
  et~al\mbox{.}}{2016}]%
        {liang2016neural}
\bibfield{author}{\bibinfo{person}{Chen Liang}, \bibinfo{person}{Jonathan
  Berant}, \bibinfo{person}{Quoc Le}, \bibinfo{person}{Kenneth~D Forbus}, {and}
  \bibinfo{person}{Ni Lao}.} \bibinfo{year}{2016}\natexlab{}.
\newblock \showarticletitle{Neural symbolic machines: Learning semantic parsers
  on freebase with weak supervision}.
\newblock \bibinfo{journal}{\emph{arXiv preprint arXiv:1611.00020}}
  (\bibinfo{year}{2016}).
\newblock


\bibitem[\protect\citeauthoryear{Lin}{Lin}{2004}]%
        {lin2004rouge}
\bibfield{author}{\bibinfo{person}{Chin-Yew Lin}.}
  \bibinfo{year}{2004}\natexlab{}.
\newblock \showarticletitle{Rouge: A package for automatic evaluation of
  summaries}. In \bibinfo{booktitle}{\emph{Text summarization branches out:
  Proceedings of the ACL-04 workshop}}, Vol.~\bibinfo{volume}{8}. Barcelona,
  Spain.
\newblock


\bibitem[\protect\citeauthoryear{Lin, Wang, Pang, Vu, and Ernst}{Lin
  et~al\mbox{.}}{2017}]%
        {lin2017program}
\bibfield{author}{\bibinfo{person}{Xi~Victoria Lin}, \bibinfo{person}{Chenglong
  Wang}, \bibinfo{person}{Deric Pang}, \bibinfo{person}{Kevin Vu}, {and}
  \bibinfo{person}{Michael~D Ernst}.} \bibinfo{year}{2017}\natexlab{}.
\newblock \bibinfo{booktitle}{\emph{Program synthesis from natural language
  using recurrent neural networks}}.
\newblock \bibinfo{type}{{T}echnical {R}eport}. \bibinfo{institution}{Technical
  Report UW-CSE-17-03-01, University of Washington Department of Computer
  Science and Engineering, Seattle, WA, USA}.
\newblock


\bibitem[\protect\citeauthoryear{Ling, Grefenstette, Hermann,
  Ko{\v{c}}isk{\`y}, Senior, Wang, and Blunsom}{Ling et~al\mbox{.}}{2016}]%
        {ling2016latent}
\bibfield{author}{\bibinfo{person}{Wang Ling}, \bibinfo{person}{Edward
  Grefenstette}, \bibinfo{person}{Karl~Moritz Hermann},
  \bibinfo{person}{Tom{\'a}{\v{s}} Ko{\v{c}}isk{\`y}}, \bibinfo{person}{Andrew
  Senior}, \bibinfo{person}{Fumin Wang}, {and} \bibinfo{person}{Phil Blunsom}.}
  \bibinfo{year}{2016}\natexlab{}.
\newblock \showarticletitle{Latent predictor networks for code generation}.
\newblock \bibinfo{journal}{\emph{arXiv preprint arXiv:1603.06744}}
  (\bibinfo{year}{2016}).
\newblock


\bibitem[\protect\citeauthoryear{Ling, Lu{\'\i}s, Marujo, Astudillo, Amir,
  Dyer, Black, and Trancoso}{Ling et~al\mbox{.}}{2015}]%
        {ling2015finding}
\bibfield{author}{\bibinfo{person}{Wang Ling}, \bibinfo{person}{Tiago
  Lu{\'\i}s}, \bibinfo{person}{Lu{\'\i}s Marujo},
  \bibinfo{person}{Ram{\'o}n~Fernandez Astudillo}, \bibinfo{person}{Silvio
  Amir}, \bibinfo{person}{Chris Dyer}, \bibinfo{person}{Alan~W Black}, {and}
  \bibinfo{person}{Isabel Trancoso}.} \bibinfo{year}{2015}\natexlab{}.
\newblock \showarticletitle{Finding function in form: Compositional character
  models for open vocabulary word representation}.
\newblock \bibinfo{journal}{\emph{arXiv preprint arXiv:1508.02096}}
  (\bibinfo{year}{2015}).
\newblock


\bibitem[\protect\citeauthoryear{Liu, Chen, Shin, Chen, and Song}{Liu
  et~al\mbox{.}}{2016a}]%
        {liu2016latent}
\bibfield{author}{\bibinfo{person}{Chang Liu}, \bibinfo{person}{Xinyun Chen},
  \bibinfo{person}{Eui~Chul Shin}, \bibinfo{person}{Mingcheng Chen}, {and}
  \bibinfo{person}{Dawn Song}.} \bibinfo{year}{2016}\natexlab{a}.
\newblock \showarticletitle{Latent attention for if-then program synthesis}. In
  \bibinfo{booktitle}{\emph{Advances in Neural Information Processing
  Systems}}. \bibinfo{pages}{4574--4582}.
\newblock


\bibitem[\protect\citeauthoryear{Liu, Wang, Shin, Gonzalez, and Song}{Liu
  et~al\mbox{.}}{2016c}]%
        {liu2016neural}
\bibfield{author}{\bibinfo{person}{Chang Liu}, \bibinfo{person}{Xin Wang},
  \bibinfo{person}{Richard Shin}, \bibinfo{person}{Joseph~E Gonzalez}, {and}
  \bibinfo{person}{Dawn Song}.} \bibinfo{year}{2016}\natexlab{c}.
\newblock \showarticletitle{Neural Code Completion}.
\newblock \bibinfo{journal}{\emph{OpenReview}} (\bibinfo{year}{2016}).
\newblock
\urldef\tempurl%
\url{https://openreview.net/pdf?id=rJbPBt9lg}
\showURL{%
\tempurl}


\bibitem[\protect\citeauthoryear{Liu, Lowe, Serban, Noseworthy, Charlin, and
  Pineau}{Liu et~al\mbox{.}}{2016b}]%
        {liu2016not}
\bibfield{author}{\bibinfo{person}{Chia-Wei Liu}, \bibinfo{person}{Ryan Lowe},
  \bibinfo{person}{Iulian~V Serban}, \bibinfo{person}{Michael Noseworthy},
  \bibinfo{person}{Laurent Charlin}, {and} \bibinfo{person}{Joelle Pineau}.}
  \bibinfo{year}{2016}\natexlab{b}.
\newblock \showarticletitle{How NOT to evaluate your dialogue system: An
  empirical study of unsupervised evaluation metrics for dialogue response
  generation}.
\newblock \bibinfo{journal}{\emph{arXiv preprint arXiv:1603.08023}}
  (\bibinfo{year}{2016}).
\newblock


\bibitem[\protect\citeauthoryear{Liu}{Liu}{2016}]%
        {liu2016towards}
\bibfield{author}{\bibinfo{person}{Han Liu}.} \bibinfo{year}{2016}\natexlab{}.
\newblock \showarticletitle{Towards better program obfuscation: optimization
  via language models}. In \bibinfo{booktitle}{\emph{Proceedings of the 38th
  International Conference on Software Engineering Companion}}. ACM,
  \bibinfo{pages}{680--682}.
\newblock


\bibitem[\protect\citeauthoryear{Lv, Zhang, Lou, Wang, Zhang, and Zhao}{Lv
  et~al\mbox{.}}{2015}]%
        {lv2015codehow}
\bibfield{author}{\bibinfo{person}{Fei Lv}, \bibinfo{person}{Hongyu Zhang},
  \bibinfo{person}{Jian-guang Lou}, \bibinfo{person}{Shaowei Wang},
  \bibinfo{person}{Dongmei Zhang}, {and} \bibinfo{person}{Jianjun Zhao}.}
  \bibinfo{year}{2015}\natexlab{}.
\newblock \showarticletitle{Codehow: Effective code search based on api
  understanding and extended boolean model (e)}. In
  \bibinfo{booktitle}{\emph{Automated Software Engineering (ASE), 2015 30th
  IEEE/ACM International Conference on}}. IEEE, \bibinfo{pages}{260--270}.
\newblock


\bibitem[\protect\citeauthoryear{Ma, Ma, Liu, Huang, Gao, and Jia}{Ma
  et~al\mbox{.}}{2014}]%
        {Ma2014ControlFO}
\bibfield{author}{\bibinfo{person}{Haoyu Ma}, \bibinfo{person}{Xinjie Ma},
  \bibinfo{person}{Weijie Liu}, \bibinfo{person}{Zhipeng Huang},
  \bibinfo{person}{Debin Gao}, {and} \bibinfo{person}{Chunfu Jia}.}
  \bibinfo{year}{2014}\natexlab{}.
\newblock \showarticletitle{Control Flow Obfuscation Using Neural Network to
  Fight Concolic Testing}. In \bibinfo{booktitle}{\emph{SecureComm}}.
\newblock


\bibitem[\protect\citeauthoryear{Maddison and Tarlow}{Maddison and
  Tarlow}{2014}]%
        {maddison2014structured}
\bibfield{author}{\bibinfo{person}{Chris Maddison} {and}
  \bibinfo{person}{Daniel Tarlow}.} \bibinfo{year}{2014}\natexlab{}.
\newblock \showarticletitle{Structured generative models of natural source
  code}. In \bibinfo{booktitle}{\emph{Proceedings of the 31st International
  Conference on Machine Learning (ICML-14)}}. \bibinfo{pages}{649--657}.
\newblock


\bibitem[\protect\citeauthoryear{Maddison, Mnih, and Teh}{Maddison
  et~al\mbox{.}}{2016}]%
        {maddison2016concrete}
\bibfield{author}{\bibinfo{person}{Chris~J Maddison}, \bibinfo{person}{Andriy
  Mnih}, {and} \bibinfo{person}{Yee~Whye Teh}.}
  \bibinfo{year}{2016}\natexlab{}.
\newblock \showarticletitle{The concrete distribution: A continuous relaxation
  of discrete random variables}.
\newblock \bibinfo{journal}{\emph{arXiv preprint arXiv:1611.00712}}
  (\bibinfo{year}{2016}).
\newblock


\bibitem[\protect\citeauthoryear{Madsen, Lhot\'{a}k, and Tip}{Madsen
  et~al\mbox{.}}{2017}]%
        {madsen2017model}
\bibfield{author}{\bibinfo{person}{Magnus Madsen}, \bibinfo{person}{Ond\v{r}ej
  Lhot\'{a}k}, {and} \bibinfo{person}{Frank Tip}.}
  \bibinfo{year}{2017}\natexlab{}.
\newblock \showarticletitle{A Model for Reasoning About JavaScript Promises}.
\newblock \bibinfo{journal}{\emph{Proc. ACM Program. Lang.}}
  \bibinfo{volume}{1}, \bibinfo{number}{OOPSLA}, Article
  \bibinfo{articleno}{86} (\bibinfo{date}{Oct.} \bibinfo{year}{2017}),
  \bibinfo{numpages}{24}~pages.
\newblock
\showISSN{2475-1421}


\bibitem[\protect\citeauthoryear{Mandt, Hoffman, and Blei}{Mandt
  et~al\mbox{.}}{2017}]%
        {mandt2017stochastic}
\bibfield{author}{\bibinfo{person}{Stephan Mandt}, \bibinfo{person}{Matthew~D
  Hoffman}, {and} \bibinfo{person}{David~M Blei}.}
  \bibinfo{year}{2017}\natexlab{}.
\newblock \showarticletitle{Stochastic gradient descent as approximate bayesian
  inference}.
\newblock \bibinfo{journal}{\emph{arXiv preprint arXiv:1704.04289}}
  (\bibinfo{year}{2017}).
\newblock


\bibitem[\protect\citeauthoryear{McCann, Bradbury, Xiong, and Socher}{McCann
  et~al\mbox{.}}{2017}]%
        {mccann2017learned}
\bibfield{author}{\bibinfo{person}{Bryan McCann}, \bibinfo{person}{James
  Bradbury}, \bibinfo{person}{Caiming Xiong}, {and} \bibinfo{person}{Richard
  Socher}.} \bibinfo{year}{2017}\natexlab{}.
\newblock \showarticletitle{Learned in translation: Contextualized word
  vectors}.
\newblock \bibinfo{journal}{\emph{arXiv preprint arXiv:1708.00107}}
  (\bibinfo{year}{2017}).
\newblock


\bibitem[\protect\citeauthoryear{McPherson, Shokri, and Shmatikov}{McPherson
  et~al\mbox{.}}{2016}]%
        {mcpherson2016defeating}
\bibfield{author}{\bibinfo{person}{Richard McPherson}, \bibinfo{person}{Reza
  Shokri}, {and} \bibinfo{person}{Vitaly Shmatikov}.}
  \bibinfo{year}{2016}\natexlab{}.
\newblock \showarticletitle{Defeating image obfuscation with deep learning}.
\newblock \bibinfo{journal}{\emph{arXiv preprint arXiv:1609.00408}}
  (\bibinfo{year}{2016}).
\newblock


\bibitem[\protect\citeauthoryear{Mehri, Kumar, Gulrajani, Kumar, Jain, Sotelo,
  Courville, and Bengio}{Mehri et~al\mbox{.}}{2016}]%
        {mehri2016samplernn}
\bibfield{author}{\bibinfo{person}{Soroush Mehri}, \bibinfo{person}{Kundan
  Kumar}, \bibinfo{person}{Ishaan Gulrajani}, \bibinfo{person}{Rithesh Kumar},
  \bibinfo{person}{Shubham Jain}, \bibinfo{person}{Jose Sotelo},
  \bibinfo{person}{Aaron Courville}, {and} \bibinfo{person}{Yoshua Bengio}.}
  \bibinfo{year}{2016}\natexlab{}.
\newblock \showarticletitle{SampleRNN: An unconditional end-to-end neural audio
  generation model}.
\newblock \bibinfo{journal}{\emph{arXiv preprint arXiv:1612.07837}}
  (\bibinfo{year}{2016}).
\newblock


\bibitem[\protect\citeauthoryear{Melis, Dyer, and Blunsom}{Melis
  et~al\mbox{.}}{2017}]%
        {melis2017state}
\bibfield{author}{\bibinfo{person}{G{\'a}bor Melis}, \bibinfo{person}{Chris
  Dyer}, {and} \bibinfo{person}{Phil Blunsom}.}
  \bibinfo{year}{2017}\natexlab{}.
\newblock \showarticletitle{On the state of the art of evaluation in neural
  language models}.
\newblock \bibinfo{journal}{\emph{arXiv preprint arXiv:1707.05589}}
  (\bibinfo{year}{2017}).
\newblock


\bibitem[\protect\citeauthoryear{Merity, Keskar, and Socher}{Merity
  et~al\mbox{.}}{2017a}]%
        {merity2017regularizing}
\bibfield{author}{\bibinfo{person}{Stephen Merity},
  \bibinfo{person}{Nitish~Shirish Keskar}, {and} \bibinfo{person}{Richard
  Socher}.} \bibinfo{year}{2017}\natexlab{a}.
\newblock \showarticletitle{Regularizing and optimizing LSTM language models}.
\newblock \bibinfo{journal}{\emph{arXiv preprint arXiv:1708.02182}}
  (\bibinfo{year}{2017}).
\newblock


\bibitem[\protect\citeauthoryear{Merity, McCann, and Socher}{Merity
  et~al\mbox{.}}{2017b}]%
        {merity2017revisiting}
\bibfield{author}{\bibinfo{person}{Stephen Merity}, \bibinfo{person}{Bryan
  McCann}, {and} \bibinfo{person}{Richard Socher}.}
  \bibinfo{year}{2017}\natexlab{b}.
\newblock \showarticletitle{Revisiting Activation Regularization for Language
  RNNs}.
\newblock \bibinfo{journal}{\emph{arXiv preprint arXiv:1708.01009}}
  (\bibinfo{year}{2017}).
\newblock


\bibitem[\protect\citeauthoryear{Mikolov, Chen, Corrado, and Dean}{Mikolov
  et~al\mbox{.}}{2013a}]%
        {mikolov2013efficient}
\bibfield{author}{\bibinfo{person}{Tomas Mikolov}, \bibinfo{person}{Kai Chen},
  \bibinfo{person}{Greg Corrado}, {and} \bibinfo{person}{Jeffrey Dean}.}
  \bibinfo{year}{2013}\natexlab{a}.
\newblock \showarticletitle{Efficient estimation of word representations in
  vector space}.
\newblock \bibinfo{journal}{\emph{arXiv preprint arXiv:1301.3781}}
  (\bibinfo{year}{2013}).
\newblock


\bibitem[\protect\citeauthoryear{Mikolov, Joulin, and Baroni}{Mikolov
  et~al\mbox{.}}{2015}]%
        {mikolov2015roadmap}
\bibfield{author}{\bibinfo{person}{Tomas Mikolov}, \bibinfo{person}{Armand
  Joulin}, {and} \bibinfo{person}{Marco Baroni}.}
  \bibinfo{year}{2015}\natexlab{}.
\newblock \showarticletitle{A roadmap towards machine intelligence}.
\newblock \bibinfo{journal}{\emph{arXiv preprint arXiv:1511.08130}}
  (\bibinfo{year}{2015}).
\newblock


\bibitem[\protect\citeauthoryear{Mikolov, Sutskever, Chen, Corrado, and
  Dean}{Mikolov et~al\mbox{.}}{2013b}]%
        {mikolov2013distributed}
\bibfield{author}{\bibinfo{person}{Tomas Mikolov}, \bibinfo{person}{Ilya
  Sutskever}, \bibinfo{person}{Kai Chen}, \bibinfo{person}{Greg~S Corrado},
  {and} \bibinfo{person}{Jeff Dean}.} \bibinfo{year}{2013}\natexlab{b}.
\newblock \showarticletitle{Distributed representations of words and phrases
  and their compositionality}. In \bibinfo{booktitle}{\emph{Advances in neural
  information processing systems}}. \bibinfo{pages}{3111--3119}.
\newblock


\bibitem[\protect\citeauthoryear{Miller, Fisch, Dodge, Karimi, Bordes, and
  Weston}{Miller et~al\mbox{.}}{2016}]%
        {miller2016key}
\bibfield{author}{\bibinfo{person}{Alexander Miller}, \bibinfo{person}{Adam
  Fisch}, \bibinfo{person}{Jesse Dodge}, \bibinfo{person}{Amir-Hossein Karimi},
  \bibinfo{person}{Antoine Bordes}, {and} \bibinfo{person}{Jason Weston}.}
  \bibinfo{year}{2016}\natexlab{}.
\newblock \showarticletitle{Key-value memory networks for directly reading
  documents}.
\newblock \bibinfo{journal}{\emph{arXiv preprint arXiv:1606.03126}}
  (\bibinfo{year}{2016}).
\newblock


\bibitem[\protect\citeauthoryear{Mishra, Rohaninejad, Chen, and Abbeel}{Mishra
  et~al\mbox{.}}{2017}]%
        {mishra2017meta}
\bibfield{author}{\bibinfo{person}{Nikhil Mishra}, \bibinfo{person}{Mostafa
  Rohaninejad}, \bibinfo{person}{Xi Chen}, {and} \bibinfo{person}{Pieter
  Abbeel}.} \bibinfo{year}{2017}\natexlab{}.
\newblock \showarticletitle{Meta-Learning with Temporal Convolutions}.
\newblock \bibinfo{journal}{\emph{arXiv preprint arXiv:1707.03141}}
  (\bibinfo{year}{2017}).
\newblock


\bibitem[\protect\citeauthoryear{Misra, Girshick, Fergus, Hebert, Gupta, and
  van~der Maaten}{Misra et~al\mbox{.}}{2017}]%
        {misra2017learning}
\bibfield{author}{\bibinfo{person}{Ishan Misra}, \bibinfo{person}{Ross
  Girshick}, \bibinfo{person}{Rob Fergus}, \bibinfo{person}{Martial Hebert},
  \bibinfo{person}{Abhinav Gupta}, {and} \bibinfo{person}{Laurens van~der
  Maaten}.} \bibinfo{year}{2017}\natexlab{}.
\newblock \showarticletitle{Learning by Asking Questions}.
\newblock \bibinfo{journal}{\emph{arXiv preprint arXiv:1712.01238}}
  (\bibinfo{year}{2017}).
\newblock


\bibitem[\protect\citeauthoryear{Miyato, Kataoka, Koyama, and Yoshida}{Miyato
  et~al\mbox{.}}{[n. d.]}]%
        {miyato2017spectral}
\bibfield{author}{\bibinfo{person}{Takeru Miyato}, \bibinfo{person}{Toshiki
  Kataoka}, \bibinfo{person}{Masanori Koyama}, {and} \bibinfo{person}{Yuichi
  Yoshida}.} \bibinfo{year}{[n. d.]}\natexlab{}.
\newblock \showarticletitle{Spectral Normalization for Generative Adversarial
  Networks}.
\newblock  (\bibinfo{year}{[n. d.]}).
\newblock


\bibitem[\protect\citeauthoryear{Mnih and Gregor}{Mnih and Gregor}{2014}]%
        {mnih2014neural}
\bibfield{author}{\bibinfo{person}{Andriy Mnih} {and} \bibinfo{person}{Karol
  Gregor}.} \bibinfo{year}{2014}\natexlab{}.
\newblock \showarticletitle{Neural variational inference and learning in belief
  networks}.
\newblock \bibinfo{journal}{\emph{arXiv preprint arXiv:1402.0030}}
  (\bibinfo{year}{2014}).
\newblock


\bibitem[\protect\citeauthoryear{Mnih and Hinton}{Mnih and Hinton}{2007}]%
        {mnih2007three}
\bibfield{author}{\bibinfo{person}{Andriy Mnih} {and} \bibinfo{person}{Geoffrey
  Hinton}.} \bibinfo{year}{2007}\natexlab{}.
\newblock \showarticletitle{Three new graphical models for statistical language
  modelling}. In \bibinfo{booktitle}{\emph{Proceedings of the 24th
  international conference on Machine learning}}. ACM,
  \bibinfo{pages}{641--648}.
\newblock


\bibitem[\protect\citeauthoryear{Mnih and Hinton}{Mnih and Hinton}{2009}]%
        {mnih2009scalable}
\bibfield{author}{\bibinfo{person}{Andriy Mnih} {and}
  \bibinfo{person}{Geoffrey~E Hinton}.} \bibinfo{year}{2009}\natexlab{}.
\newblock \showarticletitle{A scalable hierarchical distributed language
  model}. In \bibinfo{booktitle}{\emph{Advances in neural information
  processing systems}}. \bibinfo{pages}{1081--1088}.
\newblock


\bibitem[\protect\citeauthoryear{Monperrus}{Monperrus}{2018}]%
        {monperrus2018automatic}
\bibfield{author}{\bibinfo{person}{Martin Monperrus}.}
  \bibinfo{year}{2018}\natexlab{}.
\newblock \showarticletitle{Automatic software repair: a bibliography}.
\newblock \bibinfo{journal}{\emph{ACM Computing Surveys (CSUR)}}
  \bibinfo{volume}{51}, \bibinfo{number}{1} (\bibinfo{year}{2018}),
  \bibinfo{pages}{17}.
\newblock


\bibitem[\protect\citeauthoryear{Moosavi-Dezfooli, Fawzi, and
  Frossard}{Moosavi-Dezfooli et~al\mbox{.}}{2016}]%
        {moosavi2016deepfool}
\bibfield{author}{\bibinfo{person}{Seyed-Mohsen Moosavi-Dezfooli},
  \bibinfo{person}{Alhussein Fawzi}, {and} \bibinfo{person}{Pascal Frossard}.}
  \bibinfo{year}{2016}\natexlab{}.
\newblock \showarticletitle{Deepfool: a simple and accurate method to fool deep
  neural networks}. In \bibinfo{booktitle}{\emph{Proceedings of the IEEE
  Conference on Computer Vision and Pattern Recognition}}.
  \bibinfo{pages}{2574--2582}.
\newblock


\bibitem[\protect\citeauthoryear{Mou, Li, Zhang, Wang, and Jin}{Mou
  et~al\mbox{.}}{2016}]%
        {mou2016convolutional}
\bibfield{author}{\bibinfo{person}{Lili Mou}, \bibinfo{person}{Ge Li},
  \bibinfo{person}{Lu Zhang}, \bibinfo{person}{Tao Wang}, {and}
  \bibinfo{person}{Zhi Jin}.} \bibinfo{year}{2016}\natexlab{}.
\newblock \showarticletitle{Convolutional neural networks over tree structures
  for programming language processing}. In \bibinfo{booktitle}{\emph{Thirtieth
  AAAI Conference on Artificial Intelligence}}.
\newblock


\bibitem[\protect\citeauthoryear{Muggleton and De~Raedt}{Muggleton and
  De~Raedt}{1994}]%
        {muggleton1994inductive}
\bibfield{author}{\bibinfo{person}{Stephen Muggleton} {and}
  \bibinfo{person}{Luc De~Raedt}.} \bibinfo{year}{1994}\natexlab{}.
\newblock \showarticletitle{Inductive logic programming: Theory and methods}.
\newblock \bibinfo{journal}{\emph{The Journal of Logic Programming}}
  \bibinfo{volume}{19} (\bibinfo{year}{1994}), \bibinfo{pages}{629--679}.
\newblock


\bibitem[\protect\citeauthoryear{Mujika, Meier, and Steger}{Mujika
  et~al\mbox{.}}{2017}]%
        {mujika2017fast}
\bibfield{author}{\bibinfo{person}{Asier Mujika}, \bibinfo{person}{Florian
  Meier}, {and} \bibinfo{person}{Angelika Steger}.}
  \bibinfo{year}{2017}\natexlab{}.
\newblock \showarticletitle{Fast-Slow Recurrent Neural Networks}.
\newblock \bibinfo{journal}{\emph{arXiv preprint arXiv:1705.08639}}
  (\bibinfo{year}{2017}).
\newblock


\bibitem[\protect\citeauthoryear{Neel}{Neel}{2018}]%
        {kant2018recent}
\bibfield{author}{\bibinfo{person}{Kant Neel}.}
  \bibinfo{year}{2018}\natexlab{}.
\newblock \showarticletitle{Recent Advances in Neural Program Synthesis}.
\newblock \bibinfo{journal}{\emph{arXiv preprint arXiv:1802.02353}}
  (\bibinfo{year}{2018}).
\newblock


\bibitem[\protect\citeauthoryear{Neelakantan, Le, and Sutskever}{Neelakantan
  et~al\mbox{.}}{2015}]%
        {neelakantan2015neural}
\bibfield{author}{\bibinfo{person}{Arvind Neelakantan}, \bibinfo{person}{Quoc~V
  Le}, {and} \bibinfo{person}{Ilya Sutskever}.}
  \bibinfo{year}{2015}\natexlab{}.
\newblock \showarticletitle{Neural programmer: Inducing latent programs with
  gradient descent}.
\newblock \bibinfo{journal}{\emph{arXiv preprint arXiv:1511.04834}}
  (\bibinfo{year}{2015}).
\newblock


\bibitem[\protect\citeauthoryear{Nguyen, Yosinski, and Clune}{Nguyen
  et~al\mbox{.}}{2015}]%
        {nguyen2015deep}
\bibfield{author}{\bibinfo{person}{Anh Nguyen}, \bibinfo{person}{Jason
  Yosinski}, {and} \bibinfo{person}{Jeff Clune}.}
  \bibinfo{year}{2015}\natexlab{}.
\newblock \showarticletitle{Deep neural networks are easily fooled: High
  confidence predictions for unrecognizable images}. In
  \bibinfo{booktitle}{\emph{Proceedings of the IEEE Conference on Computer
  Vision and Pattern Recognition}}. \bibinfo{pages}{427--436}.
\newblock


\bibitem[\protect\citeauthoryear{Nguyen and Nguyen}{Nguyen and Nguyen}{2015}]%
        {nguyen2015graph}
\bibfield{author}{\bibinfo{person}{Anh~Tuan Nguyen} {and}
  \bibinfo{person}{Tien~N Nguyen}.} \bibinfo{year}{2015}\natexlab{}.
\newblock \showarticletitle{Graph-based statistical language model for code}.
  In \bibinfo{booktitle}{\emph{Software Engineering (ICSE), 2015 IEEE/ACM 37th
  IEEE International Conference on}}, Vol.~\bibinfo{volume}{1}. IEEE,
  \bibinfo{pages}{858--868}.
\newblock


\bibitem[\protect\citeauthoryear{Nguyen, Nguyen, Phan, and Nguyen}{Nguyen
  et~al\mbox{.}}{2017}]%
        {nguyen2017exploring}
\bibfield{author}{\bibinfo{person}{Trong~Duc Nguyen}, \bibinfo{person}{Anh~Tuan
  Nguyen}, \bibinfo{person}{Hung~Dang Phan}, {and} \bibinfo{person}{Tien~N
  Nguyen}.} \bibinfo{year}{2017}\natexlab{}.
\newblock \showarticletitle{Exploring API embedding for API usages and
  applications}. In \bibinfo{booktitle}{\emph{Software Engineering (ICSE), 2017
  IEEE/ACM 39th International Conference on}}. IEEE, \bibinfo{pages}{438--449}.
\newblock


\bibitem[\protect\citeauthoryear{Nguyen, Nguyen, Nguyen, and Nguyen}{Nguyen
  et~al\mbox{.}}{2013a}]%
        {nguyen2013ngrams}
\bibfield{author}{\bibinfo{person}{Tung~Thanh Nguyen},
  \bibinfo{person}{Anh~Tuan Nguyen}, \bibinfo{person}{Hoan~Anh Nguyen}, {and}
  \bibinfo{person}{Tien~N. Nguyen}.} \bibinfo{year}{2013}\natexlab{a}.
\newblock \showarticletitle{A Statistical Semantic Language Model for Source
  Code}. In \bibinfo{booktitle}{\emph{Proceedings of the 2013 9th Joint Meeting
  on Foundations of Software Engineering}} \emph{(\bibinfo{series}{ESEC/FSE
  2013})}. \bibinfo{publisher}{ACM}, \bibinfo{address}{New York, NY, USA},
  \bibinfo{pages}{532--542}.
\newblock


\bibitem[\protect\citeauthoryear{Nguyen, Nguyen, Nguyen, and Nguyen}{Nguyen
  et~al\mbox{.}}{2013b}]%
        {nguyen2013statistical}
\bibfield{author}{\bibinfo{person}{Tung~Thanh Nguyen},
  \bibinfo{person}{Anh~Tuan Nguyen}, \bibinfo{person}{Hoan~Anh Nguyen}, {and}
  \bibinfo{person}{Tien~N Nguyen}.} \bibinfo{year}{2013}\natexlab{b}.
\newblock \showarticletitle{A statistical semantic language model for source
  code}. In \bibinfo{booktitle}{\emph{Proceedings of the 2013 9th Joint Meeting
  on Foundations of Software Engineering}}. ACM, \bibinfo{pages}{532--542}.
\newblock


\bibitem[\protect\citeauthoryear{Oda, Fudaba, Neubig, Hata, Sakti, Toda, and
  Nakamura}{Oda et~al\mbox{.}}{2015}]%
        {oda2015learning}
\bibfield{author}{\bibinfo{person}{Yusuke Oda}, \bibinfo{person}{Hiroyuki
  Fudaba}, \bibinfo{person}{Graham Neubig}, \bibinfo{person}{Hideaki Hata},
  \bibinfo{person}{Sakriani Sakti}, \bibinfo{person}{Tomoki Toda}, {and}
  \bibinfo{person}{Satoshi Nakamura}.} \bibinfo{year}{2015}\natexlab{}.
\newblock \showarticletitle{Learning to generate pseudo-code from source code
  using statistical machine translation (t)}. In
  \bibinfo{booktitle}{\emph{Automated Software Engineering (ASE), 2015 30th
  IEEE/ACM International Conference on}}. IEEE, \bibinfo{pages}{574--584}.
\newblock


\bibitem[\protect\citeauthoryear{Oord, Dieleman, Zen, Simonyan, Vinyals,
  Graves, Kalchbrenner, Senior, and Kavukcuoglu}{Oord et~al\mbox{.}}{2016}]%
        {oord2016wavenet}
\bibfield{author}{\bibinfo{person}{Aaron van~den Oord}, \bibinfo{person}{Sander
  Dieleman}, \bibinfo{person}{Heiga Zen}, \bibinfo{person}{Karen Simonyan},
  \bibinfo{person}{Oriol Vinyals}, \bibinfo{person}{Alex Graves},
  \bibinfo{person}{Nal Kalchbrenner}, \bibinfo{person}{Andrew Senior}, {and}
  \bibinfo{person}{Koray Kavukcuoglu}.} \bibinfo{year}{2016}\natexlab{}.
\newblock \showarticletitle{Wavenet: A generative model for raw audio}.
\newblock \bibinfo{journal}{\emph{arXiv preprint arXiv:1609.03499}}
  (\bibinfo{year}{2016}).
\newblock


\bibitem[\protect\citeauthoryear{Oord, Li, Babuschkin, Simonyan, Vinyals, and
  Kavukcuoglu}{Oord et~al\mbox{.}}{2017a}]%
        {oord2017parallel}
\bibfield{author}{\bibinfo{person}{Aaron van~den Oord}, \bibinfo{person}{Yazhe
  Li}, \bibinfo{person}{Igor Babuschkin}, \bibinfo{person}{Karen Simonyan},
  \bibinfo{person}{Oriol Vinyals}, {and} \bibinfo{person}{Koray Kavukcuoglu}.}
  \bibinfo{year}{2017}\natexlab{a}.
\newblock \showarticletitle{Parallel WaveNet: Fast High-Fidelity Speech
  Synthesis}.
\newblock \bibinfo{journal}{\emph{arXiv preprint arXiv:1711.10433}}
  (\bibinfo{year}{2017}).
\newblock


\bibitem[\protect\citeauthoryear{Oord, Vinyals, and Kavukcuoglu}{Oord
  et~al\mbox{.}}{2017b}]%
        {oord2017neural}
\bibfield{author}{\bibinfo{person}{Aaron van~den Oord}, \bibinfo{person}{Oriol
  Vinyals}, {and} \bibinfo{person}{Koray Kavukcuoglu}.}
  \bibinfo{year}{2017}\natexlab{b}.
\newblock \showarticletitle{Neural Discrete Representation Learning}.
\newblock \bibinfo{journal}{\emph{arXiv preprint arXiv:1711.00937}}
  (\bibinfo{year}{2017}).
\newblock


\bibitem[\protect\citeauthoryear{Ott, Atchison, Harnack, Bergh, and
  Linstead}{Ott et~al\mbox{.}}{2018a}]%
        {ott2018deep}
\bibfield{author}{\bibinfo{person}{Jordan Ott}, \bibinfo{person}{Abigail
  Atchison}, \bibinfo{person}{Paul Harnack}, \bibinfo{person}{Adrienne Bergh},
  {and} \bibinfo{person}{Erik Linstead}.} \bibinfo{year}{2018}\natexlab{a}.
\newblock \showarticletitle{A Deep Learning Approach to Identifying Source Code
  in Images and Video}. In \bibinfo{booktitle}{\emph{Proceedings of the 15th
  International Conference on Mining Software Repositories}}
  \emph{(\bibinfo{series}{MSR '18})}. \bibinfo{publisher}{ACM},
  \bibinfo{address}{New York, NY, USA}, \bibinfo{pages}{376--386}.
\newblock
\showISBNx{978-1-4503-5716-6}


\bibitem[\protect\citeauthoryear{Ott, Atchison, Harnack, Best, Anderson,
  Firmani, and Linstead}{Ott et~al\mbox{.}}{2018b}]%
        {ott2018learning}
\bibfield{author}{\bibinfo{person}{Jordan Ott}, \bibinfo{person}{Abigail
  Atchison}, \bibinfo{person}{Paul Harnack}, \bibinfo{person}{Natalie Best},
  \bibinfo{person}{Haley Anderson}, \bibinfo{person}{Cristiano Firmani}, {and}
  \bibinfo{person}{Erik Linstead}.} \bibinfo{year}{2018}\natexlab{b}.
\newblock \showarticletitle{Learning Lexical Features of Programming Languages
  from Imagery Using Convolutional Neural Networks}. In
  \bibinfo{booktitle}{\emph{Proceedings of the 26th Conference on Program
  Comprehension}} \emph{(\bibinfo{series}{ICPC '18})}.
  \bibinfo{publisher}{ACM}, \bibinfo{address}{New York, NY, USA},
  \bibinfo{pages}{336--339}.
\newblock


\bibitem[\protect\citeauthoryear{Papernot, McDaniel, Swami, and
  Harang}{Papernot et~al\mbox{.}}{2016}]%
        {papernot2016crafting}
\bibfield{author}{\bibinfo{person}{Nicolas Papernot}, \bibinfo{person}{Patrick
  McDaniel}, \bibinfo{person}{Ananthram Swami}, {and} \bibinfo{person}{Richard
  Harang}.} \bibinfo{year}{2016}\natexlab{}.
\newblock \showarticletitle{Crafting adversarial input sequences for recurrent
  neural networks}. In \bibinfo{booktitle}{\emph{Military Communications
  Conference, MILCOM 2016-2016 IEEE}}. IEEE, \bibinfo{pages}{49--54}.
\newblock


\bibitem[\protect\citeauthoryear{Papineni, Roukos, Ward, and Zhu}{Papineni
  et~al\mbox{.}}{2002}]%
        {papineni2002bleu}
\bibfield{author}{\bibinfo{person}{Kishore Papineni}, \bibinfo{person}{Salim
  Roukos}, \bibinfo{person}{Todd Ward}, {and} \bibinfo{person}{Wei-Jing Zhu}.}
  \bibinfo{year}{2002}\natexlab{}.
\newblock \showarticletitle{BLEU: a method for automatic evaluation of machine
  translation}. In \bibinfo{booktitle}{\emph{Proceedings of the 40th annual
  meeting on association for computational linguistics}}. Association for
  Computational Linguistics, \bibinfo{pages}{311--318}.
\newblock


\bibitem[\protect\citeauthoryear{Parisotto, Mohamed, Singh, Li, Zhou, and
  Kohli}{Parisotto et~al\mbox{.}}{2016}]%
        {parisotto2016neuro}
\bibfield{author}{\bibinfo{person}{Emilio Parisotto},
  \bibinfo{person}{Abdel-rahman Mohamed}, \bibinfo{person}{Rishabh Singh},
  \bibinfo{person}{Lihong Li}, \bibinfo{person}{Dengyong Zhou}, {and}
  \bibinfo{person}{Pushmeet Kohli}.} \bibinfo{year}{2016}\natexlab{}.
\newblock \showarticletitle{Neuro-symbolic program synthesis}.
\newblock \bibinfo{journal}{\emph{arXiv preprint arXiv:1611.01855}}
  (\bibinfo{year}{2016}).
\newblock


\bibitem[\protect\citeauthoryear{Parisotto and Salakhutdinov}{Parisotto and
  Salakhutdinov}{2017}]%
        {parisotto2017neural}
\bibfield{author}{\bibinfo{person}{Emilio Parisotto} {and}
  \bibinfo{person}{Ruslan Salakhutdinov}.} \bibinfo{year}{2017}\natexlab{}.
\newblock \showarticletitle{Neural map: Structured memory for deep
  reinforcement learning}.
\newblock \bibinfo{journal}{\emph{arXiv preprint arXiv:1702.08360}}
  (\bibinfo{year}{2017}).
\newblock


\bibitem[\protect\citeauthoryear{Pascanu, Gulcehre, Cho, and Bengio}{Pascanu
  et~al\mbox{.}}{2013a}]%
        {pascanu2013construct}
\bibfield{author}{\bibinfo{person}{Razvan Pascanu}, \bibinfo{person}{Caglar
  Gulcehre}, \bibinfo{person}{Kyunghyun Cho}, {and} \bibinfo{person}{Yoshua
  Bengio}.} \bibinfo{year}{2013}\natexlab{a}.
\newblock \showarticletitle{How to construct deep recurrent neural networks}.
\newblock \bibinfo{journal}{\emph{arXiv preprint arXiv:1312.6026}}
  (\bibinfo{year}{2013}).
\newblock


\bibitem[\protect\citeauthoryear{Pascanu, Mikolov, and Bengio}{Pascanu
  et~al\mbox{.}}{2013b}]%
        {pascanu2013difficulty}
\bibfield{author}{\bibinfo{person}{Razvan Pascanu}, \bibinfo{person}{Tomas
  Mikolov}, {and} \bibinfo{person}{Yoshua Bengio}.}
  \bibinfo{year}{2013}\natexlab{b}.
\newblock \showarticletitle{On the difficulty of training recurrent neural
  networks}. In \bibinfo{booktitle}{\emph{International Conference on Machine
  Learning}}. \bibinfo{pages}{1310--1318}.
\newblock


\bibitem[\protect\citeauthoryear{Pennington, Socher, and Manning}{Pennington
  et~al\mbox{.}}{2014}]%
        {pennington2014glove}
\bibfield{author}{\bibinfo{person}{Jeffrey Pennington},
  \bibinfo{person}{Richard Socher}, {and} \bibinfo{person}{Christopher~D.
  Manning}.} \bibinfo{year}{2014}\natexlab{}.
\newblock \showarticletitle{GloVe: Global Vectors for Word Representation}. In
  \bibinfo{booktitle}{\emph{Empirical Methods in Natural Language Processing
  (EMNLP)}}. \bibinfo{pages}{1532--1543}.
\newblock


\bibitem[\protect\citeauthoryear{Perzanowski, Schultz, Adams, Marsh, and
  Bugajska}{Perzanowski et~al\mbox{.}}{2001}]%
        {perzanowski2001building}
\bibfield{author}{\bibinfo{person}{Dennis Perzanowski}, \bibinfo{person}{Alan~C
  Schultz}, \bibinfo{person}{William Adams}, \bibinfo{person}{Elaine Marsh},
  {and} \bibinfo{person}{Magda Bugajska}.} \bibinfo{year}{2001}\natexlab{}.
\newblock \showarticletitle{Building a multimodal human-robot interface}.
\newblock \bibinfo{journal}{\emph{IEEE intelligent systems}}
  \bibinfo{volume}{16}, \bibinfo{number}{1} (\bibinfo{year}{2001}),
  \bibinfo{pages}{16--21}.
\newblock


\bibitem[\protect\citeauthoryear{Peters, Ammar, Bhagavatula, and Power}{Peters
  et~al\mbox{.}}{2017}]%
        {peters2017semi}
\bibfield{author}{\bibinfo{person}{Matthew~E Peters}, \bibinfo{person}{Waleed
  Ammar}, \bibinfo{person}{Chandra Bhagavatula}, {and} \bibinfo{person}{Russell
  Power}.} \bibinfo{year}{2017}\natexlab{}.
\newblock \showarticletitle{Semi-supervised sequence tagging with bidirectional
  language models}.
\newblock \bibinfo{journal}{\emph{arXiv preprint arXiv:1705.00108}}
  (\bibinfo{year}{2017}).
\newblock


\bibitem[\protect\citeauthoryear{Polosukhin and Skidanov}{Polosukhin and
  Skidanov}{2018}]%
        {polosukhin2018neural}
\bibfield{author}{\bibinfo{person}{Illia Polosukhin} {and}
  \bibinfo{person}{Alexander Skidanov}.} \bibinfo{year}{2018}\natexlab{}.
\newblock \showarticletitle{Neural Program Search: Solving Programming Tasks
  from Description and Examples}.
\newblock \bibinfo{journal}{\emph{arXiv preprint arXiv:1802.04335}}
  (\bibinfo{year}{2018}).
\newblock


\bibitem[\protect\citeauthoryear{Ponzanelli, Bavota, Mocci, Di~Penta, Oliveto,
  Russo, Haiduc, and Lanza}{Ponzanelli et~al\mbox{.}}{2016}]%
        {ponzanelli2016codetube}
\bibfield{author}{\bibinfo{person}{Luca Ponzanelli}, \bibinfo{person}{Gabriele
  Bavota}, \bibinfo{person}{Andrea Mocci}, \bibinfo{person}{Massimiliano
  Di~Penta}, \bibinfo{person}{Rocco Oliveto}, \bibinfo{person}{Barbara Russo},
  \bibinfo{person}{Sonia Haiduc}, {and} \bibinfo{person}{Michele Lanza}.}
  \bibinfo{year}{2016}\natexlab{}.
\newblock \showarticletitle{CodeTube: extracting relevant fragments from
  software development video tutorials}. In
  \bibinfo{booktitle}{\emph{Proceedings of the 38th International Conference on
  Software Engineering Companion}}. ACM, \bibinfo{pages}{645--648}.
\newblock


\bibitem[\protect\citeauthoryear{Pouyanfar, Sadiq, Yan, Tian, Tao, Reyes, Shyu,
  Chen, and Iyengar}{Pouyanfar et~al\mbox{.}}{2018}]%
        {pouyanfar2018survey}
\bibfield{author}{\bibinfo{person}{Samira Pouyanfar}, \bibinfo{person}{Saad
  Sadiq}, \bibinfo{person}{Yilin Yan}, \bibinfo{person}{Haiman Tian},
  \bibinfo{person}{Yudong Tao}, \bibinfo{person}{Maria~Presa Reyes},
  \bibinfo{person}{Mei-Ling Shyu}, \bibinfo{person}{Shu-Ching Chen}, {and}
  \bibinfo{person}{SS Iyengar}.} \bibinfo{year}{2018}\natexlab{}.
\newblock \showarticletitle{A Survey on Deep Learning: Algorithms, Techniques,
  and Applications}.
\newblock \bibinfo{journal}{\emph{ACM Computing Surveys (CSUR)}}
  \bibinfo{volume}{51}, \bibinfo{number}{5} (\bibinfo{year}{2018}),
  \bibinfo{pages}{92}.
\newblock


\bibitem[\protect\citeauthoryear{Pradel and Sen}{Pradel and Sen}{2018}]%
        {pradel2018deepbugs}
\bibfield{author}{\bibinfo{person}{Michael Pradel} {and}
  \bibinfo{person}{Koushik Sen}.} \bibinfo{year}{2018}\natexlab{}.
\newblock \showarticletitle{DeepBugs: A Learning Approach to Name-based Bug
  Detection}.
\newblock \bibinfo{journal}{\emph{arXiv preprint arXiv:1805.11683}}
  (\bibinfo{year}{2018}).
\newblock


\bibitem[\protect\citeauthoryear{Pu, Miranda, Solar-Lezama, and Kaelbling}{Pu
  et~al\mbox{.}}{2017}]%
        {pu2017learning}
\bibfield{author}{\bibinfo{person}{Yewen Pu}, \bibinfo{person}{Zachery
  Miranda}, \bibinfo{person}{Armando Solar-Lezama}, {and}
  \bibinfo{person}{Leslie~Pack Kaelbling}.} \bibinfo{year}{2017}\natexlab{}.
\newblock \showarticletitle{Learning to select examples for program synthesis}.
\newblock \bibinfo{journal}{\emph{arXiv preprint arXiv:1711.03243}}
  (\bibinfo{year}{2017}).
\newblock


\bibitem[\protect\citeauthoryear{Quirk, Mooney, and Galley}{Quirk
  et~al\mbox{.}}{2015}]%
        {quirk2015language}
\bibfield{author}{\bibinfo{person}{Chris Quirk}, \bibinfo{person}{Raymond
  Mooney}, {and} \bibinfo{person}{Michel Galley}.}
  \bibinfo{year}{2015}\natexlab{}.
\newblock \showarticletitle{Language to code: Learning semantic parsers for
  if-this-then-that recipes}. In \bibinfo{booktitle}{\emph{Proceedings of the
  53rd Annual Meeting of the Association for Computational Linguistics and the
  7th International Joint Conference on Natural Language Processing (Volume 1:
  Long Papers)}}, Vol.~\bibinfo{volume}{1}. \bibinfo{pages}{878--888}.
\newblock


\bibitem[\protect\citeauthoryear{Rabinovich, Stern, and Klein}{Rabinovich
  et~al\mbox{.}}{2017}]%
        {rabinovich2017abstract}
\bibfield{author}{\bibinfo{person}{Maxim Rabinovich}, \bibinfo{person}{Mitchell
  Stern}, {and} \bibinfo{person}{Dan Klein}.} \bibinfo{year}{2017}\natexlab{}.
\newblock \showarticletitle{Abstract Syntax Networks for Code Generation and
  Semantic Parsing}.
\newblock \bibinfo{journal}{\emph{arXiv preprint arXiv:1704.07535}}
  (\bibinfo{year}{2017}).
\newblock


\bibitem[\protect\citeauthoryear{Rae, Hunt, Danihelka, Harley, Senior, Wayne,
  Graves, and Lillicrap}{Rae et~al\mbox{.}}{2016}]%
        {rae2016scaling}
\bibfield{author}{\bibinfo{person}{Jack Rae}, \bibinfo{person}{Jonathan~J
  Hunt}, \bibinfo{person}{Ivo Danihelka}, \bibinfo{person}{Timothy Harley},
  \bibinfo{person}{Andrew~W Senior}, \bibinfo{person}{Gregory Wayne},
  \bibinfo{person}{Alex Graves}, {and} \bibinfo{person}{Tim Lillicrap}.}
  \bibinfo{year}{2016}\natexlab{}.
\newblock \showarticletitle{Scaling memory-augmented neural networks with
  sparse reads and writes}. In \bibinfo{booktitle}{\emph{Advances in Neural
  Information Processing Systems}}. \bibinfo{pages}{3621--3629}.
\newblock


\bibitem[\protect\citeauthoryear{Rasmus, Berglund, Honkala, Valpola, and
  Raiko}{Rasmus et~al\mbox{.}}{2015}]%
        {rasmus2015semi}
\bibfield{author}{\bibinfo{person}{Antti Rasmus}, \bibinfo{person}{Mathias
  Berglund}, \bibinfo{person}{Mikko Honkala}, \bibinfo{person}{Harri Valpola},
  {and} \bibinfo{person}{Tapani Raiko}.} \bibinfo{year}{2015}\natexlab{}.
\newblock \showarticletitle{Semi-supervised learning with ladder networks}. In
  \bibinfo{booktitle}{\emph{Advances in Neural Information Processing
  Systems}}. \bibinfo{pages}{3546--3554}.
\newblock


\bibitem[\protect\citeauthoryear{Ray, Hellendoorn, Godhane, Tu, Bacchelli, and
  Devanbu}{Ray et~al\mbox{.}}{2016}]%
        {ray2016naturalness}
\bibfield{author}{\bibinfo{person}{Baishakhi Ray}, \bibinfo{person}{Vincent
  Hellendoorn}, \bibinfo{person}{Saheel Godhane}, \bibinfo{person}{Zhaopeng
  Tu}, \bibinfo{person}{Alberto Bacchelli}, {and} \bibinfo{person}{Premkumar
  Devanbu}.} \bibinfo{year}{2016}\natexlab{}.
\newblock \showarticletitle{On the" naturalness" of buggy code}. In
  \bibinfo{booktitle}{\emph{2016 IEEE/ACM 38th International Conference on
  Software Engineering (ICSE)}}. IEEE, \bibinfo{pages}{428--439}.
\newblock


\bibitem[\protect\citeauthoryear{Raychev, Bielik, and Vechev}{Raychev
  et~al\mbox{.}}{2016}]%
        {raychev2016probabilistic}
\bibfield{author}{\bibinfo{person}{Veselin Raychev}, \bibinfo{person}{Pavol
  Bielik}, {and} \bibinfo{person}{Martin Vechev}.}
  \bibinfo{year}{2016}\natexlab{}.
\newblock \showarticletitle{Probabilistic model for code with decision trees}.
  In \bibinfo{booktitle}{\emph{Proceedings of the 2016 ACM SIGPLAN
  International Conference on Object-Oriented Programming, Systems, Languages,
  and Applications}}. ACM, \bibinfo{pages}{731--747}.
\newblock


\bibitem[\protect\citeauthoryear{Raychev, Vechev, and Krause}{Raychev
  et~al\mbox{.}}{2015}]%
        {raychev2015predicting}
\bibfield{author}{\bibinfo{person}{Veselin Raychev}, \bibinfo{person}{Martin
  Vechev}, {and} \bibinfo{person}{Andreas Krause}.}
  \bibinfo{year}{2015}\natexlab{}.
\newblock \showarticletitle{Predicting program properties from big code}. In
  \bibinfo{booktitle}{\emph{ACM SIGPLAN Notices}}, Vol.~\bibinfo{volume}{50}.
  ACM, \bibinfo{pages}{111--124}.
\newblock


\bibitem[\protect\citeauthoryear{Raychev, Vechev, and Yahav}{Raychev
  et~al\mbox{.}}{2014a}]%
        {code_completion_ngrams}
\bibfield{author}{\bibinfo{person}{Veselin Raychev}, \bibinfo{person}{Martin
  Vechev}, {and} \bibinfo{person}{Eran Yahav}.}
  \bibinfo{year}{2014}\natexlab{a}.
\newblock \showarticletitle{Code Completion with Statistical Language Models}.
  In \bibinfo{booktitle}{\emph{Proceedings of the 35th ACM SIGPLAN Conference
  on Programming Language Design and Implementation}}
  \emph{(\bibinfo{series}{PLDI '14})}. \bibinfo{publisher}{ACM},
  \bibinfo{address}{New York, NY, USA}, \bibinfo{pages}{419--428}.
\newblock


\bibitem[\protect\citeauthoryear{Raychev, Vechev, and Yahav}{Raychev
  et~al\mbox{.}}{2014b}]%
        {raychev2014code}
\bibfield{author}{\bibinfo{person}{Veselin Raychev}, \bibinfo{person}{Martin
  Vechev}, {and} \bibinfo{person}{Eran Yahav}.}
  \bibinfo{year}{2014}\natexlab{b}.
\newblock \showarticletitle{Code completion with statistical language models}.
  In \bibinfo{booktitle}{\emph{ACM SIGPLAN Notices}},
  Vol.~\bibinfo{volume}{49}. ACM, \bibinfo{pages}{419--428}.
\newblock


\bibitem[\protect\citeauthoryear{Reed and De~Freitas}{Reed and
  De~Freitas}{2015}]%
        {reed2015neural}
\bibfield{author}{\bibinfo{person}{Scott Reed} {and} \bibinfo{person}{Nando
  De~Freitas}.} \bibinfo{year}{2015}\natexlab{}.
\newblock \showarticletitle{Neural programmer-interpreters}.
\newblock \bibinfo{journal}{\emph{arXiv preprint arXiv:1511.06279}}
  (\bibinfo{year}{2015}).
\newblock


\bibitem[\protect\citeauthoryear{Ren, Wang, Zhang, Lv, and Li}{Ren
  et~al\mbox{.}}{2017}]%
        {ren2017deep}
\bibfield{author}{\bibinfo{person}{Zhou Ren}, \bibinfo{person}{Xiaoyu Wang},
  \bibinfo{person}{Ning Zhang}, \bibinfo{person}{Xutao Lv}, {and}
  \bibinfo{person}{Li-Jia Li}.} \bibinfo{year}{2017}\natexlab{}.
\newblock \showarticletitle{Deep reinforcement learning-based image captioning
  with embedding reward}.
\newblock \bibinfo{journal}{\emph{arXiv preprint arXiv:1704.03899}}
  (\bibinfo{year}{2017}).
\newblock


\bibitem[\protect\citeauthoryear{Riedel, Bosnjak, and Rockt{\"a}schel}{Riedel
  et~al\mbox{.}}{2016}]%
        {riedel2016programming}
\bibfield{author}{\bibinfo{person}{Sebastian Riedel}, \bibinfo{person}{Matko
  Bosnjak}, {and} \bibinfo{person}{Tim Rockt{\"a}schel}.}
  \bibinfo{year}{2016}\natexlab{}.
\newblock \showarticletitle{Programming with a differentiable forth
  interpreter}.
\newblock \bibinfo{journal}{\emph{arXiv preprint arXiv:1605.06640}}
  (\bibinfo{year}{2016}).
\newblock


\bibitem[\protect\citeauthoryear{Rosenblatt}{Rosenblatt}{1958}]%
        {rosenblatt1958perceptron}
\bibfield{author}{\bibinfo{person}{Frank Rosenblatt}.}
  \bibinfo{year}{1958}\natexlab{}.
\newblock \showarticletitle{The perceptron: a probabilistic model for
  information storage and organization in the brain.}
\newblock \bibinfo{journal}{\emph{Psychological review}} \bibinfo{volume}{65},
  \bibinfo{number}{6} (\bibinfo{year}{1958}), \bibinfo{pages}{386}.
\newblock


\bibitem[\protect\citeauthoryear{Rothe, Lake, and Gureckis}{Rothe
  et~al\mbox{.}}{2017}]%
        {rothe2017question}
\bibfield{author}{\bibinfo{person}{Anselm Rothe}, \bibinfo{person}{Brenden~M
  Lake}, {and} \bibinfo{person}{Todd Gureckis}.}
  \bibinfo{year}{2017}\natexlab{}.
\newblock \showarticletitle{Question Asking as Program Generation}.
\newblock In \bibinfo{booktitle}{\emph{Advances in Neural Information
  Processing Systems 30}}, \bibfield{editor}{\bibinfo{person}{I.~Guyon},
  \bibinfo{person}{U.~V. Luxburg}, \bibinfo{person}{S.~Bengio},
  \bibinfo{person}{H.~Wallach}, \bibinfo{person}{R.~Fergus},
  \bibinfo{person}{S.~Vishwanathan}, {and} \bibinfo{person}{R.~Garnett}}
  (Eds.). \bibinfo{publisher}{Curran Associates, Inc.},
  \bibinfo{pages}{1046--1055}.
\newblock


\bibitem[\protect\citeauthoryear{Rush, Chopra, and Weston}{Rush
  et~al\mbox{.}}{2015}]%
        {rush2015neural}
\bibfield{author}{\bibinfo{person}{Alexander~M Rush}, \bibinfo{person}{Sumit
  Chopra}, {and} \bibinfo{person}{Jason Weston}.}
  \bibinfo{year}{2015}\natexlab{}.
\newblock \showarticletitle{A neural attention model for abstractive sentence
  summarization}.
\newblock \bibinfo{journal}{\emph{arXiv preprint arXiv:1509.00685}}
  (\bibinfo{year}{2015}).
\newblock


\bibitem[\protect\citeauthoryear{Salimans and Kingma}{Salimans and
  Kingma}{2016}]%
        {salimans2016weight}
\bibfield{author}{\bibinfo{person}{Tim Salimans} {and}
  \bibinfo{person}{Diederik~P Kingma}.} \bibinfo{year}{2016}\natexlab{}.
\newblock \showarticletitle{Weight normalization: A simple reparameterization
  to accelerate training of deep neural networks}. In
  \bibinfo{booktitle}{\emph{Advances in Neural Information Processing
  Systems}}. \bibinfo{pages}{901--909}.
\newblock


\bibitem[\protect\citeauthoryear{Santos, Campbell, Patel, Hindle, and
  Amaral}{Santos et~al\mbox{.}}{2018}]%
        {santos2018syntax}
\bibfield{author}{\bibinfo{person}{Eddie~Antonio Santos},
  \bibinfo{person}{Joshua~Charles Campbell}, \bibinfo{person}{Dhvani Patel},
  \bibinfo{person}{Abram Hindle}, {and} \bibinfo{person}{Jos{\'e}~Nelson
  Amaral}.} \bibinfo{year}{2018}\natexlab{}.
\newblock \showarticletitle{Syntax and sensibility: Using language models to
  detect and correct syntax errors}. In \bibinfo{booktitle}{\emph{2018 IEEE
  25th International Conference on Software Analysis, Evolution and
  Reengineering (SANER)}}. IEEE, \bibinfo{pages}{311--322}.
\newblock


\bibitem[\protect\citeauthoryear{Schmidhuber}{Schmidhuber}{2004}]%
        {schmidhuber2004optimal}
\bibfield{author}{\bibinfo{person}{J{\"u}rgen Schmidhuber}.}
  \bibinfo{year}{2004}\natexlab{}.
\newblock \showarticletitle{Optimal ordered problem solver}.
\newblock \bibinfo{journal}{\emph{Machine Learning}} \bibinfo{volume}{54},
  \bibinfo{number}{3} (\bibinfo{year}{2004}), \bibinfo{pages}{211--254}.
\newblock


\bibitem[\protect\citeauthoryear{Schwenk and Gauvain}{Schwenk and
  Gauvain}{2002}]%
        {schwenk2002connectionist}
\bibfield{author}{\bibinfo{person}{Holger Schwenk} {and}
  \bibinfo{person}{Jean-Luc Gauvain}.} \bibinfo{year}{2002}\natexlab{}.
\newblock \showarticletitle{Connectionist language modeling for large
  vocabulary continuous speech recognition}. In
  \bibinfo{booktitle}{\emph{Acoustics, Speech, and Signal Processing (ICASSP),
  2002 IEEE International Conference on}}, Vol.~\bibinfo{volume}{1}. IEEE,
  \bibinfo{pages}{I--765}.
\newblock


\bibitem[\protect\citeauthoryear{See, Liu, and Manning}{See
  et~al\mbox{.}}{2017}]%
        {see2017get}
\bibfield{author}{\bibinfo{person}{Abigail See}, \bibinfo{person}{Peter~J Liu},
  {and} \bibinfo{person}{Christopher~D Manning}.}
  \bibinfo{year}{2017}\natexlab{}.
\newblock \showarticletitle{Get to the point: Summarization with
  pointer-generator networks}.
\newblock \bibinfo{journal}{\emph{arXiv preprint arXiv:1704.04368}}
  (\bibinfo{year}{2017}).
\newblock


\bibitem[\protect\citeauthoryear{Semeniuta, Severyn, and Barth}{Semeniuta
  et~al\mbox{.}}{2016}]%
        {semeniuta2016recurrent}
\bibfield{author}{\bibinfo{person}{Stanislau Semeniuta},
  \bibinfo{person}{Aliaksei Severyn}, {and} \bibinfo{person}{Erhardt Barth}.}
  \bibinfo{year}{2016}\natexlab{}.
\newblock \showarticletitle{Recurrent dropout without memory loss}.
\newblock \bibinfo{journal}{\emph{arXiv preprint arXiv:1603.05118}}
  (\bibinfo{year}{2016}).
\newblock


\bibitem[\protect\citeauthoryear{Settles}{Settles}{2010}]%
        {settles2010active}
\bibfield{author}{\bibinfo{person}{Burr Settles}.}
  \bibinfo{year}{2010}\natexlab{}.
\newblock \showarticletitle{Active learning literature survey}.
\newblock \bibinfo{journal}{\emph{University of Wisconsin, Madison}}
  \bibinfo{volume}{52}, \bibinfo{number}{55-66} (\bibinfo{year}{2010}),
  \bibinfo{pages}{11}.
\newblock


\bibitem[\protect\citeauthoryear{Sharma, Tian, and Lo}{Sharma
  et~al\mbox{.}}{2015}]%
        {sharma2015nirmal}
\bibfield{author}{\bibinfo{person}{Abhishek Sharma}, \bibinfo{person}{Yuan
  Tian}, {and} \bibinfo{person}{David Lo}.} \bibinfo{year}{2015}\natexlab{}.
\newblock \showarticletitle{Nirmal: Automatic identification of software
  relevant tweets leveraging language model}. In
  \bibinfo{booktitle}{\emph{Software Analysis, Evolution and Reengineering
  (SANER), 2015 IEEE 22nd International Conference on}}. IEEE,
  \bibinfo{pages}{449--458}.
\newblock


\bibitem[\protect\citeauthoryear{Shen, Lin, Liu, Shen, and Reid}{Shen
  et~al\mbox{.}}{2017}]%
        {shen2017weakly}
\bibfield{author}{\bibinfo{person}{Tong Shen}, \bibinfo{person}{Guosheng Lin},
  \bibinfo{person}{Lingqiao Liu}, \bibinfo{person}{Chunhua Shen}, {and}
  \bibinfo{person}{Ian Reid}.} \bibinfo{year}{2017}\natexlab{}.
\newblock \showarticletitle{Weakly Supervised Semantic Segmentation Based on
  Web Image Co-segmentation}.
\newblock \bibinfo{journal}{\emph{arXiv preprint arXiv:1705.09052}}
  (\bibinfo{year}{2017}).
\newblock


\bibitem[\protect\citeauthoryear{Siegelmann and Sontag}{Siegelmann and
  Sontag}{1992}]%
        {siegelmann1992computational}
\bibfield{author}{\bibinfo{person}{Hava~T Siegelmann} {and}
  \bibinfo{person}{Eduardo~D Sontag}.} \bibinfo{year}{1992}\natexlab{}.
\newblock \showarticletitle{On the computational power of neural nets}. In
  \bibinfo{booktitle}{\emph{Proceedings of the fifth annual workshop on
  Computational learning theory}}. ACM, \bibinfo{pages}{440--449}.
\newblock


\bibitem[\protect\citeauthoryear{Skolka, Staicu, and Pradel}{Skolka
  et~al\mbox{.}}{2019}]%
        {skolka2019anything}
\bibfield{author}{\bibinfo{person}{Philippe Skolka},
  \bibinfo{person}{Cristian-Alexandru Staicu}, {and} \bibinfo{person}{Michael
  Pradel}.} \bibinfo{year}{2019}\natexlab{}.
\newblock \showarticletitle{Anything to Hide? Studying Minified and Obfuscated
  Code in the Web}. In \bibinfo{booktitle}{\emph{The World Wide Web
  Conference}}. ACM, \bibinfo{pages}{1735--1746}.
\newblock


\bibitem[\protect\citeauthoryear{Sodsong, Scholz, and Chawla}{Sodsong
  et~al\mbox{.}}{2017}]%
        {sodsong2017spark}
\bibfield{author}{\bibinfo{person}{Wasuwee Sodsong}, \bibinfo{person}{Bernhard
  Scholz}, {and} \bibinfo{person}{Sanjay Chawla}.}
  \bibinfo{year}{2017}\natexlab{}.
\newblock \showarticletitle{SPARK: static program analysis reasoning and
  retrieving knowledge}.
\newblock \bibinfo{journal}{\emph{arXiv preprint arXiv:1711.01024}}
  (\bibinfo{year}{2017}).
\newblock


\bibitem[\protect\citeauthoryear{Solar-Lezama}{Solar-Lezama}{2013}]%
        {solar2013program}
\bibfield{author}{\bibinfo{person}{Armando Solar-Lezama}.}
  \bibinfo{year}{2013}\natexlab{}.
\newblock \showarticletitle{Program sketching}.
\newblock \bibinfo{journal}{\emph{International Journal on Software Tools for
  Technology Transfer}} \bibinfo{volume}{15}, \bibinfo{number}{5-6}
  (\bibinfo{year}{2013}), \bibinfo{pages}{475--495}.
\newblock


\bibitem[\protect\citeauthoryear{Srivastava, Hinton, Krizhevsky, Sutskever, and
  Salakhutdinov}{Srivastava et~al\mbox{.}}{2014}]%
        {srivastava2014dropout}
\bibfield{author}{\bibinfo{person}{Nitish Srivastava},
  \bibinfo{person}{Geoffrey Hinton}, \bibinfo{person}{Alex Krizhevsky},
  \bibinfo{person}{Ilya Sutskever}, {and} \bibinfo{person}{Ruslan
  Salakhutdinov}.} \bibinfo{year}{2014}\natexlab{}.
\newblock \showarticletitle{Dropout: A simple way to prevent neural networks
  from overfitting}.
\newblock \bibinfo{journal}{\emph{The Journal of Machine Learning Research}}
  \bibinfo{volume}{15}, \bibinfo{number}{1} (\bibinfo{year}{2014}),
  \bibinfo{pages}{1929--1958}.
\newblock


\bibitem[\protect\citeauthoryear{Srivastava, Greff, and Schmidhuber}{Srivastava
  et~al\mbox{.}}{2015}]%
        {srivastava2015highway}
\bibfield{author}{\bibinfo{person}{Rupesh~Kumar Srivastava},
  \bibinfo{person}{Klaus Greff}, {and} \bibinfo{person}{J{\"u}rgen
  Schmidhuber}.} \bibinfo{year}{2015}\natexlab{}.
\newblock \showarticletitle{Highway networks}.
\newblock \bibinfo{journal}{\emph{arXiv preprint arXiv:1505.00387}}
  (\bibinfo{year}{2015}).
\newblock


\bibitem[\protect\citeauthoryear{Sukhbaatar, Weston, Fergus,
  et~al\mbox{.}}{Sukhbaatar et~al\mbox{.}}{2015}]%
        {sukhbaatar2015end}
\bibfield{author}{\bibinfo{person}{Sainbayar Sukhbaatar},
  \bibinfo{person}{Jason Weston}, \bibinfo{person}{Rob Fergus},
  {et~al\mbox{.}}} \bibinfo{year}{2015}\natexlab{}.
\newblock \showarticletitle{End-to-end memory networks}. In
  \bibinfo{booktitle}{\emph{Advances in neural information processing
  systems}}. \bibinfo{pages}{2440--2448}.
\newblock


\bibitem[\protect\citeauthoryear{Sun, Zhu, Mou, Xiong, Li, and Zhang}{Sun
  et~al\mbox{.}}{2019}]%
        {sun2019grammar}
\bibfield{author}{\bibinfo{person}{Zeyu Sun}, \bibinfo{person}{Qihao Zhu},
  \bibinfo{person}{Lili Mou}, \bibinfo{person}{Yingfei Xiong},
  \bibinfo{person}{Ge Li}, {and} \bibinfo{person}{Lu Zhang}.}
  \bibinfo{year}{2019}\natexlab{}.
\newblock \showarticletitle{A grammar-based structural cnn decoder for code
  generation}. In \bibinfo{booktitle}{\emph{Proceedings of the AAAI Conference
  on Artificial Intelligence}}, Vol.~\bibinfo{volume}{33}.
  \bibinfo{pages}{7055--7062}.
\newblock


\bibitem[\protect\citeauthoryear{Sutskever, Hinton, and Taylor}{Sutskever
  et~al\mbox{.}}{2009}]%
        {sutskever2009recurrent}
\bibfield{author}{\bibinfo{person}{Ilya Sutskever}, \bibinfo{person}{Geoffrey~E
  Hinton}, {and} \bibinfo{person}{Graham~W Taylor}.}
  \bibinfo{year}{2009}\natexlab{}.
\newblock \showarticletitle{The recurrent temporal restricted boltzmann
  machine}. In \bibinfo{booktitle}{\emph{Advances in Neural Information
  Processing Systems}}. \bibinfo{pages}{1601--1608}.
\newblock


\bibitem[\protect\citeauthoryear{Sutskever, Vinyals, and Le}{Sutskever
  et~al\mbox{.}}{2014}]%
        {sutskever2014sequence}
\bibfield{author}{\bibinfo{person}{Ilya Sutskever}, \bibinfo{person}{Oriol
  Vinyals}, {and} \bibinfo{person}{Quoc~V Le}.}
  \bibinfo{year}{2014}\natexlab{}.
\newblock \showarticletitle{Sequence to sequence learning with neural
  networks}. In \bibinfo{booktitle}{\emph{Advances in neural information
  processing systems}}. \bibinfo{pages}{3104--3112}.
\newblock


\bibitem[\protect\citeauthoryear{Svajlenko, Islam, Keivanloo, Roy, and
  Mia}{Svajlenko et~al\mbox{.}}{2014}]%
        {svajlenko2014towards}
\bibfield{author}{\bibinfo{person}{Jeffrey Svajlenko},
  \bibinfo{person}{Judith~F Islam}, \bibinfo{person}{Iman Keivanloo},
  \bibinfo{person}{Chanchal~K Roy}, {and} \bibinfo{person}{Mohammad~Mamun
  Mia}.} \bibinfo{year}{2014}\natexlab{}.
\newblock \showarticletitle{Towards a big data curated benchmark of
  inter-project code clones}. In \bibinfo{booktitle}{\emph{2014 IEEE
  International Conference on Software Maintenance and Evolution (ICSME)}}.
  IEEE, \bibinfo{pages}{476--480}.
\newblock


\bibitem[\protect\citeauthoryear{Tarvainen and Valpola}{Tarvainen and
  Valpola}{2017}]%
        {tarvainen2017mean}
\bibfield{author}{\bibinfo{person}{Antti Tarvainen} {and}
  \bibinfo{person}{Harri Valpola}.} \bibinfo{year}{2017}\natexlab{}.
\newblock \showarticletitle{Mean teachers are better role models:
  Weight-averaged consistency targets improve semi-supervised deep learning
  results}.
\newblock In \bibinfo{booktitle}{\emph{Advances in Neural Information
  Processing Systems 30}}, \bibfield{editor}{\bibinfo{person}{I.~Guyon},
  \bibinfo{person}{U.~V. Luxburg}, \bibinfo{person}{S.~Bengio},
  \bibinfo{person}{H.~Wallach}, \bibinfo{person}{R.~Fergus},
  \bibinfo{person}{S.~Vishwanathan}, {and} \bibinfo{person}{R.~Garnett}}
  (Eds.). \bibinfo{publisher}{Curran Associates, Inc.},
  \bibinfo{pages}{1195--1204}.
\newblock


\bibitem[\protect\citeauthoryear{Tieleman and Hinton}{Tieleman and
  Hinton}{2012}]%
        {tieleman2012lecture}
\bibfield{author}{\bibinfo{person}{Tijmen Tieleman} {and}
  \bibinfo{person}{Geoffrey Hinton}.} \bibinfo{year}{2012}\natexlab{}.
\newblock \showarticletitle{Lecture 6.5-rmsprop: Divide the gradient by a
  running average of its recent magnitude}.
\newblock \bibinfo{journal}{\emph{COURSERA: Neural networks for machine
  learning}} \bibinfo{volume}{4}, \bibinfo{number}{2} (\bibinfo{year}{2012}),
  \bibinfo{pages}{26--31}.
\newblock


\bibitem[\protect\citeauthoryear{Tomassi, Dmeiri, Wang, Bhowmick, Liu, Devanbu,
  Vasilescu, and Rubio-Gonz{\'a}lez}{Tomassi et~al\mbox{.}}{2019}]%
        {tomassi2019bugswarm}
\bibfield{author}{\bibinfo{person}{David~A Tomassi}, \bibinfo{person}{Naji
  Dmeiri}, \bibinfo{person}{Yichen Wang}, \bibinfo{person}{Antara Bhowmick},
  \bibinfo{person}{Yen-Chuan Liu}, \bibinfo{person}{Premkumar~T Devanbu},
  \bibinfo{person}{Bogdan Vasilescu}, {and} \bibinfo{person}{Cindy
  Rubio-Gonz{\'a}lez}.} \bibinfo{year}{2019}\natexlab{}.
\newblock \showarticletitle{Bugswarm: mining and continuously growing a dataset
  of reproducible failures and fixes}. In \bibinfo{booktitle}{\emph{2019
  IEEE/ACM 41st International Conference on Software Engineering (ICSE)}}.
  IEEE, \bibinfo{pages}{339--349}.
\newblock


\bibitem[\protect\citeauthoryear{Tucker, Mnih, Maddison, Lawson, and
  Sohl-Dickstein}{Tucker et~al\mbox{.}}{2017}]%
        {tucker2017rebar}
\bibfield{author}{\bibinfo{person}{George Tucker}, \bibinfo{person}{Andriy
  Mnih}, \bibinfo{person}{Chris~J Maddison}, \bibinfo{person}{John Lawson},
  {and} \bibinfo{person}{Jascha Sohl-Dickstein}.}
  \bibinfo{year}{2017}\natexlab{}.
\newblock \showarticletitle{REBAR: Low-variance, unbiased gradient estimates
  for discrete latent variable models}. In \bibinfo{booktitle}{\emph{Advances
  in Neural Information Processing Systems}}. \bibinfo{pages}{2624--2633}.
\newblock


\bibitem[\protect\citeauthoryear{van~den Oord, Kalchbrenner, Espeholt, Vinyals,
  Graves, et~al\mbox{.}}{van~den Oord et~al\mbox{.}}{2016}]%
        {van2016conditional}
\bibfield{author}{\bibinfo{person}{Aaron van~den Oord}, \bibinfo{person}{Nal
  Kalchbrenner}, \bibinfo{person}{Lasse Espeholt}, \bibinfo{person}{Oriol
  Vinyals}, \bibinfo{person}{Alex Graves}, {et~al\mbox{.}}}
  \bibinfo{year}{2016}\natexlab{}.
\newblock \showarticletitle{Conditional image generation with pixelcnn
  decoders}. In \bibinfo{booktitle}{\emph{Advances in Neural Information
  Processing Systems}}. \bibinfo{pages}{4790--4798}.
\newblock


\bibitem[\protect\citeauthoryear{Vasilescu, Casalnuovo, and Devanbu}{Vasilescu
  et~al\mbox{.}}{2017}]%
        {vasilescu2017recovering}
\bibfield{author}{\bibinfo{person}{Bogdan Vasilescu}, \bibinfo{person}{Casey
  Casalnuovo}, {and} \bibinfo{person}{Premkumar Devanbu}.}
  \bibinfo{year}{2017}\natexlab{}.
\newblock \showarticletitle{Recovering clear, natural identifiers from
  obfuscated JS names}. In \bibinfo{booktitle}{\emph{Proceedings of the 2017
  11th Joint Meeting on Foundations of Software Engineering}}. ACM,
  \bibinfo{pages}{683--693}.
\newblock


\bibitem[\protect\citeauthoryear{Vaswani, Shazeer, Parmar, Uszkoreit, Jones,
  Gomez, Kaiser, and Polosukhin}{Vaswani et~al\mbox{.}}{2017}]%
        {vaswani2017attention}
\bibfield{author}{\bibinfo{person}{Ashish Vaswani}, \bibinfo{person}{Noam
  Shazeer}, \bibinfo{person}{Niki Parmar}, \bibinfo{person}{Jakob Uszkoreit},
  \bibinfo{person}{Llion Jones}, \bibinfo{person}{Aidan~N. Gomez},
  \bibinfo{person}{Lukasz Kaiser}, {and} \bibinfo{person}{Illia Polosukhin}.}
  \bibinfo{year}{2017}\natexlab{}.
\newblock \showarticletitle{Attention is All You Need}.
\newblock \bibinfo{journal}{\emph{arXiv preprint arXiv:1706.03762}}
  (\bibinfo{year}{2017}).
\newblock


\bibitem[\protect\citeauthoryear{Vinyals, Fortunato, and Jaitly}{Vinyals
  et~al\mbox{.}}{2015}]%
        {vinyals2015pointer}
\bibfield{author}{\bibinfo{person}{Oriol Vinyals}, \bibinfo{person}{Meire
  Fortunato}, {and} \bibinfo{person}{Navdeep Jaitly}.}
  \bibinfo{year}{2015}\natexlab{}.
\newblock \showarticletitle{Pointer networks}. In
  \bibinfo{booktitle}{\emph{Advances in Neural Information Processing
  Systems}}. \bibinfo{pages}{2692--2700}.
\newblock


\bibitem[\protect\citeauthoryear{Waibel, Hanazawa, Hinton, Shikano, and
  Lang}{Waibel et~al\mbox{.}}{1988}]%
        {waibel1988phoneme}
\bibfield{author}{\bibinfo{person}{Alexander Waibel},
  \bibinfo{person}{Toshiyuki Hanazawa}, \bibinfo{person}{Geoffrey Hinton},
  \bibinfo{person}{Kiyohiro Shikano}, {and} \bibinfo{person}{K Lang}.}
  \bibinfo{year}{1988}\natexlab{}.
\newblock \showarticletitle{Phoneme recognition: neural networks vs. hidden
  Markov models vs. hidden Markov models}. In
  \bibinfo{booktitle}{\emph{Acoustics, Speech, and Signal Processing, 1988.
  ICASSP-88., 1988 International Conference on}}. IEEE,
  \bibinfo{pages}{107--110}.
\newblock


\bibitem[\protect\citeauthoryear{Wan, Zeiler, Zhang, Le~Cun, and Fergus}{Wan
  et~al\mbox{.}}{2013}]%
        {wan2013regularization}
\bibfield{author}{\bibinfo{person}{Li Wan}, \bibinfo{person}{Matthew Zeiler},
  \bibinfo{person}{Sixin Zhang}, \bibinfo{person}{Yann Le~Cun}, {and}
  \bibinfo{person}{Rob Fergus}.} \bibinfo{year}{2013}\natexlab{}.
\newblock \showarticletitle{Regularization of neural networks using
  dropconnect}. In \bibinfo{booktitle}{\emph{International Conference on
  Machine Learning}}. \bibinfo{pages}{1058--1066}.
\newblock


\bibitem[\protect\citeauthoryear{Wan, Zhao, Yang, Xu, Ying, Wu, and Yu}{Wan
  et~al\mbox{.}}{2018}]%
        {wan2018improving}
\bibfield{author}{\bibinfo{person}{Yao Wan}, \bibinfo{person}{Zhou Zhao},
  \bibinfo{person}{Min Yang}, \bibinfo{person}{Guandong Xu},
  \bibinfo{person}{Haochao Ying}, \bibinfo{person}{Jian Wu}, {and}
  \bibinfo{person}{Philip~S Yu}.} \bibinfo{year}{2018}\natexlab{}.
\newblock \showarticletitle{Improving automatic source code summarization via
  deep reinforcement learning}. In \bibinfo{booktitle}{\emph{Proceedings of the
  33rd ACM/IEEE International Conference on Automated Software Engineering}}.
  ACM, \bibinfo{pages}{397--407}.
\newblock


\bibitem[\protect\citeauthoryear{Wang, Lo, and Lawall}{Wang
  et~al\mbox{.}}{2014}]%
        {wang2014compositional}
\bibfield{author}{\bibinfo{person}{Shaowei Wang}, \bibinfo{person}{David Lo},
  {and} \bibinfo{person}{Julia Lawall}.} \bibinfo{year}{2014}\natexlab{}.
\newblock \showarticletitle{Compositional vector space models for improved bug
  localization}. In \bibinfo{booktitle}{\emph{Software Maintenance and
  Evolution (ICSME), 2014 IEEE International Conference on}}. IEEE,
  \bibinfo{pages}{171--180}.
\newblock


\bibitem[\protect\citeauthoryear{Wang, Cai, and Wei}{Wang
  et~al\mbox{.}}{2016}]%
        {wang2016deep}
\bibfield{author}{\bibinfo{person}{Yao Wang}, \bibinfo{person}{Wan-dong Cai},
  {and} \bibinfo{person}{Peng-cheng Wei}.} \bibinfo{year}{2016}\natexlab{}.
\newblock \showarticletitle{A deep learning approach for detecting malicious
  JavaScript code}.
\newblock \bibinfo{journal}{\emph{security and communication networks}}
  \bibinfo{volume}{9}, \bibinfo{number}{11} (\bibinfo{year}{2016}),
  \bibinfo{pages}{1520--1534}.
\newblock


\bibitem[\protect\citeauthoryear{Wei and Li}{Wei and Li}{2017}]%
        {wei2017supervised}
\bibfield{author}{\bibinfo{person}{Huihui Wei} {and} \bibinfo{person}{Ming
  Li}.} \bibinfo{year}{2017}\natexlab{}.
\newblock \showarticletitle{Supervised Deep Features for Software Functional
  Clone Detection by Exploiting Lexical and Syntactical Information in Source
  Code.}. In \bibinfo{booktitle}{\emph{IJCAI}}. \bibinfo{pages}{3034--3040}.
\newblock


\bibitem[\protect\citeauthoryear{Weston, Bordes, Chopra, Rush, van
  Merri{\"e}nboer, Joulin, and Mikolov}{Weston et~al\mbox{.}}{2015}]%
        {weston2015towards}
\bibfield{author}{\bibinfo{person}{Jason Weston}, \bibinfo{person}{Antoine
  Bordes}, \bibinfo{person}{Sumit Chopra}, \bibinfo{person}{Alexander~M Rush},
  \bibinfo{person}{Bart van Merri{\"e}nboer}, \bibinfo{person}{Armand Joulin},
  {and} \bibinfo{person}{Tomas Mikolov}.} \bibinfo{year}{2015}\natexlab{}.
\newblock \showarticletitle{Towards ai-complete question answering: A set of
  prerequisite toy tasks}.
\newblock \bibinfo{journal}{\emph{arXiv preprint arXiv:1502.05698}}
  (\bibinfo{year}{2015}).
\newblock


\bibitem[\protect\citeauthoryear{Weston, Chopra, and Bordes}{Weston
  et~al\mbox{.}}{2014}]%
        {weston2014memory}
\bibfield{author}{\bibinfo{person}{Jason Weston}, \bibinfo{person}{Sumit
  Chopra}, {and} \bibinfo{person}{Antoine Bordes}.}
  \bibinfo{year}{2014}\natexlab{}.
\newblock \showarticletitle{Memory networks}.
\newblock \bibinfo{journal}{\emph{arXiv preprint arXiv:1410.3916}}
  (\bibinfo{year}{2014}).
\newblock


\bibitem[\protect\citeauthoryear{Weston}{Weston}{2016}]%
        {weston2016dialog}
\bibfield{author}{\bibinfo{person}{Jason~E Weston}.}
  \bibinfo{year}{2016}\natexlab{}.
\newblock \showarticletitle{Dialog-based language learning}. In
  \bibinfo{booktitle}{\emph{Advances in Neural Information Processing
  Systems}}. \bibinfo{pages}{829--837}.
\newblock


\bibitem[\protect\citeauthoryear{White, Tufano, Vendome, and Poshyvanyk}{White
  et~al\mbox{.}}{2016}]%
        {white2016deep}
\bibfield{author}{\bibinfo{person}{Martin White}, \bibinfo{person}{Michele
  Tufano}, \bibinfo{person}{Christopher Vendome}, {and} \bibinfo{person}{Denys
  Poshyvanyk}.} \bibinfo{year}{2016}\natexlab{}.
\newblock \showarticletitle{Deep learning code fragments for code clone
  detection}. In \bibinfo{booktitle}{\emph{Proceedings of the 31st IEEE/ACM
  International Conference on Automated Software Engineering}}. ACM,
  \bibinfo{pages}{87--98}.
\newblock


\bibitem[\protect\citeauthoryear{White, Vendome, Linares-V\'{a}squez, and
  Poshyvanyk}{White et~al\mbox{.}}{2015}]%
        {white2015toward}
\bibfield{author}{\bibinfo{person}{Martin White}, \bibinfo{person}{Christopher
  Vendome}, \bibinfo{person}{Mario Linares-V\'{a}squez}, {and}
  \bibinfo{person}{Denys Poshyvanyk}.} \bibinfo{year}{2015}\natexlab{}.
\newblock \showarticletitle{Toward Deep Learning Software Repositories}. In
  \bibinfo{booktitle}{\emph{Proceedings of the 12th Working Conference on
  Mining Software Repositories}} \emph{(\bibinfo{series}{MSR '15})}.
  \bibinfo{publisher}{IEEE Press}, \bibinfo{address}{Piscataway, NJ, USA},
  \bibinfo{pages}{334--345}.
\newblock


\bibitem[\protect\citeauthoryear{Williams}{Williams}{1992}]%
        {williams1992simple}
\bibfield{author}{\bibinfo{person}{Ronald~J Williams}.}
  \bibinfo{year}{1992}\natexlab{}.
\newblock \showarticletitle{Simple statistical gradient-following algorithms
  for connectionist reinforcement learning}.
\newblock \bibinfo{journal}{\emph{Machine learning}} \bibinfo{volume}{8},
  \bibinfo{number}{3-4} (\bibinfo{year}{1992}), \bibinfo{pages}{229--256}.
\newblock


\bibitem[\protect\citeauthoryear{Wu, Schuster, Chen, Le, Norouzi, Macherey,
  Krikun, Cao, Gao, Macherey, et~al\mbox{.}}{Wu et~al\mbox{.}}{2016}]%
        {wu2016google}
\bibfield{author}{\bibinfo{person}{Yonghui Wu}, \bibinfo{person}{Mike
  Schuster}, \bibinfo{person}{Zhifeng Chen}, \bibinfo{person}{Quoc~V Le},
  \bibinfo{person}{Mohammad Norouzi}, \bibinfo{person}{Wolfgang Macherey},
  \bibinfo{person}{Maxim Krikun}, \bibinfo{person}{Yuan Cao},
  \bibinfo{person}{Qin Gao}, \bibinfo{person}{Klaus Macherey}, {et~al\mbox{.}}}
  \bibinfo{year}{2016}\natexlab{}.
\newblock \showarticletitle{Google's neural machine translation system:
  Bridging the gap between human and machine translation}.
\newblock \bibinfo{journal}{\emph{arXiv preprint arXiv:1609.08144}}
  (\bibinfo{year}{2016}).
\newblock


\bibitem[\protect\citeauthoryear{Xia, Lo, Wang, and Yang}{Xia
  et~al\mbox{.}}{2015}]%
        {xia2015should}
\bibfield{author}{\bibinfo{person}{Xin Xia}, \bibinfo{person}{David Lo},
  \bibinfo{person}{Xinyu Wang}, {and} \bibinfo{person}{Xiaohu Yang}.}
  \bibinfo{year}{2015}\natexlab{}.
\newblock \showarticletitle{Who should review this change?: Putting text and
  file location analyses together for more accurate recommendations}. In
  \bibinfo{booktitle}{\emph{Software Maintenance and Evolution (ICSME), 2015
  IEEE International Conference on}}. IEEE, \bibinfo{pages}{261--270}.
\newblock


\bibitem[\protect\citeauthoryear{Xiao, Keung, Bennin, and Mi}{Xiao
  et~al\mbox{.}}{2019}]%
        {xiao2019improving}
\bibfield{author}{\bibinfo{person}{Yan Xiao}, \bibinfo{person}{Jacky Keung},
  \bibinfo{person}{Kwabena~E Bennin}, {and} \bibinfo{person}{Qing Mi}.}
  \bibinfo{year}{2019}\natexlab{}.
\newblock \showarticletitle{Improving bug localization with word embedding and
  enhanced convolutional neural networks}.
\newblock \bibinfo{journal}{\emph{Information and Software Technology}}
  \bibinfo{volume}{105} (\bibinfo{year}{2019}), \bibinfo{pages}{17--29}.
\newblock


\bibitem[\protect\citeauthoryear{Xiao, Keung, Mi, and Bennin}{Xiao
  et~al\mbox{.}}{2017}]%
        {xiao2017improving}
\bibfield{author}{\bibinfo{person}{Yan Xiao}, \bibinfo{person}{Jacky Keung},
  \bibinfo{person}{Qing Mi}, {and} \bibinfo{person}{Kwabena~E Bennin}.}
  \bibinfo{year}{2017}\natexlab{}.
\newblock \showarticletitle{Improving bug localization with an enhanced
  convolutional neural network}. In \bibinfo{booktitle}{\emph{2017 24th
  Asia-Pacific Software Engineering Conference (APSEC)}}. IEEE,
  \bibinfo{pages}{338--347}.
\newblock


\bibitem[\protect\citeauthoryear{Xu, Ba, Kiros, Cho, Courville, Salakhudinov,
  Zemel, and Bengio}{Xu et~al\mbox{.}}{2015}]%
        {xu2015show}
\bibfield{author}{\bibinfo{person}{Kelvin Xu}, \bibinfo{person}{Jimmy Ba},
  \bibinfo{person}{Ryan Kiros}, \bibinfo{person}{Kyunghyun Cho},
  \bibinfo{person}{Aaron Courville}, \bibinfo{person}{Ruslan Salakhudinov},
  \bibinfo{person}{Rich Zemel}, {and} \bibinfo{person}{Yoshua Bengio}.}
  \bibinfo{year}{2015}\natexlab{}.
\newblock \showarticletitle{Show, attend and tell: Neural image caption
  generation with visual attention}. In \bibinfo{booktitle}{\emph{International
  Conference on Machine Learning}}. \bibinfo{pages}{2048--2057}.
\newblock


\bibitem[\protect\citeauthoryear{Xu, Liu, and Song}{Xu et~al\mbox{.}}{2017}]%
        {xu2017sqlnet}
\bibfield{author}{\bibinfo{person}{Xiaojun Xu}, \bibinfo{person}{Chang Liu},
  {and} \bibinfo{person}{Dawn Song}.} \bibinfo{year}{2017}\natexlab{}.
\newblock \showarticletitle{SQLNet: Generating Structured Queries From Natural
  Language Without Reinforcement Learning}.
\newblock \bibinfo{journal}{\emph{arXiv preprint arXiv:1711.04436}}
  (\bibinfo{year}{2017}).
\newblock


\bibitem[\protect\citeauthoryear{Yadid and Yahav}{Yadid and Yahav}{2016}]%
        {yadid2016extracting}
\bibfield{author}{\bibinfo{person}{Shir Yadid} {and} \bibinfo{person}{Eran
  Yahav}.} \bibinfo{year}{2016}\natexlab{}.
\newblock \showarticletitle{Extracting code from programming tutorial videos}.
  In \bibinfo{booktitle}{\emph{Proceedings of the 2016 ACM International
  Symposium on New Ideas, New Paradigms, and Reflections on Programming and
  Software}}. ACM, \bibinfo{pages}{98--111}.
\newblock


\bibitem[\protect\citeauthoryear{Yang, Nie, Cohen, and Lao}{Yang
  et~al\mbox{.}}{2017}]%
        {yang2017learning}
\bibfield{author}{\bibinfo{person}{Fan Yang}, \bibinfo{person}{Jiazhong Nie},
  \bibinfo{person}{William~W. Cohen}, {and} \bibinfo{person}{Ni Lao}.}
  \bibinfo{year}{2017}\natexlab{}.
\newblock \showarticletitle{Learning to Organize Knowledge with N-Gram
  Machines}.
\newblock \bibinfo{journal}{\emph{arXiv preprint arXiv:1711.06744}}
  (\bibinfo{year}{2017}).
\newblock


\bibitem[\protect\citeauthoryear{Yang}{Yang}{2016}]%
        {yang2016lie}
\bibfield{author}{\bibinfo{person}{Greg Yang}.}
  \bibinfo{year}{2016}\natexlab{}.
\newblock \showarticletitle{Lie Access Neural Turing Machine}.
\newblock \bibinfo{journal}{\emph{arXiv preprint arXiv:1602.08671}}
  (\bibinfo{year}{2016}).
\newblock


\bibitem[\protect\citeauthoryear{Ye, Xing, Foo, Li, and Kapre}{Ye
  et~al\mbox{.}}{2016}]%
        {ye2016learning}
\bibfield{author}{\bibinfo{person}{Deheng Ye}, \bibinfo{person}{Zhenchang
  Xing}, \bibinfo{person}{Chee~Yong Foo}, \bibinfo{person}{Jing Li}, {and}
  \bibinfo{person}{Nachiket Kapre}.} \bibinfo{year}{2016}\natexlab{}.
\newblock \showarticletitle{Learning to extract api mentions from informal
  natural language discussions}. In \bibinfo{booktitle}{\emph{Software
  Maintenance and Evolution (ICSME), 2016 IEEE International Conference on}}.
  IEEE, \bibinfo{pages}{389--399}.
\newblock


\bibitem[\protect\citeauthoryear{Yin, Deng, Chen, Vasilescu, and Neubig}{Yin
  et~al\mbox{.}}{2018}]%
        {yin2018learning}
\bibfield{author}{\bibinfo{person}{Pengcheng Yin}, \bibinfo{person}{Bowen
  Deng}, \bibinfo{person}{Edgar Chen}, \bibinfo{person}{Bogdan Vasilescu},
  {and} \bibinfo{person}{Graham Neubig}.} \bibinfo{year}{2018}\natexlab{}.
\newblock \showarticletitle{Learning to Mine Aligned Code and Natural Language
  Pairs from Stack Overflow}.
\newblock \bibinfo{journal}{\emph{arXiv preprint arXiv:1805.08949}}
  (\bibinfo{year}{2018}).
\newblock


\bibitem[\protect\citeauthoryear{Yin and Neubig}{Yin and Neubig}{2017}]%
        {yin2017syntactic}
\bibfield{author}{\bibinfo{person}{Pengcheng Yin} {and} \bibinfo{person}{Graham
  Neubig}.} \bibinfo{year}{2017}\natexlab{}.
\newblock \showarticletitle{A Syntactic Neural Model for General-Purpose Code
  Generation}.
\newblock \bibinfo{journal}{\emph{arXiv preprint arXiv:1704.01696}}
  (\bibinfo{year}{2017}).
\newblock


\bibitem[\protect\citeauthoryear{Zaremba and Sutskever}{Zaremba and
  Sutskever}{2015}]%
        {zaremba2015reinforcement}
\bibfield{author}{\bibinfo{person}{Wojciech Zaremba} {and}
  \bibinfo{person}{Ilya Sutskever}.} \bibinfo{year}{2015}\natexlab{}.
\newblock \showarticletitle{Reinforcement Learning Neural Turing
  Machines-Revised}.
\newblock \bibinfo{journal}{\emph{arXiv preprint arXiv:1505.00521}}
  (\bibinfo{year}{2015}).
\newblock


\bibitem[\protect\citeauthoryear{Zaremba, Sutskever, and Vinyals}{Zaremba
  et~al\mbox{.}}{2014}]%
        {zaremba2014recurrent}
\bibfield{author}{\bibinfo{person}{Wojciech Zaremba}, \bibinfo{person}{Ilya
  Sutskever}, {and} \bibinfo{person}{Oriol Vinyals}.}
  \bibinfo{year}{2014}\natexlab{}.
\newblock \showarticletitle{Recurrent neural network regularization}.
\newblock \bibinfo{journal}{\emph{arXiv preprint arXiv:1409.2329}}
  (\bibinfo{year}{2014}).
\newblock


\bibitem[\protect\citeauthoryear{Zhang, Wang, Zhang, Sun, Wang, and Liu}{Zhang
  et~al\mbox{.}}{2019}]%
        {zhang2019novel}
\bibfield{author}{\bibinfo{person}{Jian Zhang}, \bibinfo{person}{Xu Wang},
  \bibinfo{person}{Hongyu Zhang}, \bibinfo{person}{Hailong Sun},
  \bibinfo{person}{Kaixuan Wang}, {and} \bibinfo{person}{Xudong Liu}.}
  \bibinfo{year}{2019}\natexlab{}.
\newblock \showarticletitle{A novel neural source code representation based on
  abstract syntax tree}. In \bibinfo{booktitle}{\emph{Proceedings of the 41st
  International Conference on Software Engineering}}. IEEE Press,
  \bibinfo{pages}{783--794}.
\newblock


\bibitem[\protect\citeauthoryear{Zhang, Zhao, and LeCun}{Zhang
  et~al\mbox{.}}{2015}]%
        {zhang2015character}
\bibfield{author}{\bibinfo{person}{Xiang Zhang}, \bibinfo{person}{Junbo Zhao},
  {and} \bibinfo{person}{Yann LeCun}.} \bibinfo{year}{2015}\natexlab{}.
\newblock \showarticletitle{Character-level convolutional networks for text
  classification}. In \bibinfo{booktitle}{\emph{Advances in neural information
  processing systems}}. \bibinfo{pages}{649--657}.
\newblock


\bibitem[\protect\citeauthoryear{Zhong, Xiong, and Socher}{Zhong
  et~al\mbox{.}}{2017}]%
        {zhong2017seq2sql}
\bibfield{author}{\bibinfo{person}{Victor Zhong}, \bibinfo{person}{Caiming
  Xiong}, {and} \bibinfo{person}{Richard Socher}.}
  \bibinfo{year}{2017}\natexlab{}.
\newblock \showarticletitle{Seq2SQL: Generating Structured Queries from Natural
  Language using Reinforcement Learning}.
\newblock \bibinfo{journal}{\emph{arXiv preprint arXiv:1709.00103}}
  (\bibinfo{year}{2017}).
\newblock


\bibitem[\protect\citeauthoryear{Zhou, Small, Rokhlenko, and Elkan}{Zhou
  et~al\mbox{.}}{2017}]%
        {zhou2017end}
\bibfield{author}{\bibinfo{person}{Li Zhou}, \bibinfo{person}{Kevin Small},
  \bibinfo{person}{Oleg Rokhlenko}, {and} \bibinfo{person}{Charles Elkan}.}
  \bibinfo{year}{2017}\natexlab{}.
\newblock \showarticletitle{End-to-End Offline Goal-Oriented Dialog Policy
  Learning via Policy Gradient}.
\newblock \bibinfo{journal}{\emph{arXiv preprint arXiv:1712.02838}}
  (\bibinfo{year}{2017}).
\newblock


\bibitem[\protect\citeauthoryear{Zhu, Park, Isola, and Efros}{Zhu
  et~al\mbox{.}}{2017}]%
        {zhu2017unpaired}
\bibfield{author}{\bibinfo{person}{Jun-Yan Zhu}, \bibinfo{person}{Taesung
  Park}, \bibinfo{person}{Phillip Isola}, {and} \bibinfo{person}{Alexei~A
  Efros}.} \bibinfo{year}{2017}\natexlab{}.
\newblock \showarticletitle{Unpaired Image-to-Image Translation using
  Cycle-Consistent Adversarial Networks}. In \bibinfo{booktitle}{\emph{Computer
  Vision (ICCV), 2017 IEEE International Conference on}}.
\newblock


\bibitem[\protect\citeauthoryear{Zilly, Srivastava, Koutn{\'\i}k, and
  Schmidhuber}{Zilly et~al\mbox{.}}{2016}]%
        {zilly2016recurrent}
\bibfield{author}{\bibinfo{person}{Julian~Georg Zilly},
  \bibinfo{person}{Rupesh~Kumar Srivastava}, \bibinfo{person}{Jan
  Koutn{\'\i}k}, {and} \bibinfo{person}{J{\"u}rgen Schmidhuber}.}
  \bibinfo{year}{2016}\natexlab{}.
\newblock \showarticletitle{Recurrent highway networks}.
\newblock \bibinfo{journal}{\emph{arXiv preprint arXiv:1607.03474}}
  (\bibinfo{year}{2016}).
\newblock


\end{thebibliography}

\end{document}